\documentclass[%
 reprint,
 amsmath,
 amssymb,
 aps,
 nofootinbib,
 prd
]{revtex4-2}

\usepackage{graphicx}
\usepackage{dcolumn}
\usepackage{bm}

\usepackage[T1]{fontenc}
\usepackage[utf8]{inputenc}
\usepackage{amsmath}
\usepackage{mathtools} 
\usepackage{cancel} 
\usepackage{MnSymbol} 
\usepackage{graphicx,adjustbox} 
\usepackage{mathrsfs} 
\usepackage{simpler-wick} 
\usepackage{comment}
\usepackage{autobreak}
\usepackage[noend]{algpseudocode}
\usepackage{relsize}
\allowdisplaybreaks
\numberwithin{equation}{section}

\usepackage{hyperref}
\hypersetup{
	colorlinks=true,
	linktoc=page,
	citecolor=americanrose,
	linkcolor=cadmiumgreen,
	urlcolor=blue} 
\usepackage{xcolor}

\definecolor{americanrose}{rgb}{1.0, 0.01, 0.24}
\definecolor{cadmiumgreen}{rgb}{0.0, 0.42, 0.24}

\makeatletter
\newlength{\apb@width}
\newcommand{\autoparbox}[2][c]{\settowidth{\apb@width}{#2}\parbox[#1]{\apb@width}{#2}}
\newcommand{\includegraphicsbox}[2][]{\autoparbox{\includegraphics[#1]{#2}}}

\makeatletter\g@addto@macro\bfseries{\boldmath}\makeatother

\newcommand{\dd}{{d}}
\newcommand{\hd}{\hat{{d}}}

\newcommand{\hdelta}{\hat{\delta}}

\newcommand{\bu}{\bar{u}}
\newcommand{\bp}{\bar{p}}
\newcommand{\bM}{\bar{m}}
\newcommand{\eps}{\epsilon}
\newcommand{\cD}{\mathcal{D}}
\newcommand{\cO}{\mathcal{O}}
\newcommand{\cW}{\mathcal{W}}
\newcommand{\cWt}{\widetilde{\mathcal{W}}}

\newcommand{\cM}{\mathcal{M}}
\newcommand{\cN}{\mathcal{N}}
\newcommand{\cC}{\mathcal{C}}

\newcommand{\Integrand}{T}
\newcommand{\expval}[1]{\langle #1 \rangle}

\newcommand{\Res}{\mathrm{Res}}
\newcommand{\Ord}{\mathcal{O}}

\newcommand{\ampl}{\mathcal{M}}
\newcommand{\normb}{\bm{\mathit{b}}}

\newcommand{\oldornew}[1]{}

\begin{document}

\title{\boldmath Analytic One-loop Scattering Waveform in General Relativity}

\author{Giacomo Brunello$^{a,b,c}$}
\email{giacomo.brunello@sns.it}
\author{Stefano De Angelis$^{a}$}
\email{stefano.de-angelis@ipht.fr}
\author{David A. Kosower$^{a}$}
\email{david.kosower@ipht.fr}

\affiliation{$^a$Institut de Physique Théorique, CEA, CNRS, Université Paris-Saclay, F-91191 Gif-sur-Yvette cedex, France}
\affiliation{$^b$ Dipartimento di Fisica e Astronomia, Universit\`a degli Studi di Padova and INFN Sezione di Padova, Via Marzolo 8, I-35131 Padova, Italy.}
\affiliation{$^c$Scuola Normale Superiore, Piazza dei Cavalieri 7, 56126, Pisa, Italy and INFN Sezione di Pisa, Largo
Pontecorvo 3, 56127 Pisa, Italy}

\date{\today}

\begin{abstract}
    Leveraging the computational framework presented in 
    reference~\cite{Brunello:2024ibk}, we evaluate the analytic 
    scattering waveform in General Relativity to second order, 
    ${G^3 M^3}/{r b^2}$ and to all orders in velocity. This new 
    representation of the next-to-leading order waveform is 
    well-suited for numerical evaluation. Integrating the 
    [modulus square of the] waveform over the angles on the celestial sphere, we also compute the power spectrum of the radiation to order $G^4$ numerically.
\end{abstract}

\maketitle

\section{Introduction}
The emerging precision era in gravitational-wave (GW) 
physics demands a deeper theoretical understanding of 
gravitational waveforms in order to produce more accurate 
templates for matched-filtering analyses. Although the 
so-called `Multipolar Post-Minkowskian'
(MPM) formalism has been successful in 
describing gravitational-wave emission from 
a variety of sources 
analytically~\cite{Blanchet:1985sp,Blanchet:1989ki,%
Blanchet:2013haa}, recent years' developments 
originating in quantum field theory
have offered a new approach and new tools for attacking
the relativistic two-body problem. 
These include approaches based on effective field theory 
(EFT)~\cite{Goldberger:2004jt}, quantum 
scattering amplitudes, and world-line formalisms.
The new approaches have driven progress in 
both the traditional post-Newtonian (PN) and ordinary
relativistic perturbative predictions.
In the PN approach, EFT 
methods~\cite{Goldberger:2004jt,Goldberger:2009qd,%
Foffa:2013qca,Porto:2016pyg}, combined with multi-loop 
techniques~\cite{Foffa:2016rgu}, have allowed calculations
to reach 
4$\,$PN~\cite{Foffa:2019yfl,Foffa:2019rdf,Blumlein:2020pog}
and even 5$\,$PN~ \cite{Foffa:2019hrb,Blumlein:2021txe,%
Porto:2024cwd}.
In relativistic perturbation theory, the connection with 
scattering amplitudes was understood long 
ago~\cite{Iwasaki:1971vb}, but only recently has a path
to applying scattering amplitudes to gravitational calculations
been suggested~\cite{Damour:2016gwp,Damour:2017zjx}
and carried out.  The calculation of the $\Ord(G^3)$ 
corrections to the gravitational Hamiltonian using scattering 
amplitudes~\cite{Cheung:2018wkq,Bjerrum-Bohr:2018xdl} was
the opening shot to a flood of research activity. 
These corrections in turn yielded the corrections
to the emitted gravitational energy.
The development of methods based on scattering amplitudes
made possible the calculation of the $\Ord(G^3)$ corrections
\cite{Bern:2019nnu,Bern:2019crd,Parra-Martinez:2020dzs,Cheung:2020gyp,DiVecchia:2021bdo,Brandhuber:2021eyq,Kalin:2020fhe}, methods based on a
world-line quantum field theory (WQFT) approach~\cite{Kalin:2020mvi,Mogull:2020sak}
then yielded the $\Ord(G^4)$ 
corrections~\cite{Dlapa:2021npj,Bern:2021dqo,Bern:2022jvn,Dlapa:2022lmu,%
Jakobsen:2023ndj,Jakobsen:2023hig,Damgaard:2023ttc}.
Most recently, several groups have computed the
$\Ord(G^5)$ corrections at the first self-force 
order~\cite{Bern:2023ccb,Driesse:2024xad,Bern:2024adl,%
Driesse:2024feo}. These corrections to the gravitational
Hamiltonian again have led to substantial improvement
in the perturbative predictions near merger as compared
with (expensive) predictions from numerical 
relativity~\cite{Damour:2022ybd,Rettegno:2023ghr}.
The world-line methods also allowed the direct computation
of the scattering waveform at leading order (LO) for spinless
particles~\cite{Jakobsen:2021smu,Mougiakakos:2021ckm}.
These methods made possible the computation of the waveform at tree-level for spinless particles within the worldline formalism  and including spin corrections \cite{Jakobsen:2021lvp,DeAngelis:2023lvf,Brandhuber:2023hhl,Aoude:2023dui}. 
An earlier formalism 
(observables-based, Kosower-Maybee-O'Connell \textit{aka\/} 
KMOC)~\cite{Kosower:2018adc,Cristofoli:2021vyo}
links the usual quantum scattering amplitudes to classical
observables.

In this article, we use the KMOC formalism to obtain an
analytic result for the scattering gravitational waveform in 
impact parameter space.  We compute the waveform arising from
the scattering of two celestial objects to next-to-leading order (NLO) ($\Ord(G^3)$).
The waveform in momentum space was obtained 
previously~\cite{Brandhuber:2023hhy,Herderschee:2023fxh,%
Georgoudis:2023lgf,Elkhidir:2023dco,Caron-Huot:2023vxl,Bohnenblust:2023qmy};
however, the physical waveform, as measured by GW 
observatories, is related to this result only through a 
Fourier transform (FT). The FT is a non-trivial step, 
even when done numerically, as emphasized by several 
authors~\cite{Brandhuber:2023hhy,Herderschee:2023fxh,%
Bohnenblust:2023qmy}. 

In this article we extend the strategy introduced by two of
the authors~\cite{Brunello:2024ibk} to gravitational systems. 
Our central object is the frequency-domain waveform, to
which we apply scattering-amplitude techniques directly. We 
consider the Fourier transform on the same footing as the 
loop integrations, and apply reduction techniques to the
combined integrals.

We apply unitarity-based techniques for integrand 
reconstruction~\cite{Bern:1994zx,Bern:1994cg,Britto:2004nc,%
Britto:2006sj,Ossola:2006us,Anastasiou:2006gt} uniformly
to both the loop and Fourier momenta~\cite{Brunello:2024ibk}. 
We simplify the generation of integrands by performing a 
classical expansion at the integrand level, using a heavy mass 
expansion~\cite{Brandhuber:2021kpo,Brandhuber:2021eyq}.
We also extend integration-by-parts (IBP) 
identities~\cite{Tkachov:1981wb,Chetyrkin:1981qh,%
Laporta:2000dsw,Smirnov:2008iw,Lee:2012cn,Maierhofer:2017gsa,Peraro:2019svx,Wu:2023upw} to include the exponential factor in the 
Fourier transform~\cite{Matsumoto1998-2,majima2000,%
Cacciatori:2022mbi,Brunello:2023fef,Brunello:2024ibk}.  We
then use IBPs in a standard fashion to reduce the integrated
form of the amplitude to a sum of master integrals (MIs) 
alongside their coefficients. 

We are able to express the waveform using a novel basis of 28
functions.  Remarkably, while our computations are for
spinless compact objects, the same basis suffices to express 
waveforms of spinning compact objects.

\def\delh{\hat\delta}
\textit{Notation\/} As is standard in scattering amplitudes,
we work in the mostly minus signature $(+,-,-,-)$, with
relativistic units $c=1$. As usual, we define the coupling
$\kappa := \sqrt{32 \pi G}$, where $G$ is Newton's constant. We
adopt a $\pi$-absorbing \textit{hat\/} 
notation~\cite{Kosower:2018adc} for
the integral measure,
\begin{equation}
\hd^m q := \frac{\dd^mq}{(2\pi)^m}\,,
\end{equation}
and for delta functions,
\begin{equation}
    \hdelta^{(m)}(\cdot ):= (2 \pi)^m \delta^{(m)}(\cdot)\,.
\end{equation} 
We also abbreviate loop-integral ($\ell$) and
combined Fourier- ($q$) and loop-integral measures,
defining,
\begin{equation}
\begin{aligned}
\int_{\hat{\ell}} &:= e^{\eps \gamma_E} \int \hd^D \ell\,,
\\
\int_{\hat q,\hat{\ell}} &:= e^{\eps \gamma_E} \int \hd^D q\,\hd^D \ell\,, 
\end{aligned}
\end{equation}
where the exponential factor is intended to remove 
explicit appearances of $\gamma_E$ in finite parts
of integrals.
The amplitudes have graviton momenta outgoing. 
For in- and out-states, scalar momenta are taken to be incoming and outgoing, respectively.  We take $\hbar=1$ in the initial quantum 
expressions, and leave implicit its restoration and the taking 
of $\hbar\rightarrow0$ to obtain the classical limit.  
We leave the
change from massive momenta to wavenumbers in the classical limit
implicit as well.

\section{Framework}
Following the KMOC 
approach~\cite{Kosower:2018adc,Cristofoli:2021vyo,Kosower:2022yvp}, 
classical observables such as the gravitational waveform are 
defined as the classical limits of quantum observables.
The latter are given by
the difference between the expectation value of a quantum operator $\mathcal{O}$ in the final 
$|\psi\rangle_{\rm out}$ and initial $|\psi\rangle_{\rm in}$
states. 
The final states are given by the unitarity evolution of the 
initial states via the $S$-matrix,
\begin{equation}
  |\psi\rangle_{\rm out} = S\,|\psi\rangle_{\rm in} \, . \label{eq:final_state}
\end{equation}
The $S$-matrix can be expressed in terms of the transfer matrix $T$ as:
\begin{equation}
    S = 1 + i\, T \,  .
\end{equation}
In our work we will define scattering amplitudes $\cM( i \to f)$ as the matrix elements of $T$:
\begin{equation}
    \cM(i\to f)\,\delh^D(k_i-k_f)
    = \langle f \vert\,  T\, \vert i \rangle \,  .  
\end{equation}
Thus, following the definition of~\eqref{eq:final_state}, a generic observable is written as:
\begin{equation}
    \expval{\Delta\cO} = \, \expval{\cO}_{\rm out} - 
    \expval{\cO}_{\rm in} \ =  \, _{\rm in}\langle \psi | 
    S^\dagger [\cO, S] | \psi\rangle_{\rm in}\, .
\end{equation}
Equivalently, we could compute expectation values in 
quantum field theory (QFT) using the Schwinger--Keldysh 
formalism~\cite{Schwinger:1960qe,Keldysh:1964ud}. 
This formalism is equivalent to the KMOC approach based on
scattering amplitude computations, and is more commonly used 
within the WQFT approach~\cite{Jakobsen:2022psy,Kalin:2022hph}. 
The two formalisms are expected to produce the same results. 
Indeed, we may observe that several terms in the KMOC 
formulation combine to produce the retarded propagators
characteristic of Schwinger--Keldysh 
approach~\cite{Caron-Huot:2023vxl}.

In this article, we will use the KMOC approach to compute
the radiation waveform emerging from gravitational scattering 
of a two-body system.  We build the appropriate initial state
out of two-particle plane-wave states
$|p_1, p_2 \rangle_{\rm in}$.  We take a superposition
given by on-shell wavefunctions $\phi_i(p_i)$,
\begin{equation}
    |\psi \rangle_{\rm in} = 
    \int\!\! \prod_{i=1}^2 \dd\Phi(p_i) \phi_i(p_i)
    \, e^{i b_i \cdot p_i}\, | p_1, p_2 \rangle_{\rm in}\,,
\end{equation}
where  $\dd \Phi (p_i)$ is the Lorentz-invariant on-shell 
phase-space (LIPS) measure, and the vectors $b_{1,2}^\mu$ 
encode the transverse separations
of the two wavepackets with respect to the chosen origin.
The impact parameter is the difference of the separation
vectors, $b^\mu= b_1^\mu - b_2^\mu$; it is space-like.
Strongly localizing the wavefunctions $\phi(p_i)$ 
around a classical
value of the momenta~\cite{Kosower:2018adc} will
take the classical limit of the observable,
suppressing quantum corrections.

The gravitational waveform is given by the expectation 
value of 
the gravitational field operator in the far future as a 
function of the (retarded) time and the angles on the celestial 
sphere. In our case, we may choose,
\begin{align}
    \mathcal{O}_{\cW} & = \kappa \,\varepsilon_{\lambda}^{\mu\nu} h_{\mu\nu}\,,
\end{align}
where $\lambda$ denotes the helicity configuration of the waves
which we are interested in observing, and $h_{\mu\nu}$ is the
quantum operator corresponding to linearized
deviations from a background metric,
linearized metric field 
($g_{\mu\nu}= \eta_{\mu\nu} + 
h^{\text{cl}}_{\mu\nu}$).  As we approach an infinite distance from
the source, this is equivalent to the Riemann (and also
Weyl) tensor observables considered in 
Ref.~\cite{Cristofoli:2021vyo},
though it is also sensitive to gravitational memory. 
It is also equivalent to computing the Newman--Penrose observable
$\Psi_4$, all other Newman--Penrose quantities being subleading
in $1/r$.
Non-linearities in the metric are asymptotically suppressed 
given the enormous distance $r$ of the observer from the source.
This is by far the largest scale in our calculation.  We may
also note that the operator $\mathcal{O}_{\cW}$ is invariant 
under diffeomorphisms which vanish when 
$r \to \infty$.\footnote{The waveform is not invariant under 
large diffeomorphisms, which introduce an angular-dependent phase shift \cite{Veneziano:2022zwh,Georgoudis:2023eke,Bini:2024rsy}. This phase shift will be irrelevant for collaborations of terrestrial or near-Earth observatories, given the attostopic size of the angles involved.}

Our setup is characterized by the following hierarchy of scales,
\begin{equation}
    \label{eq:hierarchy}
    \frac{1}{m_i} \ll G m_i \ll |b| \ll r\ ,
\end{equation}
where the first corresponds to the classical limit, and the 
second to the validity of relativistic perturbation theory.
The expansion parameters for observables of interest are 
the quantities ${G m_i}/{|b|}$.  We take these two to be
parametrically equivalent. In intermediate steps in our 
computations, 
we may see the appearance of 
terms which go as $G m_i^2$; these are sometimes called
\textit{classically singular terms}. Such terms will cancel in 
the final expression for physical observables\footnote{If we 
perform our computations using the Schwinger--Keldysh formalism 
instead of the scattering-amplitudes approach, classically 
singular terms do not appear even at intermediate steps.
This simplification comes at the cost of computing integrals 
with retarded and advanced propagators rather than Feynman, or anti-Feynman ones.}. 
At leading order in the ${1}/{r}$ expansion, the full
metric includes the backgrounds of the scattering objects
in addition to the waveform we compute.  Only on-shell 
modes contribute to the waveform. Expressing the waveform 
operator in terms of creation and annihilation operators, and 
using a asymptotic approximation we 
find~\cite{Cristofoli:2021vyo} the following expression
for the time-domain waveform,
\def\unitn{\hat{\bm{\mathit n}}}
\begin{equation}
    \label{eq:timedomain}
    \cWt_h (u,\unitn) = \frac{1}{4\pi r} 
    \!\int_0^\infty\! \hd \omega \left[ e^{- i \omega u} 
    \cW_{h} (\omega,\unitn) + \mathrm{c.c.}\right]\,,
\end{equation}
where $u=t-r$ is the \textit{retarded time} with $t$ the time as
measured by the observer and $\cW_{h} (\omega,\vec{n})$ 
the frequency-domain 
waveform\footnote{Here, we combined a factor of
$-i$ from the definition 
of the waveform with an $i$ from rewriting the $S$-matrix 
operator in terms of the transfer matrix. Note that $\cW_h (\omega,\unitn)$
has dimension $[\text{energy}]^{-2}$ in four dimensions.}:
\def\onshellMeasure{\mu}
\begin{equation}
    \label{eq:frequencydomain}
    \begin{split}
        \cW_{h} (\omega,\unitn)&  =
        \\ \kappa \!\int\! \dd\onshellMeasure
        \,\Big[&
        \ampl(p_1 p_2\to p_1^\prime p_2^\prime k^{-h}) 
        \\& -i \ampl\big(p_1 p_2 \rightarrow X k^{-h}\big) \otimes_X^{(+)} \ampl^*\big(p_1^{\prime} p_2^{\prime} \rightarrow X\big) \Big]\,;\hspace*{-5mm}
    \end{split}
\end{equation}
$k^\mu = \omega n^{\mu} = \omega (1,\unitn)$ (with $\unitn$ a
unit spatial vector pointing in the direction of the observer at
spatial infinity), and $h=\pm 2$ is the emitted graviton helicity;
the operator $\otimes_X^{(+)}$ denoting a sum over all intermediate
states $X$; and the on-shell measure is defined as,
\begin{equation}
    \label{eq:on_shell_measure}
    \begin{split}
        \dd \onshellMeasure & = 
        \prod_{i=1}^{2}\hd^D\! q_i \, \hdelta(2 p_i \cdot q_i - q_i^2)\, e^{i b_i \cdot q_i} \, \hdelta^D (q_1+q_2 - k)     \\
            & = \hd^D\! q \, \hdelta(2\bar{p}_1\cdot q) \,\hdelta(2\bar{p}_2\cdot (k-q))
            e^{i k \cdot b_2 + i b\cdot q}\, ,
    \end{split}
\end{equation}
where $q_i^\mu = p_i^\mu - p_i^{\prime \mu}$ is the \textit{mismatch} between the incoming and outgoing wavepackets; in the second line of Eq.~\eqref{eq:on_shell_measure}, we have introduced shifted momenta,
\begin{equation}
    \bar{p}^\mu_i = p^\mu_i + \frac{q^\mu_i}{2}\ .
\end{equation}
The complex conjugation in Eq.~\eqref{eq:timedomain} is needed 
to ensure that the waveform in the \textit{time domain} is real, 
once we strip off the polarization tensor $\varepsilon_{\mu\nu}$.
We write the polarization tensor in double-copy form, as a 
product of spin-1 polarization vectors,
\begin{equation}
\varepsilon^{(h)}_{\mu\nu} = 
\varepsilon^{(h/2)}_{\mu}\varepsilon^{(h/2)}_{\nu}\,.
\end{equation}

The first term in the second line of 
Eq.~\eqref{eq:frequencydomain} represents an in--out 
observable, \textit{i.e.} a $2\to 3$ scattering amplitude.
The second term is needed to subtract classically-singular
contributions and to restore the correct causality properties.
The latter are dictated by retarded propagators, ensuring that
overall we obtain an in--in 
observable~\cite{Caron-Huot:2023vxl}. We emphasize that 
Eq.~\eqref{eq:timedomain} is valid strictly in four dimensions, 
whereas the frequency domain 
expression~\eqref{eq:frequencydomain} is valid in generic 
dimension $D$.

In the following calculations, we will perform both the loop
and Fourier integrals in $D=4-2\eps$.  We then take the limit 
$\eps\to 0$. We treat Fourier and loop momenta on the same footing, 
allowing us to express frequency-space waveforms in terms of a  
minimal basis of functions which we denote 
\textit{combined master integrals\/} (CMIs): 
\begin{align}
\label{eq:FL_decomposition}
    \cW_{h} (\omega,\unitn) & = \; \sum_{i=1}^n c_i\; I_i \,. 
\end{align}
Here, $n$ is the number of combined master integrals. 
It turns out that the basis described here is not limited to 
scalar scattering but also suffices to express waveforms for spinning objects. 

\paragraph*{Kinematics} Expressions for
the five-point process depend on the incoming and outgoing 
momenta of the two massive particles, the emitted graviton
momentum, and the impact parameter. We can parametrize the 
massive momenta as
\begin{equation}
    p_i^\mu = m_i u_i^\mu \,,
    \label{MomentumParametrization}
\end{equation}
where the $u_i^\mu$ are the classical four-velocities of the two
particles. As emphasized above in the definition of the Fourier 
measure~\eqref{eq:on_shell_measure}, it is convenient to consider the classical limit of the in--in observable around 
shifted momenta $\bar{p}_i^\mu$, which we can parametrize
analogously to Eq.~\eqref{MomentumParametrization},
\begin{equation}
    \bar{p}_i^\mu = \bar{m}_i \bar{u}_i^\mu = m_i u_i^\mu + \frac{q_i^\mu}{2}\,.
    \label{BarredMomentum}
\end{equation}
In following sections, we shall work with barred variables. After 
computing the Fourier transform, the difference between $u_i^\mu$
and $\bu_i^\mu$ will amount to a correction of order 
$\frac{1}{m |b|}$, which is negligible according to the 
hierarchy~\eqref{eq:hierarchy}.  We will keep the barred variables 
throughout the computation for clarity, and drop the bar in the 
final result.

The frequency-domain waveform depends on five independent 
Lorentz invariants constructed from the 
four-vectors $\{u_1^\mu, u_2^\mu, b^\mu, k^\mu\}$,
\begin{equation}
    \gamma = u_1 \cdot u_2\,, \quad w_1 = u_1 \cdot k\,, \quad w_2 = u_2 \cdot k\,, \quad b^2\,, 
    \quad b \cdot k\,.
\end{equation}
The vectors also satisfy the following relations:
\begin{align}
    u_i^2 &= 1\,, & u_i \cdot b &= 0\,, & k^2 &= 0\,.
\end{align}
We also define a norm,
\begin{equation}
    \normb = \sqrt{-b^2}\,.
\end{equation}

\section{Computational steps}
\begin{figure}[!t]
    \centering
    \includegraphicsbox[scale=1.5]{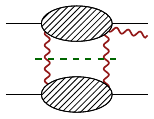} \>
    \includegraphicsbox[scale=1.5]{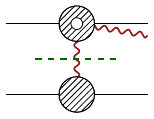} \>
    \caption{
    Generalized unitarity cuts probing the singularities at $q_i^2=0$ at one-loop. 
    Left: double graviton exchange. Right: single graviton exchange 
    with the one-loop gravitational Compton amplitude on one side of the cut. 
    This figure has been copied from Ref.~\cite{Brunello:2024ibk}.}
    \label{fig:1loop_cuts}
\end{figure}

The computation of the analytic one-loop gravitational waveform 
proceeds in several steps, previously 
presented~\cite{Brunello:2024ibk} by two of the authors,
\begin{enumerate}
    \item \textit{Integrand computation.} 
    We select only contributions that can yield terms
    non-analytic in the $q_i^2$.  The Fourier transform from 
    $q$-space to impact-parameter space will turn analytic 
    terms into local interactions, which will not contribute to 
    the observable. Such terms may be extracted using 
    generalized unitarity cuts~\cite{Bern:1994zx,Bern:1994zx} 
    in the corresponding channels, as shown in 
    Fig.~\eqref{fig:1loop_cuts}.  This corresponds to 
    \textit{sewing} lower-loop and lower-point amplitudes. The 
    latter amplitudes are obtained by first taking the 
    heavy-mass expansion~\cite{Brandhuber:2021kpo,%
    Brandhuber:2021eyq} on both sides of each cut.  This is 
    suggested by the classical limit in 
    Eq.~\eqref{eq:hierarchy}~\cite{Damgaard:2019lfh}, which 
    significantly reduces the maximum rank of the tensors in 
    the power counting.
    \item \textit{Mapping in--out to in--in.} As mentioned in 
    the previous section, both terms inside the brackets in 
    Eq.~\eqref{eq:frequencydomain}
    contain terms singular in the classical limit.  These
    cancel in the final result, after the terms are combined.
    On the other hand, we know that the Feynman rules of the two terms are identical up to $i\varepsilon$s. Therefore, in principle, it is sufficient to understand how propagators combine when we sum over appropriate $i\varepsilon$ coefficients in a simplified context, \textit{e.g.} a scalar theory, and assign the corresponding prescriptions to the integrand previously determined. This strategy has been used in Ref.~\cite{Caron-Huot:2023vxl} to show for the one-loop classical waveform that the map from the in--out formulation to the in--in one for an observable corresponds to substituting \textit{retarded propagators} for Feynman propagators (which in the classical limit would have become principal-valued linearized propagators).
    \item \textit{Tensor and IBP reduction.} In order to avoid 
    the appearance of unphysical poles in intermediate steps, 
    we treat Fourier and loop momenta on the same footing. For
    example, tensor decompositions can be performed as though
    the integrand was that of a two-loop integral; the exponential 
    factor does not play a role.  In contrast, IBP relations are
    modified, and we will find fewer of them than for a pure 
    two-loop integral. As we shall see, IBP relations for
    combined Fourier-loop integrals are linear combinations 
    of IBP relations for the loop integrals. 
    \item \textit{Loop and Fourier integrals.} Once the waveform is expressed in terms of a basis of CMIs, we
    must evaluate them.  We do this loop by loop, first
    computing the one-loop integrals, and then their Fourier transforms.  At higher order, it will be worth exploring using differential equations (DEs) directly on the CMIs.
\end{enumerate}

Let us now discuss some of the technical details of the 
steps above.

\paragraph{Integrand and Generalized Unitarity.}
We can construct the integrand for combined Fourier and loop 
integrand using generalized unitarity. The quantum amplitude
has several (generalized) unitarity cuts, involving the massive 
states and/or the gravitons.  In the classical limit, 
singularities corresponding to cuts of massive lines are pushed 
to infinity in the $q_i^2$. The relevant cuts for the Fourier 
transform are then only those associated to on-shell massless 
particles, shown in Fig.~\ref{fig:1loop_cuts}.

\begin{figure}[!t]
    \centering
    \includegraphicsbox[scale=1]{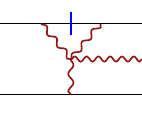} \>
    \includegraphicsbox[scale=1]{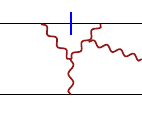} \>
    \caption{
    When considering one-particle and two-particle generalized unitarity cut in general relativity, there are overlapping contributions due to the nonlinear nature of General Relativity. These diagrams give non-zero contributions in taking residues and discontinuities at $q_i^2=0$.
    }
    \label{fig:1loop_doublecount}
\end{figure}

At one loop the building blocks needed to construct the
integrand are the tree-level amplitudes with up to three 
gravitons and the one-loop gravitational Compton amplitude. 
The former have been computed in Ref.~\cite{Brandhuber:2021kpo} within the heavy-mass expansion.  This expansion offers
a power-counting in the masses, allowing us to distinguish
genuine higher-order corrections from classically singular
terms (see also Refs.~\cite{Bern:2019crd,Bjerrum-Bohr:2021wwt} 
for equivalent tree-level amplitudes). The sewing of the 
Compton amplitude is performed prior to the loop integration, 
in order to avoid multiple counting of contributions probed 
both by the one-particle and the two-particle cuts 
(\textit{e.g.\/} as in Fig.~\ref{fig:1loop_doublecount}). 
We can obtain a gauge-invariant form of the Compton integrand 
by combining generalized unitarity cuts in two different
channels using a permutation-symmetric ansatz.
We give more details on integrand construction for the Compton
and the five-point integrands 
in Appendices~\ref{app:compton}~and~\ref{app:integrand}. 
Both our integrands are linear in the linearized Riemann tensor
of the external gravitons, 
\begin{equation}
    R_{ \mu\nu\rho\sigma} = k_{ \mu} k_{ \nu} \varepsilon_{ \rho \sigma} - k_{ \nu} k_{ \rho} \varepsilon_{ \mu \sigma} - k_{ \mu} k_{ \sigma} \varepsilon_{ \nu \rho} + k_{ \rho} k_{ \sigma} \varepsilon_{ \mu \nu}\ .
\end{equation}
This ensures gauge invariance at 
each step of the computation.
For the internal graviton state sums we use,
\begin{equation}
    \sum_h \varepsilon^{(h)}_{\mu \nu} (k) 
    \varepsilon^{(-h)}_{\rho \sigma} (-k) = 
    \frac{P_{\mu \rho} P_{\nu \sigma} + 
    P_{\mu \sigma} P_{\nu \rho}}{2} - 
    \frac{P_{\mu \nu} P_{\rho \sigma}}{D_s-2}\,,
    \label{eq:graviton_sum}
\end{equation}
where,
\begin{equation}
    P_{\mu \nu} = \eta_{\mu \nu} - \frac{k_\mu r_\nu + k_\nu r_\mu}{k \cdot r} + r^2 \frac{k_\mu k_\nu}{(k \cdot r)^2}\,,
\end{equation}
in which $r_\mu$ is a reference vector in axial quantization
and $D_s-2$ is the number of spin degrees of freedom of the
graviton in dimensional regularization. We consider two 
distinct regularization schemes: the `t Hooft-Veltman (HV) 
scheme\footnote{Recall that the HV scheme keeps external
momenta four-dimensional, while the conventional scheme
keeps them $D$-dimensional. 
Restricting the external kinematics to four dimensions 
simplifies tensor reductions.} with $D_s = D = 4 - 2\eps$ 
and the four-dimensional helicity (FDH) \cite{Bern:1991aq}
or dimensional-reduction (DR) schemes 
with $D_s = 4$ \cite{Siegel:1979wq}.
The latter two are equivalent at one loop. 

The integrand is provided in the ancillary file 
\texttt{integrand.m}. We expect the tensor reduction of
the integrand constructed as above to yield results
equivalent to those obtained in 
Refs.~\cite{Brandhuber:2023hhy,Herderschee:2023fxh,%
Bohnenblust:2023qmy}. Indeed, we find that the difference 
between our results and those of the latter references is
polynomial in both $q_1^2$ and $q_2^2$ (for example, 
see the re-organization of the waveform in terms of 
transcendental functions free of spurious poles in 
Ref.~\cite{Bini:2024rsy}). 
Such terms would give local contributions in impact-parameter
space.  We are able to remove all such terms, which are of no
interest to observers located very far from the source.
The integrand computation can be performed entirely within the
heavy-mass approximation, without needing the subtle contributions
from the so-called \textit{snail} diagrams; such terms were 
computed in Refs.~\cite{Brandhuber:2023hhy,Herderschee:2023fxh} through subtle forward limits as some of the relevant contributions could not be captured through two-particle cuts over internal graviton legs. In our approach, these contributions are captured by the single-graviton cuts shown in 
the right-hand diagram in Fig.~\ref{fig:1loop_cuts}.

\paragraph{Integrand reduction.}
The frequency-space one-loop contributions to the
NLO waveform are analogous
to a two-loop integral,
\begin{equation}
    \label{eq:integrand}
    \begin{split}
        \cW_h^{(1)} (\omega,\unitn) = 
        \frac{e^{i \omega \unitn\cdot b_2}}{4\pi r}
        \int_{\hat{q},\hat{\ell}}
        e^{i b\cdot q}\, \Integrand^{(1)}\ ,
    \end{split}
\end{equation}\\
where we treat the Fourier transform to impact-parameter space 
on an equal footing with the loop-momentum integration. The 
integrand appearing in Eq.~\eqref{eq:integrand} contains tensor 
structures in which the linearized Riemann tensor is contracted 
either with the loop or the Fourier momentum. 
We perform the tensor reduction following 
Ref.~\cite{Anastasiou:2023koq} where the four-dimensional
space of external momenta is spanned by the vectors 
$u_1^\mu$, $u_2^\mu$, $b^\mu$ and $k^\mu$. 
We need to reduce tensors with up to four powers of
an integrated momentum. Whenever one of these momenta
is contracted with an external momentum (rather than with
the Riemann tensor), the projectors defined in 
Ref.~\cite{Anastasiou:2023koq} simply reproduce
the given momentum.
We provide the explicit form of the required projectors in Appendix~\ref{app:tensor}.

\begin{figure}[!t]
    \centering
    \includegraphicsbox[scale=1]{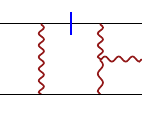} \>
    \includegraphicsbox[scale=1]{figures/triu1comp2} \>
    \includegraphicsbox[scale=1]{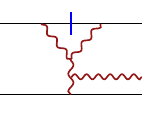} \>
    \caption{
    Integrals arising from the application of IBP identities 
    to the one-loop amplitude for combined Fourier and loop
    integrals can be reorganized into one of these
    three topologies, ordered as in Eq.~\eqref{eq:subtopo_fl}. 
    The integral family of Eq.~\eqref{eq:topo_fl} includes 
    additional master integrals, but they do not contribute to 
    the final result. 
    }
    \label{fig:fl_topologies}
\end{figure}

After tensor reduction, we obtain a sum of integrals with
nontrivial numerators but no free indices.  Each integral
belongs generalized integral family,
\begin{equation}
    I_{a_1\ldots a_{11}}[\cN(q,\ell)] =
    \int_{\hat{q},\hat{\ell}}\ \frac{e^{D_1} \,\cN(q,\ell)}
    {\prod_{i=1}^{11}D_i^{a_i}}\ ,
\end{equation}
where the invariants $D_i$ are,
\begin{equation}
\begin{aligned}
     & D_1  = i q\cdot b\,,\quad D_2 =   q^2\,,\quad
    D_3  =   (q-k)^2\,,\quad 
    \\ & D_4 =  \bu_1 \cdot q\,,\quad
    D_5 =  \bu_2 \cdot (k-q)\,,\quad     
    D_6 =    \bu_1\cdot \ell\,,\quad
    \\&
    D_7 =  \bu_2\cdot \ell\,,\quad D_8 =  \ell^2\,,
    D_9 =  (\ell+q)^2\,,\quad
    \\ & D_{10} =  (\ell+q- k )^2\,,\quad
    D_{11} = i  b \cdot \ell\,.\\
\end{aligned}
\end{equation}
Of these invariants, $D_{1,11}$ can appear only as numerators; $D_{4,5}$ appear as delta functions because of the 
LIPS measure; and either $D_6$ or $D_7$ are localized 
on-shell by the classical limit.  We can therefore
restrict attention to two subfamilies, the first:
\begin{widetext}
    \begin{equation}
        I^{(1)}_{a_1,a_2, a_3, 1, 1, 1,  a_7, \ldots, a_{11}}
        [\cN(q,\ell)] = 
        \! \int_{\hat{q},\hat{\ell}}\! e^{D_1} \hdelta(D_4) 
        \hdelta(D_5) \hdelta(D_6) 
        \frac{D_1^{-a_1}D_{11}^{-a_{11}}}
        {D_2^{a_2}D_3^{a_3}D_7^{a_7}D_8^{a_8}D_9^{a_9}
        D_{10}^{a_{10}}}\,,
        \label{eq:topo_fl}
    \end{equation}
\end{widetext}
\vspace*{3mm}
with the second $I^{(2)}$ obtained from $I^{(1)}$ by exchanging 
$D_6\leftrightarrow D_7$. These are \textit{twisted period 
integrals\/}~\cite{Brunello:2023fef}.  It is possible to 
decompose them into a finite basis of independent integrals, 
the so-called \textit{master integrals}, 
either via intersection 
theory~\cite{Mastrolia:2018uzb,Brunello:2023rpq,%
Brunello:2023fef} or via generalized IBP
identities~\cite{Brunello:2024ibk}:
  \begin{equation}
 \int_{\hat{q}, \hat{\ell}} 
 \frac{\partial}{\partial r^\mu }
 \biggl(\frac{e^{D_1}\,v^\mu}{\prod_{i=1}^{11} D_i^{a_i}}\biggr)=
 0\,,\quad r^\mu \in \{\ell^\mu, q^\mu  \} 
\end{equation}   
Localized propagators ($D_{4,5}$ and either $D_6$ or $D_7$)
can be treated as denominators via reverse 
unitarity~\cite{Anastasiou:2002yz,Anastasiou:2003gr,%
Herrmann:2021lqe}. The IBP relations for the combined 
Fourier and loop integrals can be written in terms of 
the IBP relations of ordinary integrals (without exponentials),
\begin{widetext}
\begin{equation}
    \text{IBP}_{a_1,\ldots, a_{11}}[\cN(q,\ell)] = 
    \!\int_{\hat{q}, \hat{\ell}} 
    \cN(q,\ell) \frac{\partial}{\partial r^\mu }
    \biggl(\frac{v^\mu}{\prod_{i=1}^{11} D_i^{a_i}}\biggr)\,,\quad
    r^\mu \in \{\ell^\mu, q^\mu  \} 
\end{equation}
via,\footnote{Should we consider total derivatives with 
respect to the loop momentum, the second and third terms in 
Eq.~\eqref{eq:gen_ibps} 
cancel and we recover standard IBP 
relations, as expected. In Eq.~\eqref{eq:gen_ibps}, 
$v^\mu$ must be the same for all terms.}
\begin{equation}
    \label{eq:gen_ibps}
    \text{IBP}_{a_1,\ldots, a_{11}}[1-D_1] + 
    \text{IBP}_{a_1-1,\ldots, a_{11}}[1]  = 0 \ ,
\end{equation}
\end{widetext}
The generalized IBP relations can be generated systematically
using one of the standard IBP codes; we have used
\textsf{LiteRed}~\cite{Lee:2012cn}. 
We solve the generalized IBP system using 
\textsf{LiteRed} and finite-field
techniques as implemented in 
\textsf{FiniteFlow\/}~\cite{Peraro:2019svx} (which follows the Laporta 
algorithm~\cite{Laporta:2000dsw}).
Each of the two integral subfamilies $I^{(1,2)}$ can be
decomposed in terms of $76$ combined master integrals.  Only
$28$ of these integrals actually appear in the result for
the waveform.
This simplification can be understood in terms of the 
Feynman diagrams contributing to the process. All terms 
in the integrand are daughters of the following 
three topologies (see Fig.~\ref{fig:fl_topologies}),
\begin{figure}[!t]
    \centering
    \includegraphicsbox[scale=1]{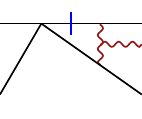} \>
    \includegraphicsbox[scale=1]{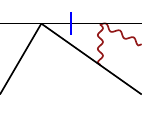} \>
    \includegraphicsbox[scale=1]{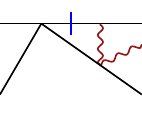} \>
    \caption{
    One-loop diagrams with a pinched graviton propagator $D_8$ 
    arise when applying IBPs solely to the loop integration,
    but vanish in the final result after both Fourier and loop
    IBP reduction.
    }
    \label{fig:fl_pinched}
\end{figure}
\begin{widetext}
    \begin{equation}
        \begin{split}
            I^{(1)}_{a_1a_2 a_3 1 1 1  a_7 \dots a_{11}} & =  
            \int_{\hat{q},\hat{\ell}} 
            \hdelta(D_4) \hdelta(D_5) \hdelta(D_6) 
            \frac{e^{D_1} D_1^{-a_1} D_2^{-a_2} D_3^{-a_3} 
            D_{11}^{-a_{11}}}
            {(D_7+i\epsilon)^{a_7} D_8^{a_8} D_9^{a_9} 
            D_{10}^{a_{10}}}\,,
            \\
            I^{(1)}_{a_1a_2 a_3 1 1 1  a_7 \dots a_{11}} & =  
            \int_{\hat{q},\hat{\ell}} 
            \hdelta(D_4) \hdelta(D_5) \hdelta(D_6) 
            \frac{e^{D_1} D_1^{-a_1} D_2^{-a_2} D_7^{-a_7} 
            D_{11}^{-a_{11}}}
            {D_3^{a_3} D_8^{a_8} D_9^{a_9} D_{10}^{a_{10}}}\,,
            \\
            I^{(1)}_{a_1a_2 a_3 1 1 1  a_7 \dots a_{11}} & = 
            \int_{\hat{q},\hat{\ell}} 
            \hdelta(D_4) \hdelta(D_5) \hdelta(D_6) 
            \frac{e^{D_1} D_1^{-a_1} D_7^{-a_7} 
            D_{10}^{-a_{10}} D_{11}^{-a_{11}}}
            {D_2^{a_2} D_3^{a_3} D_8^{a_8} D_9^{a_9} }\,.
        \end{split}
        \label{eq:subtopo_fl}
    \end{equation}
\end{widetext}

We emphasize that triangle and box loop integrals with pinched 
$D_8$ (see Fig.~\ref{fig:fl_pinched}) never appear in the
combined Fourier and loop
approach, whereas they would were we to follow a loop-only 
reduction, as was done in 
Refs.~\cite{Brandhuber:2023hhy,Herderschee:2023fxh,%
Bohnenblust:2023qmy}. 
Indeed, in such an approach, although their final contributions
would be polynomial in $q_1^2$ and $q_2^2$ (and hence local
and of no interest), they would appear multiplied by rational
functions of the $q_i^2$.  One would therefore expect them
to contribute nontrivially to the waveform. 
We choose the CMIs so that their rational 
coefficients are finite as $\eps \to 0$, using the algorithm presented in  Ref.~\cite{DeAngelis:2025agn}. We split them into 
two sets: 16 `base' integrals and 12 `derivative' integrals 
(indicated below by underlining) which can be obtained 
from the former by taking a derivative with respect to the 
impact parameter $b$. 
\def\derivlabel#1{\underline{#1}}
In particular, we have
\begin{align*}
    \bullet &~I^{(1)}_{00111100110}\,;\quad
    \derivlabel{I^{(1)}_{-10111100110}} \nonumber\\
    \bullet &~I^{(1)}_{00011101100}\,,\quad 
    I^{(1)}_{00111101100}\,,\quad I^{(1)}_{0-1111101100}\,, \nonumber\\
    &~I^{(1)}_{00211101100}\,;
    \quad\derivlabel{I^{(1)}_{-10011101100}}\,,\quad
    \derivlabel{I^{(1)}_{-10111101100}} \nonumber\\
    \bullet &~I^{(1)}_{00011101010}\,,\quad 
    I^{(1)}_{00111101010}\,;\quad
    \derivlabel{I^{(1)}_{-10011101010}}\,,\nonumber \\
    &~\derivlabel{I^{(1)}_{-10111101010}} \nonumber\\
    \bullet &~I^{(1)}_{00011101110}\,,\quad 
    I^{(1)}_{001111-11110}\,,\quad I^{(1)}_{0011110111-1}\,, \nonumber\\
    &~I^{(1)}_{00111101110}\,,\quad I^{(1)}_{0-1111101110}\,, \ 
    \quad I^{(1)}_{00111102110}\,; \nonumber\\
    &~\derivlabel{I^{(1)}_{-10011101110}}\,,\quad
    \derivlabel{I^{(1)}_{-101111-11110}}\,,\quad
    \derivlabel{I^{(1)}_{-1011110111-1}}\,, \nonumber\\
    &~\derivlabel{I^{(1)}_{-10111101110}} \\
    \bullet &~I^{(1)}_{00011111100}\,,\quad
    I^{(1)}_{00011111010}\,,\quad I^{(1)}_{00011111110}\,; \nonumber\\
    &~\derivlabel{I^{(1)}_{-10011111100}}\,,\quad 
    \derivlabel{I^{(1)}_{-10011111010}}\,,\quad
    \derivlabel{I^{(1)}_{-10011111110}}\,. \nonumber\\
\end{align*}
Each bullet corresponds to a different topology 
in the combined momenta.
The result in terms of combined MIs multiplied by their rational coefficients is provided in the ancillary file \texttt{reduced\_integrand.m}.

\paragraph{Integration.}
We have yet to explore a systematic approach to the combined
integrals using differential equations.  Instead,
we have performed the integrations sequentially.  We first
evaluated the loop integrals using canonical 
differential equations, following Ref.~\cite{Brunello:2024ibk}.
We computed the Fourier transform afterwards. 
We record the required loop integrals in 
Appendix~\ref{app:loopintegrals}. 
We find it helpful to split the second integration into 
two pieces, performing the Fourier transforms of the infrared-
divergent and -finite terms separately.
The Fourier transform does not introduce additional 
divergences; we must however treat $\eps/\eps$ terms carefully,
as they are physically relevant at one-loop~\cite{Bini:2024rsy}.
Accordingly, we need a proper $D$-dimensional measure for the 
Fourier integrals to compute them correctly.
In contrast, the Fourier transform of the finite part can be 
performed using the four-dimensional measure as introduced in 
Ref.~\cite{Cristofoli:2021vyo}.  We provide details of the 
integration in Appendix~\ref{app:Fouriertransform}.
We provide the analytic expression for the required integrals in the ancillary file \texttt{fl\_integrals.m}. The expressions given in the file are the integrands of one-fold exponential integrals, whose precise construction is explained in the comments.

\section{Comparisons with Previous Results}
We can separate terms in the one-loop contributions to
the waveform into
infrared-divergent ones, the corresponding logarithms, 
so-called \emph{tail terms}~\cite{Blanchet:1987wq}, 
and finite ones,
\begin{equation}
    \cW_h^{(1)} = \cW_{h,\text{IR}}^{(1)} 
    +\cW_{h,\text{tail}}^{(1)}
    +\cW_{h,\text{fin}}^{(1)}\,,
\end{equation}
where,
\begin{itemize}
    \item $\cW_{h,\text{IR}}^{(1)}$ is the infrared-divergent 
    part.  It is proportional to the tree-level 
    frequency-domain waveform~\cite{Weinberg:1965nx,Caron-Huot:2023vxl} (presented in 
    Appendix~\ref{app:treewaveform}):
    %
    \begin{equation}
        \cW_{h,\text{IR}}^{(1)} =  i \frac{\kappa^2 s_W}{\epsilon} \cW_{h}^{(0)} \ , 
    \end{equation}
    where the prefactor $s_W$ is
    \begin{equation}
        s_W = \frac{m_1 w_1 + m_2 w_2}{32\pi} \Gamma_\alpha\ , 
    \end{equation}
    in which,
    \begin{equation}
        \Gamma_\alpha = 1-\alpha \frac{\gamma \bigl(\gamma^2-\frac{3}{2}\bigr)}{\bigl(\gamma^2-1\bigr)^{3/2}} \ ,
    \end{equation}
    with $\alpha = -1,0$ as in 
    Appendix~\ref{app:loopintegrals}. The values of $\alpha$ 
    correspond to two different choices of classical boundary 
    conditions:
    \begin{itemize}
        \item $\alpha=-1$ corresponds to an 
        \textit{asymmetric} choice of frame between incoming 
        and outgoing momenta.  Here, 
        $p^\mu_{i, \text{in}} = p^\mu_{i}$ and 
        $p^\mu_{i, \text{out}} = p^\mu_{i} - \Delta p^\mu_i$, 
        where $\Delta p^\mu_i$ is the impulse on the 
        $i^{\textrm{th}}$ particle. This is the frame 
        typically used in the KMOC formalism \cite{Kosower:2018adc,Cristofoli:2021vyo}.
        \item $\alpha=0$ corresponds to the \textit{symmetric} 
        frame for which 
        $p^\mu_{i, \text{in}} = p^\mu_{i} + 
        {\Delta p^\mu_i}/{2}$ and 
        $p^\mu_{i, \text{out}} = p^\mu_{i} - 
        {\Delta p^\mu_i}/{2}$. This is the frame usually 
        chosen in so-called `multipolar post-Minkowskian' 
        computations, where classical boundary conditions are fixed at 
        $t=0$, rather than at $t=-\infty$~\cite{Bini:2023fiz}. At NLO, this frame is related to the previous one 
        through dropping the contribution of the cut terms in 
        the KMOC formalism in the classical limit.  Their
        contribution is then equivalent to performing the 
        corresponding rotation~\cite{Georgoudis:2023eke}.
    \end{itemize}
    The infrared divergences exponentiate to an overall phase 
    in the waveform in frequency 
    space~\cite{Weinberg:1965nx,Caron-Huot:2023vxl},
\begin{widetext}
\begin{equation}
\label{eq:exponentiated_IR_divs}
    \cW_{h} (\omega, \unitn) = 
        e^{-{i\kappa^2 s_W}/{\eps}} \cW_{h, \text{DR}} (\omega, \unitn)\, ,
\end{equation}
\end{widetext}
where DR in the subscript stands for dimensional regularisation. Inside the right-hand side, we keep terms of higher order in $\epsilon$.
    The infrared-divergent phase encodes information about 
    different time delays due to the long-range nature of the 
    gravitational interactions in four dimensions (for a clear 
    discussion of the physical interpretation of this 
    divergences, we refer to Ref.~\cite{Caron-Huot:2023vxl}). 
    On the other hand, an overall phase is not relevant 
    to gravitational-wave observations. Indeed, in the time
    domain, it corresponds to a constant shift of the 
    arrival time of the gravitational wave.  This is
    \textit{a priori\/} non measurable.  When extracting a
    signal from the background noise, observers apply a filter
    is applied with different time shifts to search for the 
    signal at an \textit{a priori\/} undetermined time 
    (see the discussion in Ref.~\cite{Owen:1998dk}).
    
    \item $\cW_{h,\text{tail}}^{(1)}$ contains the `tail' 
    terms. These express the effect of rescattering of the emitted  gravitational waves by the background generated by the two-body system at large distances. This includes a modification of the group velocity~\cite{Blanchet:1993ec}. 
    Given our standard use of infinite interaction time\footnote{That is ignoring frequency band limits and noise floor limits of any realistic detectors.}, these effects manifest themselves in an infrared divergence. The infrared divergence exponentiates and it is accompanied by a finite logarithm (which also exponentiates). At fixed order we have logarithms; at NLO we have a single logarithm proportional to the lowest-order waveform:
    \begin{equation}
    \begin{aligned}
        \cW_{h,\text{tail}}^{(1)} = 
        & i \kappa^2 s_W \log\biggl( 
        \frac{\omega^2}{4\pi\, \mu^2}\biggr)
        \cW_{h}^{(0)}\,,
    \end{aligned}
    \end{equation}
    where $\mu^2$ is an (infrared) scale. 
    The precise definition of this scale
    does not affect observables, as different choices of 
    $\mu^2$ correspond to changes in the phase of the 
    waveform, \textit{i.e.\/} to a finite redefinition of the observer's clock reference time.
    \item $\cW_{h,\text{fin}}^{(1)}$ are new 
    contributions which are not proportional to the
    lowest-order ones. We focus mainly on these terms in
    frequency-domain plots below. In Eq.~\eqref{eq:exponentiated_IR_divs}, $\cW_{h,\text{DR}}$ is computed to all orders in $\eps$. When we expand Eq.~\eqref{eq:exponentiated_IR_divs}, we will pick up finite contributions which arise from the cancellations of the singular expansion of the phase against $\eps$'s in $\cW_{h,\text{DR}}$. These terms should be excluded from the observable waveform. We can expand the $\cW_{h,\text{DR}}$ in $\eps$,
    \begin{equation}
        \cW_{h, \text{DR}} = \cW_{h, \text{obs}} + \eps\, \cW_{h, \eps} + \cO(\eps^2) \ ,
    \end{equation}
    where $\cW_{h, \text{obs}}$ is free of $\eps$-dependence. The $\eps/\eps$ term should be excluded from the observable waveform, which is then,
    \begin{equation}
        \label{eq:obs_1loop}
        \cW^{(1)}_{h, \text{obs}} = \cW_{h,\text{fin}}^{(1)} - i \kappa^2 s_W \cW_{h, \eps} \ .
    \end{equation}
\end{itemize}

\subsection{Comparison with other calculations}

Several groups have computed the fully relativistic scalar NLO waveform in momentum space using amplitudes techniques \cite{Brandhuber:2023hhy,Herderschee:2023fxh,Georgoudis:2023lgf}. Two of these groups \cite{Brandhuber:2023hhy,Herderschee:2023fxh} also used the KMOC formalism. In addition, Ref.~\cite{Bohnenblust:2023qmy,Bohnenblust:2025gir} also computed the NLO waveform for spinning particles, using the WQFT formalism \cite{Jakobsen:2022psy}. The results in Refs.~\cite{Brandhuber:2023hhy,Herderschee:2023fxh,Georgoudis:2023lgf} were previously compared and found to agree. We have compared our results with those of Ref.~\cite{Brandhuber:2023hhy}, after loop integration, but before the FT. We find complete agreement up to local terms (such as polynomial terms in $q_1^2$ and $q_2^2$), which are irrelevant in the classical limit. We would not expect local terms to agree, because our construction of the integrand drops them.

\subsection{The center-of-momentum frame}
\begin{figure*}[!t]
    \centering
    \includegraphics[width=0.325\textwidth]
    {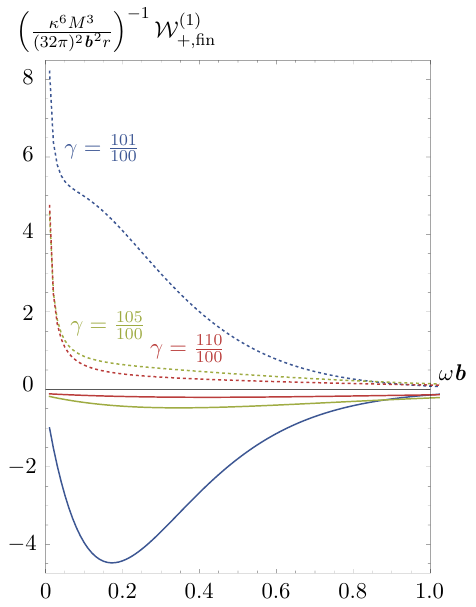} 
    \includegraphics[width=0.325\textwidth]
    {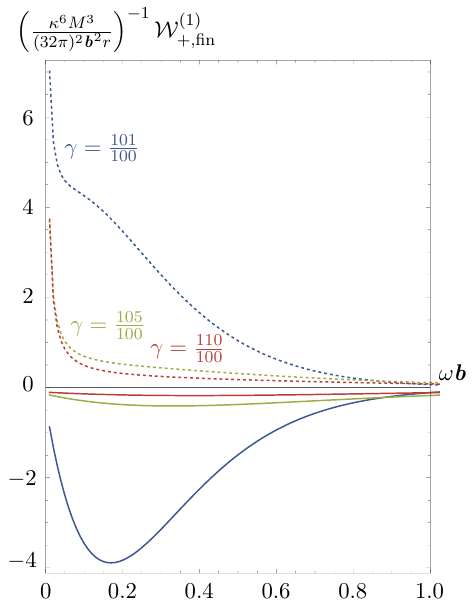} 
    \includegraphics[width=0.325\textwidth]
    {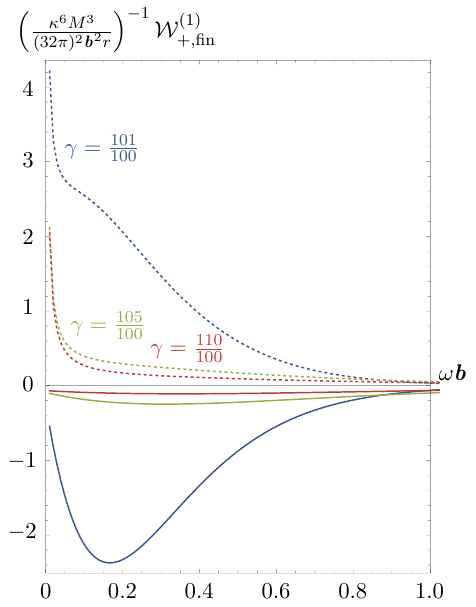} 
    \caption{Plots of the finite part of the 
    NLO frequency-domain waveform
    $\cW_{h,\text{fin}}^{(1)}$, for the $[+]$ polarization. Each plot 
    show different values of the boost $\gamma$ with fixed
    symmetric mass ratio $\nu$, which are, respectively $\{ \frac{1}{4}, \frac{2}{9},\frac{5}{36}\}$. The plots are at the fixed angle $(\theta,\phi)= \left(\frac{7 \pi}{5},\frac{7\pi}{10} \right)$. Solid lines represent the real part, while dashed lines represent the imaginary part.}
    \label{fig:plot_NLO_frequency}
\end{figure*}
We can generate frequency-domain plots for
the finite part 
$\cW_{h,\text{fin}}^{(1)}$.
We choose the symmetric frame discussed above for our
numerical evaluation. Define four orthonormal vectors: 
a time-like $e_0^\mu$, and three space-like $e^\mu_i$ 
($i = 1, 2, 3$). The (adjusted) momenta of the two bodies are,
\begin{equation}
    p_1^\mu = E_1\, e_0^\mu + P_{\text{cm}}\, e_2^\mu\ , \quad p_2^\mu = E_2\, e_0^\mu - P_{\text{cm}}\, e_2^\mu \ , 
\end{equation}
with $E_i = \sqrt{m_i^2+P_{\text{cm}}^2}$, and $P_{\text{cm}}$ the momentum in the symmetric frame. 
We can rewrite these quantities in terms of the masses and the 
Lorentz boost introduced earlier,
\begin{equation}
    P_{\text{cm}} = \frac{m_1 m_2 \sqrt{\gamma^2-1}}{E} \ , \quad E = E_1 + E_2 = \sqrt{m_1^2 + m_2^2+ 2 m_1m_2 \gamma} \ . 
\end{equation}
We choose the impact parameters $b_1$ and $b_2$ to be,
\begin{equation}
    b_1^\mu  =  \frac{E_2}{E}b^\mu \, , \quad b_2^\mu = -\frac{E_1}{E}b^\mu\,, \quad b^\mu = \normb\, e_1^\mu \,.
\end{equation}
Because the observer is very far away from the scattering,
the momentum $k^\mu$ of the emitted graviton is essentially
fixed to be along the direction towards the observer.
Thus, its momentum and polarization tensor 
$\varepsilon^{\mu \nu} = \varepsilon^{\mu} \varepsilon^{\nu}$ 
are given in terms of,
\begin{equation}
\begin{aligned}
    k^\mu &= \omega \left[e_0^\mu + n^\mu(\theta,\phi)\right] \,, \\
    \varepsilon_{(\pm)}^\mu &= \frac{1}{\sqrt{2}} \left(\partial_\theta n^\mu(\theta,\phi) \mp \frac{i}{\text{sin}\,\theta}\partial_\phi  n^\mu(\theta,\phi)\right) \,,
\end{aligned}
\end{equation}
where $n^\mu(\theta,\phi)$ is the unit space-like vector pointing in the direction of the observation:
\begin{equation}
    n^\mu(\theta,\phi) = \text{sin}\theta  \, \text{cos}\phi\, e_1^\mu\,+\,\text{sin}\theta  \, \text{sin}\phi\, e_2^\mu\,+\text{cos}\theta\, e_3^\mu \, . 
\end{equation}
In our case, we are interested in computing the 
waveforms for the $[+],[\times]$ polarizations of the 
graviton; these are defined in terms of the helicities as
follows,
\begin{equation}
\begin{split}
    \varepsilon_{[+]}^{\mu\nu} &= \frac{1}{2}\left( \varepsilon_{(-)}^\mu\varepsilon_{(-)}^\nu +
    \varepsilon_{(+)}^\mu\varepsilon_{(+)}^\nu \right)\,,\\
    \varepsilon_{[\times]}^{\mu\nu} &= 
    \frac{1}{2i}\left( \varepsilon_{(-)}^\mu\varepsilon_{(-)}^\nu - \varepsilon_{(+)}^\mu\varepsilon_{(+)}^\nu \right)\,,
\end{split} 
\end{equation}
We can write our result in terms of the total mass of the 
system $M$ and the symmetric mass ratio $\nu$,
\begin{equation}
    M = (m_1+m_2) \,, \quad \nu = \frac{m_1 m_2}{(m_1+m_2)^2}\,;
\end{equation}
we use the latter as one of the parameters in the plots.

\subsection{Visualization of results}
In Fig.~\ref{fig:plot_NLO_frequency} we show the NLO corrections
to the frequency domain waveform at fixed angle 
$\left(\theta=\frac{7 \pi}{10},\phi=\frac{7 \pi}{5}\right)$
for different values of the symmetric mass ratio $\nu$ and of
the Lorentz factor $\gamma$. 
One of the interesting aspects of our approach is
the simplification of the Fourier transform to 
impact-parameter space, as it is now written as a 
one-dimensional integral over an exponentially-damped 
integrand involving at worst logarithms and square roots. 
This allows for fast numerical evaluation, avoiding
spurious singularities on the integration contour. 
Indeed, the combined Fourier and loop IBP relations
trade the would-be spurious poles in $q_i^2$ 
for Gram determinants outside the 
integration. There is only a small price to pay: close to certain 
kinematic points, the numerical integration requires higher 
precision, as cancellations occur between different MIs and 
their rational coefficients.
For example, at $\theta=0,\pi$ we have spurious poles
corresponding to kinematic configurations where $b\cdot k = 0$.

\subsection{Post-Newtonian Expansion}
In order to compare our result with the GR results of Ref.~\cite{Bini:2023fiz,Bini:2024rsy}, we can expand our result in terms of the dimensionless 
velocity parameter $p_\infty$ after changing variables,
\begin{equation}
    \gamma = \sqrt{1+p_\infty^2}\,, \quad 
    \omega = p_\infty\, \Omega_\infty\,, \label{eq:pn_expansion}
\end{equation}
with $\Omega_\infty$ the frequency of the emitted graviton.
In this work, we have kept so far the state-counting
dimensional regularization parameter $D_s$ arbitrary. In the
following, we will set \( D_s = 4 \) for the graviton state
counting, \textit{i.e.\/} we will work within the so-called 
FDH or DR schemes, which are equivalent at one loop. This includes performing the FT in $D=4-2\epsilon$ dimensions. Would
we have chosen instead \( D_s = D \), additional 
\( \eps/\eps \) contributions would appear. These 
arise from the \(\eps\)-expansion of both the finite and 
infrared-divergent parts of the amplitude. Such terms 
would ultimately cancel when including corrections 
higher order in $\eps$ to the tree-level waveform.

To perform the comparison, we work with $\alpha=0$, and in the center-of-momentum frame. We expand our result using Eq.~\eqref{eq:pn_expansion}. We choose numerical values for the angles, the impact parameter, the frequency and the masses. We choose integer values for the frequency, the impact parameter and the masses. Two particular combinations of the polar angles appear,
\begin{equation}
    z = e^{i \phi} \cot \theta \ , \quad \bar{z} = e^{-i \phi} \cot \theta\ .
\end{equation}
We choose integer values for $z$ and $\bar{z}$ as well. This ensures that all arguments of logarithms and Bessel functions are integral, and all coefficients are rational. This in turn gives us confidence that cancellations are treated exactly, 
without needing to worry about precision loss in the numerical evaluation.  Indeed, while the NLO waveform starts at order $1/p_\infty^3$, in intermediate stages we find terms of order $1/p_\infty^{41}$ (or possibly even higher). All the unwanted higher powers cancel exactly, and we are left to compare terms of order $1/p_\infty^3$ through $p_\infty^1$ with the GR result. 
We have chosen\footnote{The particular choice of the angular variable $z$ corresponds to $\theta \simeq \frac{\pi}{9}$ and $\phi \simeq \frac{\pi}{3}$.}
\begin{equation}
    \normb \omega = 3\ , \quad \frac{m_2 - m_1}{m_1 + m_2} =  \frac{1}{6}\ , \quad z = 3 + 5i\ ,
\end{equation}
finding perfect agreement with the terms given in Ref.~\cite{Bini:2024rsy}, complemented by the more recent results presented in Ref.~\cite{Heissenberg:2025fcr}. We compared the observable waveform $\cW_{h, \text{obs}}^{(1)}$ of Eq.~\eqref{eq:obs_1loop} with the GR result. The latter is not computed in dimensional regularization, and accordingly has no $\eps$ dependence. However, it has a dependence on the frame-dependent scale $b_0$~\cite{Bini:2024rsy}. We have translated the $\eps$ dependence to a dependence on $\mu$ in a standard fashion. We must now relate the scale $\mu$ to $1/b_0$.  The check up to order $p_\infty^1$ does not include subtleties related to the angular-dependent redefinition of the retarded time, interpreted as the supertranslation identified by Veneziano and Vilkovisky~\cite{Veneziano:2022zwh} in Refs.~\cite{Georgoudis:2023eke,Bini:2024rsy,Elkhidir:2024izo}. The first term affected by this mapping is at order $p_\infty^2$. Analytic comparison at higher orders in $p_\infty$ expansion with MPM result would be an interesting direction to pursue, for example implementing strategies recently explored in Ref.~\cite{Brunello:2025cot}.

\section{The Power Spectrum}


The differential power spectrum (or energy spectrum) carried
by gravitational waves gives the total energy emitted per unit 
frequency during the scattering process. Starting from the 
frequency-domain waveform $\cW_h(\omega,\theta,\varphi)$, 
it can be defined as follows,
\begin{equation}
    \frac{\dd E }{\dd \omega} = \frac{2}{\kappa^2} \!\int\! \dd\Omega \sum_h \left|\; \omega \, \cW_h(\omega,\Omega) \, \right|^2\,,
\end{equation}
where the solid-angle integral covers the entire celestial 
sphere and the sum runs over the two graviton helicities. This 
observable is gauge-invariant and infrared-finite and provides 
a clean connection between theoretical predictions and 
measurable quantities. The spectrum enters directly in the 
computation of the gravitational-wave flux and, ultimately, in 
the signal-to-noise ratio used by matched-filtering 
analysis pipelines.
It has been presented analytically to high order in the PN expansion 
based on the expansion of the leading-order relativistic waveform \cite{Jakobsen:2021smu,Bini:2024tft}. 
Here we are interested in the numerical computation of 
the NLO ($\Ord(G^4)$) power spectrum.
We can perform the angular integral numerically by dividing
up the celestial sphere into finite-size elements. We employed 
the Lebedev sparse-grid quadrature method~\cite{Lebedev:1976}.
We used the Lebedev rules on the unit sphere from the dataset
directory \href{https://people.sc.fsu.edu/~jburkardt/datasets/sphere_lebedev_rule/sphere_lebedev_rule.html}{SPHERE\_LEBEDEV\_RULE}. For a chosen grid of points 
$L_N=\{(\theta_{i},\varphi_{i})\}_{i=1}^{P}$ with associated
weights $W_{i}^{(N)}$ we write:
\begin{equation}
    \!\int\! \dd\Omega\; f(\theta,\varphi) \simeq \ (4\pi)\sum_{\theta_i \varphi_i \in L_N} W_i^{(N)}(\theta_i,\varphi_i) f(\theta_i,\varphi_i)\,.
\end{equation}
We have used $N= 17$, corresponding to 110 points in $L_N$.
The Lebedev rule of order $N$ will integrate a spherical
harmonic $Y_{\ell m}(\theta,\phi)$ for $\ell\le N$ exactly.
\subsection{LO Power Spectrum}
\begin{figure*}[!t]
    \centering
    \includegraphics[width=0.325\textwidth]
    {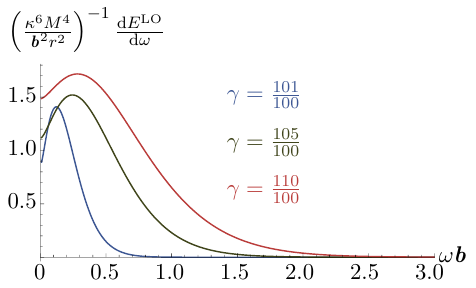} 
    \includegraphics[width=0.325\textwidth]
    {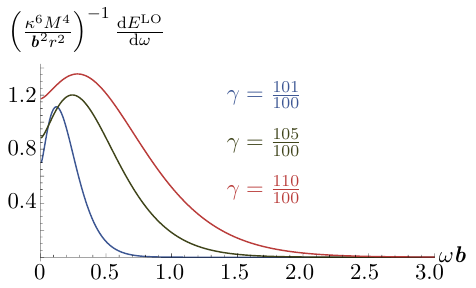} 
    \includegraphics[width=0.325\textwidth]
    {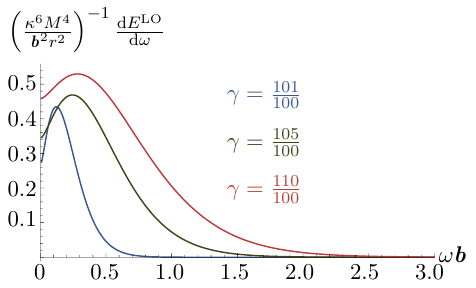} 
    \caption{Plots of the gravitational power spectrum at LO
    ${\dd E^{\text{LO}}}/{\dd \omega}$ as a function of the energy $\omega$. Each plot 
    shows different values of the boost $\gamma$ with fixed
    symmetric mass ratios $\nu$, which are, respectively $\{ \frac{1}{4}, \frac{2}{9},\frac{5}{36}\}$.  }
    \label{fig:plot_LO_flux}
\end{figure*}
\begin{figure*}[!ht]
    \centering
    \includegraphics[width=0.325\textwidth]
    {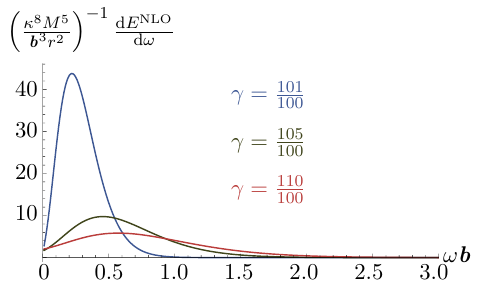} 
    \includegraphics[width=0.325\textwidth]
    {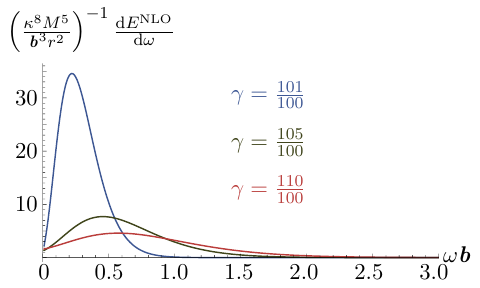} 
    \includegraphics[width=0.325\textwidth]
    {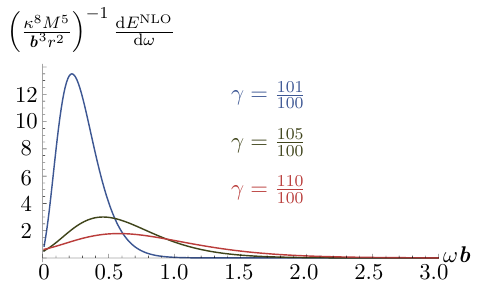} 
    \caption{Plots of the NLO corrections to the gravitational power spectrum
    ${\dd E^{\text{NLO}}}/{\dd \omega}$ as a function of the energy $\omega$. Each plot 
    shows different values of the boost $\gamma$ with fixed
    symmetric mass ratios $\nu$, which are, respectively $\{ \frac{1}{4}, \frac{2}{9},\frac{5}{36}\}$.  }
    \label{fig:plot_NLO_flux_m}
\end{figure*}
\begin{figure*}[!ht]
    \centering
    \includegraphics[width=0.325\textwidth]
    {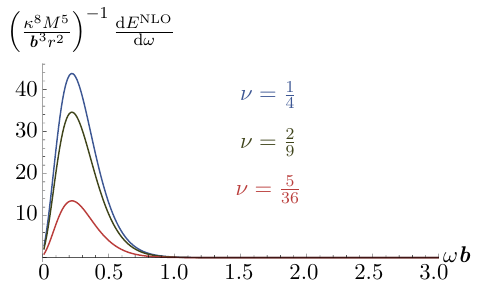} 
    \includegraphics[width=0.325\textwidth]
    {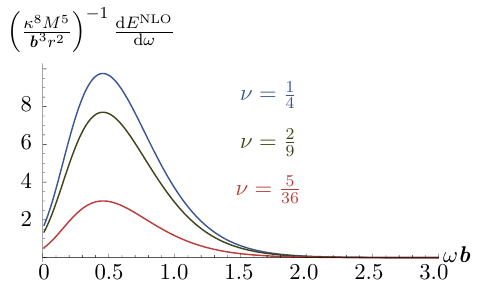} 
    \includegraphics[width=0.325\textwidth]
    {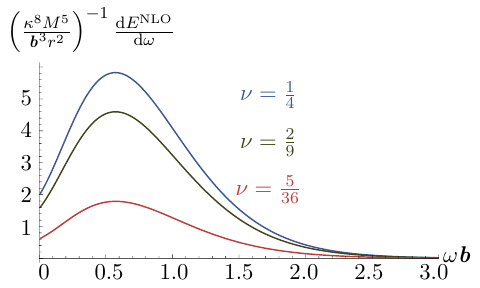} 
    \caption{Plots of the NLO corrections to the gravitational 
    power spectrum ${\dd E^{\text{NLO}}}/{\dd \omega}$ as a function of the energy $\omega$. Each plot 
    shows different values of the symmetric mass ratio $\nu$ with fixed values of the
    boost  $\gamma$, which are, respectively $\{ \frac{101}{100}, \frac{105}{100},\frac{110}{100}\}$.  }
    \label{fig:plot_NLO_flux_y}
\end{figure*}

Using the tree-level waveform given in Appendix~\ref{app:treewaveform}, we can compute the 
LO flux as follows,
\begin{equation}
    \frac{dE^{\text{LO}} }{d\omega} =   \frac{2}{\kappa^2} \!\int\! \dd\Omega \; \omega^2 \sum_{h=\pm} \big\vert \cW_h^{(0)}(\omega,\Omega) \big\vert^2 \,.
\end{equation}
We can evaluate the flux numerically using the integrals in 
Eq.~\eqref{eq:integrals_tree_num}, and Lebedev quadrature 
for angular integrations. In Fig.~\ref{fig:plot_LO_flux} we show the relativistic
spectrum computed numerically 
for different velocities.
From the small-velocity expansion of the tree-level waveform instead, we can obtain an analytic result of the LO [$\cO(G^3)$] power spectrum in powers of $p_\infty$. In this limit, all frequency-domain integrals are reduced to the Bessel functions 
$K_0(\normb\,\Omega),K_1(\normb\,\Omega)$. 
We obtained an analytic expression for the LO spectrum up to order $p_\infty^{8}$, which is in agreement with previous 
results~\cite{Jakobsen:2021smu,Bini:2024tft}. 

\subsection{NLO Power Spectrum}
We obtain the NLO contributions to the
power spectrum from the interference of
the LO and NLO frequency-space waveforms,
\begin{equation}
    \frac{d E^{\text{NLO}} }{d\omega} = 
    \frac{2}{\kappa^2}\!\int\! \dd\Omega \, \omega^2 
    \sum_{h=\pm} \big[\cW_h^{(1)}  \cW_{h}^{(0)*}+
    \cW_{h}^{(0)}  \cW_h^{(1)*}\big]\,.
\end{equation}
In Figs.~\ref{fig:plot_NLO_flux_m}~and~\ref{fig:plot_NLO_flux_y} we show the NLO power spectrum 
for different velocities and mass ratios. In all cases, the results for the spectrum are in good agreement with that obtained from the small-velocity expansion of the waveform in Ref.~\cite{Bini:2024rsy}, within the expected error from the truncation. Moreover, we have checked that the numerical integration of the spectrum reproduces the analytic result for the total radiated energy at $\Ord(G^4)$ given in Ref.~\cite{Dlapa:2022lmu} to a very good approximation.

\section{Conclusions}

In this article, we presented the first analytic expression
for the gravitational waveform in impact-parameter space
to next-to-leading order ($\Ord(\kappa^{5/2})$) for
scalar particles.  We computed it directly from scattering 
amplitudes via the observables-based (KMOC) formalism. 
We treat Fourier and loop integrations on the same footing, 
which offers a novel strategy that extends unitarity-based 
integrand construction and integration-by-parts (IBP) 
reduction techniques to accommodate the Fourier exponential.

We evaluate the resulting integrals analytically,
by combining canonical differential equations with a suitable 
contour deformation strategy. This hybrid method allows us to 
avoid spurious poles in the Fourier transform
 while maintaining 
analytic control over all master integrals contributing to the 
frequency-domain waveform.

We express the waveform in terms of a minimal basis of 28 
combined master integrals. This set of integrals is not
limited to emission from scalar scattering, but also
suffices for emission from systems of spinning bodies.

We have verified that our results are consistent with
those given previously in the literature~\cite{Brandhuber:2023hhy,Herderschee:2023fxh,Georgoudis:2023lgf,Bohnenblust:2023qmy,Bini:2024rsy} by separating infrared-divergent and tail contributions 
and confirmed agreement with known expressions in the 
small-velocity limit.
We also evaluated the NLO power spectrum numerically,
using Lebedev quadrature to integrate over the celestial sphere.
Our results demonstrate the power of amplitude-based 
techniques in computing classical observables and pave the way 
for further analytic studies of gravitational radiation 
at higher orders and including spin effects.

\acknowledgments
We would like to thank Lara Bohnenblust and Harald Ita for sharing the data of the plots in Ref.~\cite{Bohnenblust:2023qmy} for numerical comparison. SDA would like to thank Donato Bini and Thibault Damour for insightful discussions; and Vincent He, Aidan Herderschee, Julio Parra-Martinez, Radu Roiban, and Fei Teng for collaborations on related projects. GB would like to thank Vsevolod Chestnov and Pierpaolo Mastrolia for valuable discussions.
This research was supported by the European Research Council (ERC) under the European Union's research and innovation program grant agreement ERC-AdG-885414 (`Ampl2Einstein'). G.B. research is partially supported by the Italian MIUR under contract 20223ANFHR (PRIN2022), by the Università Italo-Francese, under grant Vinci, and by the Amplitudes INFN scientific initiative.

\bibliographystyle{apsrev4-1}
\bibliography{binary.bib}

\begin{thebibliography}{108}%
\makeatletter
\providecommand \@ifxundefined [1]{%
 \@ifx{#1\undefined}
}%
\providecommand \@ifnum [1]{%
 \ifnum #1\expandafter \@firstoftwo
 \else \expandafter \@secondoftwo
 \fi
}%
\providecommand \@ifx [1]{%
 \ifx #1\expandafter \@firstoftwo
 \else \expandafter \@secondoftwo
 \fi
}%
\providecommand \natexlab [1]{#1}%
\providecommand \enquote  [1]{``#1''}%
\providecommand \bibnamefont  [1]{#1}%
\providecommand \bibfnamefont [1]{#1}%
\providecommand \citenamefont [1]{#1}%
\providecommand \href@noop [0]{\@secondoftwo}%
\providecommand \href [0]{\begingroup \@sanitize@url \@href}%
\providecommand \@href[1]{\@@startlink{#1}\@@href}%
\providecommand \@@href[1]{\endgroup#1\@@endlink}%
\providecommand \@sanitize@url [0]{\catcode `\\12\catcode `\$12\catcode `\&12\catcode `\#12\catcode `\^12\catcode `\_12\catcode `\%12\relax}%
\providecommand \@@startlink[1]{}%
\providecommand \@@endlink[0]{}%
\providecommand \url  [0]{\begingroup\@sanitize@url \@url }%
\providecommand \@url [1]{\endgroup\@href {#1}{\urlprefix }}%
\providecommand \urlprefix  [0]{URL }%
\providecommand \Eprint [0]{\href }%
\providecommand \doibase [0]{http://dx.doi.org/}%
\providecommand \selectlanguage [0]{\@gobble}%
\providecommand \bibinfo  [0]{\@secondoftwo}%
\providecommand \bibfield  [0]{\@secondoftwo}%
\providecommand \translation [1]{[#1]}%
\providecommand \BibitemOpen [0]{}%
\providecommand \bibitemStop [0]{}%
\providecommand \bibitemNoStop [0]{.\EOS\space}%
\providecommand \EOS [0]{\spacefactor3000\relax}%
\providecommand \BibitemShut  [1]{\csname bibitem#1\endcsname}%
\let\auto@bib@innerbib\@empty
\bibitem [{\citenamefont {Brunello}\ and\ \citenamefont {De~Angelis}(2024)}]{Brunello:2024ibk}%
  \BibitemOpen
  \bibfield  {author} {\bibinfo {author} {\bibfnamefont {G.}~\bibnamefont {Brunello}}\ and\ \bibinfo {author} {\bibfnamefont {S.}~\bibnamefont {De~Angelis}},\ }\href {\doibase 10.1007/JHEP07(2024)062} {\bibfield  {journal} {\bibinfo  {journal} {JHEP}\ }\textbf {\bibinfo {volume} {07}},\ \bibinfo {pages} {062} (\bibinfo {year} {2024})},\ \Eprint {http://arxiv.org/abs/2403.08009} {arXiv:2403.08009 [hep-th]} \BibitemShut {NoStop}%
\bibitem [{\citenamefont {Blanchet}\ and\ \citenamefont {Damour}(1986)}]{Blanchet:1985sp}%
  \BibitemOpen
  \bibfield  {author} {\bibinfo {author} {\bibfnamefont {L.}~\bibnamefont {Blanchet}}\ and\ \bibinfo {author} {\bibfnamefont {T.}~\bibnamefont {Damour}},\ }\href {\doibase 10.1098/rsta.1986.0125} {\bibfield  {journal} {\bibinfo  {journal} {Phil. Trans. Roy. Soc. Lond. A}\ }\textbf {\bibinfo {volume} {320}},\ \bibinfo {pages} {379} (\bibinfo {year} {1986})}\BibitemShut {NoStop}%
\bibitem [{\citenamefont {Blanchet}\ and\ \citenamefont {Damour}(1989)}]{Blanchet:1989ki}%
  \BibitemOpen
  \bibfield  {author} {\bibinfo {author} {\bibfnamefont {L.}~\bibnamefont {Blanchet}}\ and\ \bibinfo {author} {\bibfnamefont {T.}~\bibnamefont {Damour}},\ }\href@noop {} {\bibfield  {journal} {\bibinfo  {journal} {Ann. Inst. H. Poincare Phys. Theor.}\ }\textbf {\bibinfo {volume} {50}},\ \bibinfo {pages} {377} (\bibinfo {year} {1989})}\BibitemShut {NoStop}%
\bibitem [{\citenamefont {Blanchet}(2014)}]{Blanchet:2013haa}%
  \BibitemOpen
  \bibfield  {author} {\bibinfo {author} {\bibfnamefont {L.}~\bibnamefont {Blanchet}},\ }\href {\doibase 10.12942/lrr-2014-2} {\bibfield  {journal} {\bibinfo  {journal} {Living Rev. Rel.}\ }\textbf {\bibinfo {volume} {17}},\ \bibinfo {pages} {2} (\bibinfo {year} {2014})},\ \Eprint {http://arxiv.org/abs/1310.1528} {arXiv:1310.1528 [gr-qc]} \BibitemShut {NoStop}%
\bibitem [{\citenamefont {Goldberger}\ and\ \citenamefont {Rothstein}(2006)}]{Goldberger:2004jt}%
  \BibitemOpen
  \bibfield  {author} {\bibinfo {author} {\bibfnamefont {W.~D.}\ \bibnamefont {Goldberger}}\ and\ \bibinfo {author} {\bibfnamefont {I.~Z.}\ \bibnamefont {Rothstein}},\ }\href {\doibase 10.1103/PhysRevD.73.104029} {\bibfield  {journal} {\bibinfo  {journal} {Phys. Rev. D}\ }\textbf {\bibinfo {volume} {73}},\ \bibinfo {pages} {104029} (\bibinfo {year} {2006})},\ \Eprint {http://arxiv.org/abs/hep-th/0409156} {arXiv:hep-th/0409156} \BibitemShut {NoStop}%
\bibitem [{\citenamefont {Goldberger}\ and\ \citenamefont {Ross}(2010)}]{Goldberger:2009qd}%
  \BibitemOpen
  \bibfield  {author} {\bibinfo {author} {\bibfnamefont {W.~D.}\ \bibnamefont {Goldberger}}\ and\ \bibinfo {author} {\bibfnamefont {A.}~\bibnamefont {Ross}},\ }\href {\doibase 10.1103/PhysRevD.81.124015} {\bibfield  {journal} {\bibinfo  {journal} {Phys. Rev. D}\ }\textbf {\bibinfo {volume} {81}},\ \bibinfo {pages} {124015} (\bibinfo {year} {2010})},\ \Eprint {http://arxiv.org/abs/0912.4254} {arXiv:0912.4254 [gr-qc]} \BibitemShut {NoStop}%
\bibitem [{\citenamefont {Foffa}\ and\ \citenamefont {Sturani}(2014)}]{Foffa:2013qca}%
  \BibitemOpen
  \bibfield  {author} {\bibinfo {author} {\bibfnamefont {S.}~\bibnamefont {Foffa}}\ and\ \bibinfo {author} {\bibfnamefont {R.}~\bibnamefont {Sturani}},\ }\href {\doibase 10.1088/0264-9381/31/4/043001} {\bibfield  {journal} {\bibinfo  {journal} {Class. Quant. Grav.}\ }\textbf {\bibinfo {volume} {31}},\ \bibinfo {pages} {043001} (\bibinfo {year} {2014})},\ \Eprint {http://arxiv.org/abs/1309.3474} {arXiv:1309.3474 [gr-qc]} \BibitemShut {NoStop}%
\bibitem [{\citenamefont {Porto}(2016)}]{Porto:2016pyg}%
  \BibitemOpen
  \bibfield  {author} {\bibinfo {author} {\bibfnamefont {R.~A.}\ \bibnamefont {Porto}},\ }\href {\doibase 10.1016/j.physrep.2016.04.003} {\bibfield  {journal} {\bibinfo  {journal} {Phys. Rept.}\ }\textbf {\bibinfo {volume} {633}},\ \bibinfo {pages} {1} (\bibinfo {year} {2016})},\ \Eprint {http://arxiv.org/abs/1601.04914} {arXiv:1601.04914 [hep-th]} \BibitemShut {NoStop}%
\bibitem [{\citenamefont {Foffa}\ \emph {et~al.}(2017)\citenamefont {Foffa}, \citenamefont {Mastrolia}, \citenamefont {Sturani},\ and\ \citenamefont {Sturm}}]{Foffa:2016rgu}%
  \BibitemOpen
  \bibfield  {author} {\bibinfo {author} {\bibfnamefont {S.}~\bibnamefont {Foffa}}, \bibinfo {author} {\bibfnamefont {P.}~\bibnamefont {Mastrolia}}, \bibinfo {author} {\bibfnamefont {R.}~\bibnamefont {Sturani}}, \ and\ \bibinfo {author} {\bibfnamefont {C.}~\bibnamefont {Sturm}},\ }\href {\doibase 10.1103/PhysRevD.95.104009} {\bibfield  {journal} {\bibinfo  {journal} {Phys. Rev. D}\ }\textbf {\bibinfo {volume} {95}},\ \bibinfo {pages} {104009} (\bibinfo {year} {2017})},\ \Eprint {http://arxiv.org/abs/1612.00482} {arXiv:1612.00482 [gr-qc]} \BibitemShut {NoStop}%
\bibitem [{\citenamefont {Foffa}\ \emph {et~al.}(2019{\natexlab{a}})\citenamefont {Foffa}, \citenamefont {Porto}, \citenamefont {Rothstein},\ and\ \citenamefont {Sturani}}]{Foffa:2019yfl}%
  \BibitemOpen
  \bibfield  {author} {\bibinfo {author} {\bibfnamefont {S.}~\bibnamefont {Foffa}}, \bibinfo {author} {\bibfnamefont {R.~A.}\ \bibnamefont {Porto}}, \bibinfo {author} {\bibfnamefont {I.}~\bibnamefont {Rothstein}}, \ and\ \bibinfo {author} {\bibfnamefont {R.}~\bibnamefont {Sturani}},\ }\href {\doibase 10.1103/PhysRevD.100.024048} {\bibfield  {journal} {\bibinfo  {journal} {Phys. Rev. D}\ }\textbf {\bibinfo {volume} {100}},\ \bibinfo {pages} {024048} (\bibinfo {year} {2019}{\natexlab{a}})},\ \Eprint {http://arxiv.org/abs/1903.05118} {arXiv:1903.05118 [gr-qc]} \BibitemShut {NoStop}%
\bibitem [{\citenamefont {Foffa}\ and\ \citenamefont {Sturani}(2019)}]{Foffa:2019rdf}%
  \BibitemOpen
  \bibfield  {author} {\bibinfo {author} {\bibfnamefont {S.}~\bibnamefont {Foffa}}\ and\ \bibinfo {author} {\bibfnamefont {R.}~\bibnamefont {Sturani}},\ }\href {\doibase 10.1103/PhysRevD.100.024047} {\bibfield  {journal} {\bibinfo  {journal} {Phys. Rev. D}\ }\textbf {\bibinfo {volume} {100}},\ \bibinfo {pages} {024047} (\bibinfo {year} {2019})},\ \Eprint {http://arxiv.org/abs/1903.05113} {arXiv:1903.05113 [gr-qc]} \BibitemShut {NoStop}%
\bibitem [{\citenamefont {Bl\"umlein}\ \emph {et~al.}(2020)\citenamefont {Bl\"umlein}, \citenamefont {Maier}, \citenamefont {Marquard},\ and\ \citenamefont {Sch\"afer}}]{Blumlein:2020pog}%
  \BibitemOpen
  \bibfield  {author} {\bibinfo {author} {\bibfnamefont {J.}~\bibnamefont {Bl\"umlein}}, \bibinfo {author} {\bibfnamefont {A.}~\bibnamefont {Maier}}, \bibinfo {author} {\bibfnamefont {P.}~\bibnamefont {Marquard}}, \ and\ \bibinfo {author} {\bibfnamefont {G.}~\bibnamefont {Sch\"afer}},\ }\href {\doibase 10.1016/j.nuclphysb.2020.115041} {\bibfield  {journal} {\bibinfo  {journal} {Nucl. Phys. B}\ }\textbf {\bibinfo {volume} {955}},\ \bibinfo {pages} {115041} (\bibinfo {year} {2020})},\ \Eprint {http://arxiv.org/abs/2003.01692} {arXiv:2003.01692 [gr-qc]} \BibitemShut {NoStop}%
\bibitem [{\citenamefont {Foffa}\ \emph {et~al.}(2019{\natexlab{b}})\citenamefont {Foffa}, \citenamefont {Mastrolia}, \citenamefont {Sturani}, \citenamefont {Sturm},\ and\ \citenamefont {Torres~Bobadilla}}]{Foffa:2019hrb}%
  \BibitemOpen
  \bibfield  {author} {\bibinfo {author} {\bibfnamefont {S.}~\bibnamefont {Foffa}}, \bibinfo {author} {\bibfnamefont {P.}~\bibnamefont {Mastrolia}}, \bibinfo {author} {\bibfnamefont {R.}~\bibnamefont {Sturani}}, \bibinfo {author} {\bibfnamefont {C.}~\bibnamefont {Sturm}}, \ and\ \bibinfo {author} {\bibfnamefont {W.~J.}\ \bibnamefont {Torres~Bobadilla}},\ }\href {\doibase 10.1103/PhysRevLett.122.241605} {\bibfield  {journal} {\bibinfo  {journal} {Phys. Rev. Lett.}\ }\textbf {\bibinfo {volume} {122}},\ \bibinfo {pages} {241605} (\bibinfo {year} {2019}{\natexlab{b}})},\ \Eprint {http://arxiv.org/abs/1902.10571} {arXiv:1902.10571 [gr-qc]} \BibitemShut {NoStop}%
\bibitem [{\citenamefont {Bl\"umlein}\ \emph {et~al.}(2022)\citenamefont {Bl\"umlein}, \citenamefont {Maier}, \citenamefont {Marquard},\ and\ \citenamefont {Sch\"afer}}]{Blumlein:2021txe}%
  \BibitemOpen
  \bibfield  {author} {\bibinfo {author} {\bibfnamefont {J.}~\bibnamefont {Bl\"umlein}}, \bibinfo {author} {\bibfnamefont {A.}~\bibnamefont {Maier}}, \bibinfo {author} {\bibfnamefont {P.}~\bibnamefont {Marquard}}, \ and\ \bibinfo {author} {\bibfnamefont {G.}~\bibnamefont {Sch\"afer}},\ }\href {\doibase 10.1016/j.nuclphysb.2022.115900} {\bibfield  {journal} {\bibinfo  {journal} {Nucl. Phys. B}\ }\textbf {\bibinfo {volume} {983}},\ \bibinfo {pages} {115900} (\bibinfo {year} {2022})},\ \bibinfo {note} {[Erratum: Nucl.Phys.B 985, 115991 (2022)]},\ \Eprint {http://arxiv.org/abs/2110.13822} {arXiv:2110.13822 [gr-qc]} \BibitemShut {NoStop}%
\bibitem [{\citenamefont {Porto}\ \emph {et~al.}(2025)\citenamefont {Porto}, \citenamefont {Riva},\ and\ \citenamefont {Yang}}]{Porto:2024cwd}%
  \BibitemOpen
  \bibfield  {author} {\bibinfo {author} {\bibfnamefont {R.~A.}\ \bibnamefont {Porto}}, \bibinfo {author} {\bibfnamefont {M.~M.}\ \bibnamefont {Riva}}, \ and\ \bibinfo {author} {\bibfnamefont {Z.}~\bibnamefont {Yang}},\ }\href {\doibase 10.1007/JHEP04(2025)050} {\bibfield  {journal} {\bibinfo  {journal} {JHEP}\ }\textbf {\bibinfo {volume} {04}},\ \bibinfo {pages} {050} (\bibinfo {year} {2025})},\ \Eprint {http://arxiv.org/abs/2409.05860} {arXiv:2409.05860 [gr-qc]} \BibitemShut {NoStop}%
\bibitem [{\citenamefont {Iwasaki}(1971)}]{Iwasaki:1971vb}%
  \BibitemOpen
  \bibfield  {author} {\bibinfo {author} {\bibfnamefont {Y.}~\bibnamefont {Iwasaki}},\ }\href {\doibase 10.1143/PTP.46.1587} {\bibfield  {journal} {\bibinfo  {journal} {Prog. Theor. Phys.}\ }\textbf {\bibinfo {volume} {46}},\ \bibinfo {pages} {1587} (\bibinfo {year} {1971})}\BibitemShut {NoStop}%
\bibitem [{\citenamefont {Damour}(2016)}]{Damour:2016gwp}%
  \BibitemOpen
  \bibfield  {author} {\bibinfo {author} {\bibfnamefont {T.}~\bibnamefont {Damour}},\ }\href {\doibase 10.1103/PhysRevD.94.104015} {\bibfield  {journal} {\bibinfo  {journal} {Phys. Rev. D}\ }\textbf {\bibinfo {volume} {94}},\ \bibinfo {pages} {104015} (\bibinfo {year} {2016})},\ \Eprint {http://arxiv.org/abs/1609.00354} {arXiv:1609.00354 [gr-qc]} \BibitemShut {NoStop}%
\bibitem [{\citenamefont {Damour}(2018)}]{Damour:2017zjx}%
  \BibitemOpen
  \bibfield  {author} {\bibinfo {author} {\bibfnamefont {T.}~\bibnamefont {Damour}},\ }\href {\doibase 10.1103/PhysRevD.97.044038} {\bibfield  {journal} {\bibinfo  {journal} {Phys. Rev. D}\ }\textbf {\bibinfo {volume} {97}},\ \bibinfo {pages} {044038} (\bibinfo {year} {2018})},\ \Eprint {http://arxiv.org/abs/1710.10599} {arXiv:1710.10599 [gr-qc]} \BibitemShut {NoStop}%
\bibitem [{\citenamefont {Cheung}\ \emph {et~al.}(2018)\citenamefont {Cheung}, \citenamefont {Rothstein},\ and\ \citenamefont {Solon}}]{Cheung:2018wkq}%
  \BibitemOpen
  \bibfield  {author} {\bibinfo {author} {\bibfnamefont {C.}~\bibnamefont {Cheung}}, \bibinfo {author} {\bibfnamefont {I.~Z.}\ \bibnamefont {Rothstein}}, \ and\ \bibinfo {author} {\bibfnamefont {M.~P.}\ \bibnamefont {Solon}},\ }\href {\doibase 10.1103/PhysRevLett.121.251101} {\bibfield  {journal} {\bibinfo  {journal} {Phys. Rev. Lett.}\ }\textbf {\bibinfo {volume} {121}},\ \bibinfo {pages} {251101} (\bibinfo {year} {2018})},\ \Eprint {http://arxiv.org/abs/1808.02489} {arXiv:1808.02489 [hep-th]} \BibitemShut {NoStop}%
\bibitem [{\citenamefont {Bjerrum-Bohr}\ \emph {et~al.}(2018)\citenamefont {Bjerrum-Bohr}, \citenamefont {Damgaard}, \citenamefont {Festuccia}, \citenamefont {Plant\'e},\ and\ \citenamefont {Vanhove}}]{Bjerrum-Bohr:2018xdl}%
  \BibitemOpen
  \bibfield  {author} {\bibinfo {author} {\bibfnamefont {N.~E.~J.}\ \bibnamefont {Bjerrum-Bohr}}, \bibinfo {author} {\bibfnamefont {P.~H.}\ \bibnamefont {Damgaard}}, \bibinfo {author} {\bibfnamefont {G.}~\bibnamefont {Festuccia}}, \bibinfo {author} {\bibfnamefont {L.}~\bibnamefont {Plant\'e}}, \ and\ \bibinfo {author} {\bibfnamefont {P.}~\bibnamefont {Vanhove}},\ }\href {\doibase 10.1103/PhysRevLett.121.171601} {\bibfield  {journal} {\bibinfo  {journal} {Phys. Rev. Lett.}\ }\textbf {\bibinfo {volume} {121}},\ \bibinfo {pages} {171601} (\bibinfo {year} {2018})},\ \Eprint {http://arxiv.org/abs/1806.04920} {arXiv:1806.04920 [hep-th]} \BibitemShut {NoStop}%
\bibitem [{\citenamefont {Bern}\ \emph {et~al.}(2019{\natexlab{a}})\citenamefont {Bern}, \citenamefont {Cheung}, \citenamefont {Roiban}, \citenamefont {Shen}, \citenamefont {Solon},\ and\ \citenamefont {Zeng}}]{Bern:2019nnu}%
  \BibitemOpen
  \bibfield  {author} {\bibinfo {author} {\bibfnamefont {Z.}~\bibnamefont {Bern}}, \bibinfo {author} {\bibfnamefont {C.}~\bibnamefont {Cheung}}, \bibinfo {author} {\bibfnamefont {R.}~\bibnamefont {Roiban}}, \bibinfo {author} {\bibfnamefont {C.-H.}\ \bibnamefont {Shen}}, \bibinfo {author} {\bibfnamefont {M.~P.}\ \bibnamefont {Solon}}, \ and\ \bibinfo {author} {\bibfnamefont {M.}~\bibnamefont {Zeng}},\ }\href {\doibase 10.1103/PhysRevLett.122.201603} {\bibfield  {journal} {\bibinfo  {journal} {Phys. Rev. Lett.}\ }\textbf {\bibinfo {volume} {122}},\ \bibinfo {pages} {201603} (\bibinfo {year} {2019}{\natexlab{a}})},\ \Eprint {http://arxiv.org/abs/1901.04424} {arXiv:1901.04424 [hep-th]} \BibitemShut {NoStop}%
\bibitem [{\citenamefont {Bern}\ \emph {et~al.}(2019{\natexlab{b}})\citenamefont {Bern}, \citenamefont {Cheung}, \citenamefont {Roiban}, \citenamefont {Shen}, \citenamefont {Solon},\ and\ \citenamefont {Zeng}}]{Bern:2019crd}%
  \BibitemOpen
  \bibfield  {author} {\bibinfo {author} {\bibfnamefont {Z.}~\bibnamefont {Bern}}, \bibinfo {author} {\bibfnamefont {C.}~\bibnamefont {Cheung}}, \bibinfo {author} {\bibfnamefont {R.}~\bibnamefont {Roiban}}, \bibinfo {author} {\bibfnamefont {C.-H.}\ \bibnamefont {Shen}}, \bibinfo {author} {\bibfnamefont {M.~P.}\ \bibnamefont {Solon}}, \ and\ \bibinfo {author} {\bibfnamefont {M.}~\bibnamefont {Zeng}},\ }\href {\doibase 10.1007/JHEP10(2019)206} {\bibfield  {journal} {\bibinfo  {journal} {JHEP}\ }\textbf {\bibinfo {volume} {10}},\ \bibinfo {pages} {206} (\bibinfo {year} {2019}{\natexlab{b}})},\ \Eprint {http://arxiv.org/abs/1908.01493} {arXiv:1908.01493 [hep-th]} \BibitemShut {NoStop}%
\bibitem [{\citenamefont {Parra-Martinez}\ \emph {et~al.}(2020)\citenamefont {Parra-Martinez}, \citenamefont {Ruf},\ and\ \citenamefont {Zeng}}]{Parra-Martinez:2020dzs}%
  \BibitemOpen
  \bibfield  {author} {\bibinfo {author} {\bibfnamefont {J.}~\bibnamefont {Parra-Martinez}}, \bibinfo {author} {\bibfnamefont {M.~S.}\ \bibnamefont {Ruf}}, \ and\ \bibinfo {author} {\bibfnamefont {M.}~\bibnamefont {Zeng}},\ }\href {\doibase 10.1007/JHEP11(2020)023} {\bibfield  {journal} {\bibinfo  {journal} {JHEP}\ }\textbf {\bibinfo {volume} {11}},\ \bibinfo {pages} {023} (\bibinfo {year} {2020})},\ \Eprint {http://arxiv.org/abs/2005.04236} {arXiv:2005.04236 [hep-th]} \BibitemShut {NoStop}%
\bibitem [{\citenamefont {Cheung}\ and\ \citenamefont {Solon}(2020)}]{Cheung:2020gyp}%
  \BibitemOpen
  \bibfield  {author} {\bibinfo {author} {\bibfnamefont {C.}~\bibnamefont {Cheung}}\ and\ \bibinfo {author} {\bibfnamefont {M.~P.}\ \bibnamefont {Solon}},\ }\href {\doibase 10.1007/JHEP06(2020)144} {\bibfield  {journal} {\bibinfo  {journal} {JHEP}\ }\textbf {\bibinfo {volume} {06}},\ \bibinfo {pages} {144} (\bibinfo {year} {2020})},\ \Eprint {http://arxiv.org/abs/2003.08351} {arXiv:2003.08351 [hep-th]} \BibitemShut {NoStop}%
\bibitem [{\citenamefont {Di~Vecchia}\ \emph {et~al.}(2021)\citenamefont {Di~Vecchia}, \citenamefont {Heissenberg}, \citenamefont {Russo},\ and\ \citenamefont {Veneziano}}]{DiVecchia:2021bdo}%
  \BibitemOpen
  \bibfield  {author} {\bibinfo {author} {\bibfnamefont {P.}~\bibnamefont {Di~Vecchia}}, \bibinfo {author} {\bibfnamefont {C.}~\bibnamefont {Heissenberg}}, \bibinfo {author} {\bibfnamefont {R.}~\bibnamefont {Russo}}, \ and\ \bibinfo {author} {\bibfnamefont {G.}~\bibnamefont {Veneziano}},\ }\href {\doibase 10.1007/JHEP07(2021)169} {\bibfield  {journal} {\bibinfo  {journal} {JHEP}\ }\textbf {\bibinfo {volume} {07}},\ \bibinfo {pages} {169} (\bibinfo {year} {2021})},\ \Eprint {http://arxiv.org/abs/2104.03256} {arXiv:2104.03256 [hep-th]} \BibitemShut {NoStop}%
\bibitem [{\citenamefont {Brandhuber}\ \emph {et~al.}(2021{\natexlab{a}})\citenamefont {Brandhuber}, \citenamefont {Chen}, \citenamefont {Travaglini},\ and\ \citenamefont {Wen}}]{Brandhuber:2021eyq}%
  \BibitemOpen
  \bibfield  {author} {\bibinfo {author} {\bibfnamefont {A.}~\bibnamefont {Brandhuber}}, \bibinfo {author} {\bibfnamefont {G.}~\bibnamefont {Chen}}, \bibinfo {author} {\bibfnamefont {G.}~\bibnamefont {Travaglini}}, \ and\ \bibinfo {author} {\bibfnamefont {C.}~\bibnamefont {Wen}},\ }\href {\doibase 10.1007/JHEP10(2021)118} {\bibfield  {journal} {\bibinfo  {journal} {JHEP}\ }\textbf {\bibinfo {volume} {10}},\ \bibinfo {pages} {118} (\bibinfo {year} {2021}{\natexlab{a}})},\ \Eprint {http://arxiv.org/abs/2108.04216} {arXiv:2108.04216 [hep-th]} \BibitemShut {NoStop}%
\bibitem [{\citenamefont {K\"alin}\ \emph {et~al.}(2020)\citenamefont {K\"alin}, \citenamefont {Liu},\ and\ \citenamefont {Porto}}]{Kalin:2020fhe}%
  \BibitemOpen
  \bibfield  {author} {\bibinfo {author} {\bibfnamefont {G.}~\bibnamefont {K\"alin}}, \bibinfo {author} {\bibfnamefont {Z.}~\bibnamefont {Liu}}, \ and\ \bibinfo {author} {\bibfnamefont {R.~A.}\ \bibnamefont {Porto}},\ }\href {\doibase 10.1103/PhysRevLett.125.261103} {\bibfield  {journal} {\bibinfo  {journal} {Phys. Rev. Lett.}\ }\textbf {\bibinfo {volume} {125}},\ \bibinfo {pages} {261103} (\bibinfo {year} {2020})},\ \Eprint {http://arxiv.org/abs/2007.04977} {arXiv:2007.04977 [hep-th]} \BibitemShut {NoStop}%
\bibitem [{\citenamefont {K\"alin}\ and\ \citenamefont {Porto}(2020)}]{Kalin:2020mvi}%
  \BibitemOpen
  \bibfield  {author} {\bibinfo {author} {\bibfnamefont {G.}~\bibnamefont {K\"alin}}\ and\ \bibinfo {author} {\bibfnamefont {R.~A.}\ \bibnamefont {Porto}},\ }\href {\doibase 10.1007/JHEP11(2020)106} {\bibfield  {journal} {\bibinfo  {journal} {JHEP}\ }\textbf {\bibinfo {volume} {11}},\ \bibinfo {pages} {106} (\bibinfo {year} {2020})},\ \Eprint {http://arxiv.org/abs/2006.01184} {arXiv:2006.01184 [hep-th]} \BibitemShut {NoStop}%
\bibitem [{\citenamefont {Mogull}\ \emph {et~al.}(2021)\citenamefont {Mogull}, \citenamefont {Plefka},\ and\ \citenamefont {Steinhoff}}]{Mogull:2020sak}%
  \BibitemOpen
  \bibfield  {author} {\bibinfo {author} {\bibfnamefont {G.}~\bibnamefont {Mogull}}, \bibinfo {author} {\bibfnamefont {J.}~\bibnamefont {Plefka}}, \ and\ \bibinfo {author} {\bibfnamefont {J.}~\bibnamefont {Steinhoff}},\ }\href {\doibase 10.1007/JHEP02(2021)048} {\bibfield  {journal} {\bibinfo  {journal} {JHEP}\ }\textbf {\bibinfo {volume} {02}},\ \bibinfo {pages} {048} (\bibinfo {year} {2021})},\ \Eprint {http://arxiv.org/abs/2010.02865} {arXiv:2010.02865 [hep-th]} \BibitemShut {NoStop}%
\bibitem [{\citenamefont {Dlapa}\ \emph {et~al.}(2022)\citenamefont {Dlapa}, \citenamefont {K\"alin}, \citenamefont {Liu},\ and\ \citenamefont {Porto}}]{Dlapa:2021npj}%
  \BibitemOpen
  \bibfield  {author} {\bibinfo {author} {\bibfnamefont {C.}~\bibnamefont {Dlapa}}, \bibinfo {author} {\bibfnamefont {G.}~\bibnamefont {K\"alin}}, \bibinfo {author} {\bibfnamefont {Z.}~\bibnamefont {Liu}}, \ and\ \bibinfo {author} {\bibfnamefont {R.~A.}\ \bibnamefont {Porto}},\ }\href {\doibase 10.1016/j.physletb.2022.137203} {\bibfield  {journal} {\bibinfo  {journal} {Phys. Lett. B}\ }\textbf {\bibinfo {volume} {831}},\ \bibinfo {pages} {137203} (\bibinfo {year} {2022})},\ \Eprint {http://arxiv.org/abs/2106.08276} {arXiv:2106.08276 [hep-th]} \BibitemShut {NoStop}%
\bibitem [{\citenamefont {Bern}\ \emph {et~al.}(2021)\citenamefont {Bern}, \citenamefont {Parra-Martinez}, \citenamefont {Roiban}, \citenamefont {Ruf}, \citenamefont {Shen}, \citenamefont {Solon},\ and\ \citenamefont {Zeng}}]{Bern:2021dqo}%
  \BibitemOpen
  \bibfield  {author} {\bibinfo {author} {\bibfnamefont {Z.}~\bibnamefont {Bern}}, \bibinfo {author} {\bibfnamefont {J.}~\bibnamefont {Parra-Martinez}}, \bibinfo {author} {\bibfnamefont {R.}~\bibnamefont {Roiban}}, \bibinfo {author} {\bibfnamefont {M.~S.}\ \bibnamefont {Ruf}}, \bibinfo {author} {\bibfnamefont {C.-H.}\ \bibnamefont {Shen}}, \bibinfo {author} {\bibfnamefont {M.~P.}\ \bibnamefont {Solon}}, \ and\ \bibinfo {author} {\bibfnamefont {M.}~\bibnamefont {Zeng}},\ }\href {\doibase 10.1103/PhysRevLett.126.171601} {\bibfield  {journal} {\bibinfo  {journal} {Phys. Rev. Lett.}\ }\textbf {\bibinfo {volume} {126}},\ \bibinfo {pages} {171601} (\bibinfo {year} {2021})},\ \Eprint {http://arxiv.org/abs/2101.07254} {arXiv:2101.07254 [hep-th]} \BibitemShut {NoStop}%
\bibitem [{\citenamefont {Bern}\ \emph {et~al.}(2022)\citenamefont {Bern}, \citenamefont {Parra-Martinez}, \citenamefont {Roiban}, \citenamefont {Ruf}, \citenamefont {Shen}, \citenamefont {Solon},\ and\ \citenamefont {Zeng}}]{Bern:2022jvn}%
  \BibitemOpen
  \bibfield  {author} {\bibinfo {author} {\bibfnamefont {Z.}~\bibnamefont {Bern}}, \bibinfo {author} {\bibfnamefont {J.}~\bibnamefont {Parra-Martinez}}, \bibinfo {author} {\bibfnamefont {R.}~\bibnamefont {Roiban}}, \bibinfo {author} {\bibfnamefont {M.~S.}\ \bibnamefont {Ruf}}, \bibinfo {author} {\bibfnamefont {C.-H.}\ \bibnamefont {Shen}}, \bibinfo {author} {\bibfnamefont {M.~P.}\ \bibnamefont {Solon}}, \ and\ \bibinfo {author} {\bibfnamefont {M.}~\bibnamefont {Zeng}},\ }\href {\doibase 10.22323/1.416.0051} {\bibfield  {journal} {\bibinfo  {journal} {PoS}\ }\textbf {\bibinfo {volume} {LL2022}},\ \bibinfo {pages} {051} (\bibinfo {year} {2022})}\BibitemShut {NoStop}%
\bibitem [{\citenamefont {Dlapa}\ \emph {et~al.}(2023)\citenamefont {Dlapa}, \citenamefont {K\"alin}, \citenamefont {Liu}, \citenamefont {Neef},\ and\ \citenamefont {Porto}}]{Dlapa:2022lmu}%
  \BibitemOpen
  \bibfield  {author} {\bibinfo {author} {\bibfnamefont {C.}~\bibnamefont {Dlapa}}, \bibinfo {author} {\bibfnamefont {G.}~\bibnamefont {K\"alin}}, \bibinfo {author} {\bibfnamefont {Z.}~\bibnamefont {Liu}}, \bibinfo {author} {\bibfnamefont {J.}~\bibnamefont {Neef}}, \ and\ \bibinfo {author} {\bibfnamefont {R.~A.}\ \bibnamefont {Porto}},\ }\href {\doibase 10.1103/PhysRevLett.130.101401} {\bibfield  {journal} {\bibinfo  {journal} {Phys. Rev. Lett.}\ }\textbf {\bibinfo {volume} {130}},\ \bibinfo {pages} {101401} (\bibinfo {year} {2023})},\ \Eprint {http://arxiv.org/abs/2210.05541} {arXiv:2210.05541 [hep-th]} \BibitemShut {NoStop}%
\bibitem [{\citenamefont {Jakobsen}\ \emph {et~al.}(2023{\natexlab{a}})\citenamefont {Jakobsen}, \citenamefont {Mogull}, \citenamefont {Plefka}, \citenamefont {Sauer},\ and\ \citenamefont {Xu}}]{Jakobsen:2023ndj}%
  \BibitemOpen
  \bibfield  {author} {\bibinfo {author} {\bibfnamefont {G.~U.}\ \bibnamefont {Jakobsen}}, \bibinfo {author} {\bibfnamefont {G.}~\bibnamefont {Mogull}}, \bibinfo {author} {\bibfnamefont {J.}~\bibnamefont {Plefka}}, \bibinfo {author} {\bibfnamefont {B.}~\bibnamefont {Sauer}}, \ and\ \bibinfo {author} {\bibfnamefont {Y.}~\bibnamefont {Xu}},\ }\href {\doibase 10.1103/PhysRevLett.131.151401} {\bibfield  {journal} {\bibinfo  {journal} {Phys. Rev. Lett.}\ }\textbf {\bibinfo {volume} {131}},\ \bibinfo {pages} {151401} (\bibinfo {year} {2023}{\natexlab{a}})},\ \Eprint {http://arxiv.org/abs/2306.01714} {arXiv:2306.01714 [hep-th]} \BibitemShut {NoStop}%
\bibitem [{\citenamefont {Jakobsen}\ \emph {et~al.}(2023{\natexlab{b}})\citenamefont {Jakobsen}, \citenamefont {Mogull}, \citenamefont {Plefka},\ and\ \citenamefont {Sauer}}]{Jakobsen:2023hig}%
  \BibitemOpen
  \bibfield  {author} {\bibinfo {author} {\bibfnamefont {G.~U.}\ \bibnamefont {Jakobsen}}, \bibinfo {author} {\bibfnamefont {G.}~\bibnamefont {Mogull}}, \bibinfo {author} {\bibfnamefont {J.}~\bibnamefont {Plefka}}, \ and\ \bibinfo {author} {\bibfnamefont {B.}~\bibnamefont {Sauer}},\ }\href {\doibase 10.1103/PhysRevLett.131.241402} {\bibfield  {journal} {\bibinfo  {journal} {Phys. Rev. Lett.}\ }\textbf {\bibinfo {volume} {131}},\ \bibinfo {pages} {241402} (\bibinfo {year} {2023}{\natexlab{b}})},\ \Eprint {http://arxiv.org/abs/2308.11514} {arXiv:2308.11514 [hep-th]} \BibitemShut {NoStop}%
\bibitem [{\citenamefont {Damgaard}\ \emph {et~al.}(2023)\citenamefont {Damgaard}, \citenamefont {Hansen}, \citenamefont {Plant\'e},\ and\ \citenamefont {Vanhove}}]{Damgaard:2023ttc}%
  \BibitemOpen
  \bibfield  {author} {\bibinfo {author} {\bibfnamefont {P.~H.}\ \bibnamefont {Damgaard}}, \bibinfo {author} {\bibfnamefont {E.~R.}\ \bibnamefont {Hansen}}, \bibinfo {author} {\bibfnamefont {L.}~\bibnamefont {Plant\'e}}, \ and\ \bibinfo {author} {\bibfnamefont {P.}~\bibnamefont {Vanhove}},\ }\href {\doibase 10.1007/JHEP09(2023)183} {\bibfield  {journal} {\bibinfo  {journal} {JHEP}\ }\textbf {\bibinfo {volume} {09}},\ \bibinfo {pages} {183} (\bibinfo {year} {2023})},\ \Eprint {http://arxiv.org/abs/2307.04746} {arXiv:2307.04746 [hep-th]} \BibitemShut {NoStop}%
\bibitem [{\citenamefont {Bern}\ \emph {et~al.}(2024{\natexlab{a}})\citenamefont {Bern}, \citenamefont {Herrmann}, \citenamefont {Roiban}, \citenamefont {Ruf}, \citenamefont {Smirnov}, \citenamefont {Smirnov},\ and\ \citenamefont {Zeng}}]{Bern:2023ccb}%
  \BibitemOpen
  \bibfield  {author} {\bibinfo {author} {\bibfnamefont {Z.}~\bibnamefont {Bern}}, \bibinfo {author} {\bibfnamefont {E.}~\bibnamefont {Herrmann}}, \bibinfo {author} {\bibfnamefont {R.}~\bibnamefont {Roiban}}, \bibinfo {author} {\bibfnamefont {M.~S.}\ \bibnamefont {Ruf}}, \bibinfo {author} {\bibfnamefont {A.~V.}\ \bibnamefont {Smirnov}}, \bibinfo {author} {\bibfnamefont {V.~A.}\ \bibnamefont {Smirnov}}, \ and\ \bibinfo {author} {\bibfnamefont {M.}~\bibnamefont {Zeng}},\ }\href {\doibase 10.1103/PhysRevLett.132.251601} {\bibfield  {journal} {\bibinfo  {journal} {Phys. Rev. Lett.}\ }\textbf {\bibinfo {volume} {132}},\ \bibinfo {pages} {251601} (\bibinfo {year} {2024}{\natexlab{a}})},\ \Eprint {http://arxiv.org/abs/2305.08981} {arXiv:2305.08981 [hep-th]} \BibitemShut {NoStop}%
\bibitem [{\citenamefont {Driesse}\ \emph {et~al.}(2024)\citenamefont {Driesse}, \citenamefont {Jakobsen}, \citenamefont {Mogull}, \citenamefont {Plefka}, \citenamefont {Sauer},\ and\ \citenamefont {Usovitsch}}]{Driesse:2024xad}%
  \BibitemOpen
  \bibfield  {author} {\bibinfo {author} {\bibfnamefont {M.}~\bibnamefont {Driesse}}, \bibinfo {author} {\bibfnamefont {G.~U.}\ \bibnamefont {Jakobsen}}, \bibinfo {author} {\bibfnamefont {G.}~\bibnamefont {Mogull}}, \bibinfo {author} {\bibfnamefont {J.}~\bibnamefont {Plefka}}, \bibinfo {author} {\bibfnamefont {B.}~\bibnamefont {Sauer}}, \ and\ \bibinfo {author} {\bibfnamefont {J.}~\bibnamefont {Usovitsch}},\ }\href {\doibase 10.1103/PhysRevLett.132.241402} {\bibfield  {journal} {\bibinfo  {journal} {Phys. Rev. Lett.}\ }\textbf {\bibinfo {volume} {132}},\ \bibinfo {pages} {241402} (\bibinfo {year} {2024})},\ \Eprint {http://arxiv.org/abs/2403.07781} {arXiv:2403.07781 [hep-th]} \BibitemShut {NoStop}%
\bibitem [{\citenamefont {Bern}\ \emph {et~al.}(2024{\natexlab{b}})\citenamefont {Bern}, \citenamefont {Herrmann}, \citenamefont {Roiban}, \citenamefont {Ruf}, \citenamefont {Smirnov}, \citenamefont {Smirnov},\ and\ \citenamefont {Zeng}}]{Bern:2024adl}%
  \BibitemOpen
  \bibfield  {author} {\bibinfo {author} {\bibfnamefont {Z.}~\bibnamefont {Bern}}, \bibinfo {author} {\bibfnamefont {E.}~\bibnamefont {Herrmann}}, \bibinfo {author} {\bibfnamefont {R.}~\bibnamefont {Roiban}}, \bibinfo {author} {\bibfnamefont {M.~S.}\ \bibnamefont {Ruf}}, \bibinfo {author} {\bibfnamefont {A.~V.}\ \bibnamefont {Smirnov}}, \bibinfo {author} {\bibfnamefont {V.~A.}\ \bibnamefont {Smirnov}}, \ and\ \bibinfo {author} {\bibfnamefont {M.}~\bibnamefont {Zeng}},\ }\href {\doibase 10.1007/JHEP10(2024)023} {\bibfield  {journal} {\bibinfo  {journal} {JHEP}\ }\textbf {\bibinfo {volume} {10}},\ \bibinfo {pages} {023} (\bibinfo {year} {2024}{\natexlab{b}})},\ \Eprint {http://arxiv.org/abs/2406.01554} {arXiv:2406.01554 [hep-th]} \BibitemShut {NoStop}%
\bibitem [{\citenamefont {Driesse}\ \emph {et~al.}(2025)\citenamefont {Driesse}, \citenamefont {Jakobsen}, \citenamefont {Klemm}, \citenamefont {Mogull}, \citenamefont {Nega}, \citenamefont {Plefka}, \citenamefont {Sauer},\ and\ \citenamefont {Usovitsch}}]{Driesse:2024feo}%
  \BibitemOpen
  \bibfield  {author} {\bibinfo {author} {\bibfnamefont {M.}~\bibnamefont {Driesse}}, \bibinfo {author} {\bibfnamefont {G.~U.}\ \bibnamefont {Jakobsen}}, \bibinfo {author} {\bibfnamefont {A.}~\bibnamefont {Klemm}}, \bibinfo {author} {\bibfnamefont {G.}~\bibnamefont {Mogull}}, \bibinfo {author} {\bibfnamefont {C.}~\bibnamefont {Nega}}, \bibinfo {author} {\bibfnamefont {J.}~\bibnamefont {Plefka}}, \bibinfo {author} {\bibfnamefont {B.}~\bibnamefont {Sauer}}, \ and\ \bibinfo {author} {\bibfnamefont {J.}~\bibnamefont {Usovitsch}},\ }\href {\doibase 10.1038/s41586-025-08984-2} {\bibfield  {journal} {\bibinfo  {journal} {Nature}\ }\textbf {\bibinfo {volume} {641}},\ \bibinfo {pages} {603} (\bibinfo {year} {2025})},\ \Eprint {http://arxiv.org/abs/2411.11846} {arXiv:2411.11846 [hep-th]} \BibitemShut {NoStop}%
\bibitem [{\citenamefont {Damour}\ and\ \citenamefont {Rettegno}(2023)}]{Damour:2022ybd}%
  \BibitemOpen
  \bibfield  {author} {\bibinfo {author} {\bibfnamefont {T.}~\bibnamefont {Damour}}\ and\ \bibinfo {author} {\bibfnamefont {P.}~\bibnamefont {Rettegno}},\ }\href {\doibase 10.1103/PhysRevD.107.064051} {\bibfield  {journal} {\bibinfo  {journal} {Phys. Rev. D}\ }\textbf {\bibinfo {volume} {107}},\ \bibinfo {pages} {064051} (\bibinfo {year} {2023})},\ \Eprint {http://arxiv.org/abs/2211.01399} {arXiv:2211.01399 [gr-qc]} \BibitemShut {NoStop}%
\bibitem [{\citenamefont {Rettegno}\ \emph {et~al.}(2023)\citenamefont {Rettegno}, \citenamefont {Pratten}, \citenamefont {Thomas}, \citenamefont {Schmidt},\ and\ \citenamefont {Damour}}]{Rettegno:2023ghr}%
  \BibitemOpen
  \bibfield  {author} {\bibinfo {author} {\bibfnamefont {P.}~\bibnamefont {Rettegno}}, \bibinfo {author} {\bibfnamefont {G.}~\bibnamefont {Pratten}}, \bibinfo {author} {\bibfnamefont {L.~M.}\ \bibnamefont {Thomas}}, \bibinfo {author} {\bibfnamefont {P.}~\bibnamefont {Schmidt}}, \ and\ \bibinfo {author} {\bibfnamefont {T.}~\bibnamefont {Damour}},\ }\href {\doibase 10.1103/PhysRevD.108.124016} {\bibfield  {journal} {\bibinfo  {journal} {Phys. Rev. D}\ }\textbf {\bibinfo {volume} {108}},\ \bibinfo {pages} {124016} (\bibinfo {year} {2023})},\ \Eprint {http://arxiv.org/abs/2307.06999} {arXiv:2307.06999 [gr-qc]} \BibitemShut {NoStop}%
\bibitem [{\citenamefont {Jakobsen}\ \emph {et~al.}(2021)\citenamefont {Jakobsen}, \citenamefont {Mogull}, \citenamefont {Plefka},\ and\ \citenamefont {Steinhoff}}]{Jakobsen:2021smu}%
  \BibitemOpen
  \bibfield  {author} {\bibinfo {author} {\bibfnamefont {G.~U.}\ \bibnamefont {Jakobsen}}, \bibinfo {author} {\bibfnamefont {G.}~\bibnamefont {Mogull}}, \bibinfo {author} {\bibfnamefont {J.}~\bibnamefont {Plefka}}, \ and\ \bibinfo {author} {\bibfnamefont {J.}~\bibnamefont {Steinhoff}},\ }\href {\doibase 10.1103/PhysRevLett.126.201103} {\bibfield  {journal} {\bibinfo  {journal} {Phys. Rev. Lett.}\ }\textbf {\bibinfo {volume} {126}},\ \bibinfo {pages} {201103} (\bibinfo {year} {2021})},\ \Eprint {http://arxiv.org/abs/2101.12688} {arXiv:2101.12688 [gr-qc]} \BibitemShut {NoStop}%
\bibitem [{\citenamefont {Mougiakakos}\ \emph {et~al.}(2021)\citenamefont {Mougiakakos}, \citenamefont {Riva},\ and\ \citenamefont {Vernizzi}}]{Mougiakakos:2021ckm}%
  \BibitemOpen
  \bibfield  {author} {\bibinfo {author} {\bibfnamefont {S.}~\bibnamefont {Mougiakakos}}, \bibinfo {author} {\bibfnamefont {M.~M.}\ \bibnamefont {Riva}}, \ and\ \bibinfo {author} {\bibfnamefont {F.}~\bibnamefont {Vernizzi}},\ }\href {\doibase 10.1103/PhysRevD.104.024041} {\bibfield  {journal} {\bibinfo  {journal} {Phys. Rev. D}\ }\textbf {\bibinfo {volume} {104}},\ \bibinfo {pages} {024041} (\bibinfo {year} {2021})},\ \Eprint {http://arxiv.org/abs/2102.08339} {arXiv:2102.08339 [gr-qc]} \BibitemShut {NoStop}%
\bibitem [{\citenamefont {Jakobsen}\ \emph {et~al.}(2022{\natexlab{a}})\citenamefont {Jakobsen}, \citenamefont {Mogull}, \citenamefont {Plefka},\ and\ \citenamefont {Steinhoff}}]{Jakobsen:2021lvp}%
  \BibitemOpen
  \bibfield  {author} {\bibinfo {author} {\bibfnamefont {G.~U.}\ \bibnamefont {Jakobsen}}, \bibinfo {author} {\bibfnamefont {G.}~\bibnamefont {Mogull}}, \bibinfo {author} {\bibfnamefont {J.}~\bibnamefont {Plefka}}, \ and\ \bibinfo {author} {\bibfnamefont {J.}~\bibnamefont {Steinhoff}},\ }\href {\doibase 10.1103/PhysRevLett.128.011101} {\bibfield  {journal} {\bibinfo  {journal} {Phys. Rev. Lett.}\ }\textbf {\bibinfo {volume} {128}},\ \bibinfo {pages} {011101} (\bibinfo {year} {2022}{\natexlab{a}})},\ \Eprint {http://arxiv.org/abs/2106.10256} {arXiv:2106.10256 [hep-th]} \BibitemShut {NoStop}%
\bibitem [{\citenamefont {De~Angelis}\ \emph {et~al.}(2024)\citenamefont {De~Angelis}, \citenamefont {Novichkov},\ and\ \citenamefont {Gonzo}}]{DeAngelis:2023lvf}%
  \BibitemOpen
  \bibfield  {author} {\bibinfo {author} {\bibfnamefont {S.}~\bibnamefont {De~Angelis}}, \bibinfo {author} {\bibfnamefont {P.~P.}\ \bibnamefont {Novichkov}}, \ and\ \bibinfo {author} {\bibfnamefont {R.}~\bibnamefont {Gonzo}},\ }\href {\doibase 10.1103/PhysRevD.110.L041502} {\bibfield  {journal} {\bibinfo  {journal} {Phys. Rev. D}\ }\textbf {\bibinfo {volume} {110}},\ \bibinfo {pages} {L041502} (\bibinfo {year} {2024})},\ \Eprint {http://arxiv.org/abs/2309.17429} {arXiv:2309.17429 [hep-th]} \BibitemShut {NoStop}%
\bibitem [{\citenamefont {Brandhuber}\ \emph {et~al.}(2024)\citenamefont {Brandhuber}, \citenamefont {Brown}, \citenamefont {Chen}, \citenamefont {Gowdy},\ and\ \citenamefont {Travaglini}}]{Brandhuber:2023hhl}%
  \BibitemOpen
  \bibfield  {author} {\bibinfo {author} {\bibfnamefont {A.}~\bibnamefont {Brandhuber}}, \bibinfo {author} {\bibfnamefont {G.~R.}\ \bibnamefont {Brown}}, \bibinfo {author} {\bibfnamefont {G.}~\bibnamefont {Chen}}, \bibinfo {author} {\bibfnamefont {J.}~\bibnamefont {Gowdy}}, \ and\ \bibinfo {author} {\bibfnamefont {G.}~\bibnamefont {Travaglini}},\ }\href {\doibase 10.1007/JHEP02(2024)026} {\bibfield  {journal} {\bibinfo  {journal} {JHEP}\ }\textbf {\bibinfo {volume} {02}},\ \bibinfo {pages} {026} (\bibinfo {year} {2024})},\ \Eprint {http://arxiv.org/abs/2310.04405} {arXiv:2310.04405 [hep-th]} \BibitemShut {NoStop}%
\bibitem [{\citenamefont {Aoude}\ \emph {et~al.}(2024)\citenamefont {Aoude}, \citenamefont {Haddad}, \citenamefont {Heissenberg},\ and\ \citenamefont {Helset}}]{Aoude:2023dui}%
  \BibitemOpen
  \bibfield  {author} {\bibinfo {author} {\bibfnamefont {R.}~\bibnamefont {Aoude}}, \bibinfo {author} {\bibfnamefont {K.}~\bibnamefont {Haddad}}, \bibinfo {author} {\bibfnamefont {C.}~\bibnamefont {Heissenberg}}, \ and\ \bibinfo {author} {\bibfnamefont {A.}~\bibnamefont {Helset}},\ }\href {\doibase 10.1103/PhysRevD.109.036007} {\bibfield  {journal} {\bibinfo  {journal} {Phys. Rev. D}\ }\textbf {\bibinfo {volume} {109}},\ \bibinfo {pages} {036007} (\bibinfo {year} {2024})},\ \Eprint {http://arxiv.org/abs/2310.05832} {arXiv:2310.05832 [hep-th]} \BibitemShut {NoStop}%
\bibitem [{\citenamefont {Kosower}\ \emph {et~al.}(2019)\citenamefont {Kosower}, \citenamefont {Maybee},\ and\ \citenamefont {O'Connell}}]{Kosower:2018adc}%
  \BibitemOpen
  \bibfield  {author} {\bibinfo {author} {\bibfnamefont {D.~A.}\ \bibnamefont {Kosower}}, \bibinfo {author} {\bibfnamefont {B.}~\bibnamefont {Maybee}}, \ and\ \bibinfo {author} {\bibfnamefont {D.}~\bibnamefont {O'Connell}},\ }\href {\doibase 10.1007/JHEP02(2019)137} {\bibfield  {journal} {\bibinfo  {journal} {JHEP}\ }\textbf {\bibinfo {volume} {02}},\ \bibinfo {pages} {137} (\bibinfo {year} {2019})},\ \Eprint {http://arxiv.org/abs/1811.10950} {arXiv:1811.10950 [hep-th]} \BibitemShut {NoStop}%
\bibitem [{\citenamefont {Cristofoli}\ \emph {et~al.}(2022)\citenamefont {Cristofoli}, \citenamefont {Gonzo}, \citenamefont {Kosower},\ and\ \citenamefont {O'Connell}}]{Cristofoli:2021vyo}%
  \BibitemOpen
  \bibfield  {author} {\bibinfo {author} {\bibfnamefont {A.}~\bibnamefont {Cristofoli}}, \bibinfo {author} {\bibfnamefont {R.}~\bibnamefont {Gonzo}}, \bibinfo {author} {\bibfnamefont {D.~A.}\ \bibnamefont {Kosower}}, \ and\ \bibinfo {author} {\bibfnamefont {D.}~\bibnamefont {O'Connell}},\ }\href {\doibase 10.1103/PhysRevD.106.056007} {\bibfield  {journal} {\bibinfo  {journal} {Phys. Rev. D}\ }\textbf {\bibinfo {volume} {106}},\ \bibinfo {pages} {056007} (\bibinfo {year} {2022})},\ \Eprint {http://arxiv.org/abs/2107.10193} {arXiv:2107.10193 [hep-th]} \BibitemShut {NoStop}%
\bibitem [{\citenamefont {Brandhuber}\ \emph {et~al.}(2023)\citenamefont {Brandhuber}, \citenamefont {Brown}, \citenamefont {Chen}, \citenamefont {De~Angelis}, \citenamefont {Gowdy},\ and\ \citenamefont {Travaglini}}]{Brandhuber:2023hhy}%
  \BibitemOpen
  \bibfield  {author} {\bibinfo {author} {\bibfnamefont {A.}~\bibnamefont {Brandhuber}}, \bibinfo {author} {\bibfnamefont {G.~R.}\ \bibnamefont {Brown}}, \bibinfo {author} {\bibfnamefont {G.}~\bibnamefont {Chen}}, \bibinfo {author} {\bibfnamefont {S.}~\bibnamefont {De~Angelis}}, \bibinfo {author} {\bibfnamefont {J.}~\bibnamefont {Gowdy}}, \ and\ \bibinfo {author} {\bibfnamefont {G.}~\bibnamefont {Travaglini}},\ }\href {\doibase 10.1007/JHEP06(2023)048} {\bibfield  {journal} {\bibinfo  {journal} {JHEP}\ }\textbf {\bibinfo {volume} {06}},\ \bibinfo {pages} {048} (\bibinfo {year} {2023})},\ \Eprint {http://arxiv.org/abs/2303.06111} {arXiv:2303.06111 [hep-th]} \BibitemShut {NoStop}%
\bibitem [{\citenamefont {Herderschee}\ \emph {et~al.}(2023)\citenamefont {Herderschee}, \citenamefont {Roiban},\ and\ \citenamefont {Teng}}]{Herderschee:2023fxh}%
  \BibitemOpen
  \bibfield  {author} {\bibinfo {author} {\bibfnamefont {A.}~\bibnamefont {Herderschee}}, \bibinfo {author} {\bibfnamefont {R.}~\bibnamefont {Roiban}}, \ and\ \bibinfo {author} {\bibfnamefont {F.}~\bibnamefont {Teng}},\ }\href {\doibase 10.1007/JHEP06(2023)004} {\bibfield  {journal} {\bibinfo  {journal} {JHEP}\ }\textbf {\bibinfo {volume} {06}},\ \bibinfo {pages} {004} (\bibinfo {year} {2023})},\ \Eprint {http://arxiv.org/abs/2303.06112} {arXiv:2303.06112 [hep-th]} \BibitemShut {NoStop}%
\bibitem [{\citenamefont {Georgoudis}\ \emph {et~al.}(2023)\citenamefont {Georgoudis}, \citenamefont {Heissenberg},\ and\ \citenamefont {Vazquez-Holm}}]{Georgoudis:2023lgf}%
  \BibitemOpen
  \bibfield  {author} {\bibinfo {author} {\bibfnamefont {A.}~\bibnamefont {Georgoudis}}, \bibinfo {author} {\bibfnamefont {C.}~\bibnamefont {Heissenberg}}, \ and\ \bibinfo {author} {\bibfnamefont {I.}~\bibnamefont {Vazquez-Holm}},\ }\href {\doibase 10.1007/JHEP06(2023)126} {\bibfield  {journal} {\bibinfo  {journal} {JHEP}\ }\textbf {\bibinfo {volume} {06}},\ \bibinfo {pages} {126} (\bibinfo {year} {2023})},\ \Eprint {http://arxiv.org/abs/2303.07006} {arXiv:2303.07006 [hep-th]} \BibitemShut {NoStop}%
\bibitem [{\citenamefont {Elkhidir}\ \emph {et~al.}(2024)\citenamefont {Elkhidir}, \citenamefont {O'Connell}, \citenamefont {Sergola},\ and\ \citenamefont {Vazquez-Holm}}]{Elkhidir:2023dco}%
  \BibitemOpen
  \bibfield  {author} {\bibinfo {author} {\bibfnamefont {A.}~\bibnamefont {Elkhidir}}, \bibinfo {author} {\bibfnamefont {D.}~\bibnamefont {O'Connell}}, \bibinfo {author} {\bibfnamefont {M.}~\bibnamefont {Sergola}}, \ and\ \bibinfo {author} {\bibfnamefont {I.~A.}\ \bibnamefont {Vazquez-Holm}},\ }\href {\doibase 10.1007/JHEP07(2024)272} {\bibfield  {journal} {\bibinfo  {journal} {JHEP}\ }\textbf {\bibinfo {volume} {07}},\ \bibinfo {pages} {272} (\bibinfo {year} {2024})},\ \Eprint {http://arxiv.org/abs/2303.06211} {arXiv:2303.06211 [hep-th]} \BibitemShut {NoStop}%
\bibitem [{\citenamefont {Caron-Huot}\ \emph {et~al.}(2024)\citenamefont {Caron-Huot}, \citenamefont {Giroux}, \citenamefont {Hannesdottir},\ and\ \citenamefont {Mizera}}]{Caron-Huot:2023vxl}%
  \BibitemOpen
  \bibfield  {author} {\bibinfo {author} {\bibfnamefont {S.}~\bibnamefont {Caron-Huot}}, \bibinfo {author} {\bibfnamefont {M.}~\bibnamefont {Giroux}}, \bibinfo {author} {\bibfnamefont {H.~S.}\ \bibnamefont {Hannesdottir}}, \ and\ \bibinfo {author} {\bibfnamefont {S.}~\bibnamefont {Mizera}},\ }\href {\doibase 10.1007/JHEP01(2024)139} {\bibfield  {journal} {\bibinfo  {journal} {JHEP}\ }\textbf {\bibinfo {volume} {01}},\ \bibinfo {pages} {139} (\bibinfo {year} {2024})},\ \Eprint {http://arxiv.org/abs/2308.02125} {arXiv:2308.02125 [hep-th]} \BibitemShut {NoStop}%
\bibitem [{\citenamefont {Bohnenblust}\ \emph {et~al.}(2024)\citenamefont {Bohnenblust}, \citenamefont {Ita}, \citenamefont {Kraus},\ and\ \citenamefont {Schlenk}}]{Bohnenblust:2023qmy}%
  \BibitemOpen
  \bibfield  {author} {\bibinfo {author} {\bibfnamefont {L.}~\bibnamefont {Bohnenblust}}, \bibinfo {author} {\bibfnamefont {H.}~\bibnamefont {Ita}}, \bibinfo {author} {\bibfnamefont {M.}~\bibnamefont {Kraus}}, \ and\ \bibinfo {author} {\bibfnamefont {J.}~\bibnamefont {Schlenk}},\ }\href {\doibase 10.1007/JHEP11(2024)109} {\bibfield  {journal} {\bibinfo  {journal} {JHEP}\ }\textbf {\bibinfo {volume} {11}},\ \bibinfo {pages} {109} (\bibinfo {year} {2024})},\ \Eprint {http://arxiv.org/abs/2312.14859} {arXiv:2312.14859 [hep-th]} \BibitemShut {NoStop}%
\bibitem [{\citenamefont {Bern}\ \emph {et~al.}(1994)\citenamefont {Bern}, \citenamefont {Dixon}, \citenamefont {Dunbar},\ and\ \citenamefont {Kosower}}]{Bern:1994zx}%
  \BibitemOpen
  \bibfield  {author} {\bibinfo {author} {\bibfnamefont {Z.}~\bibnamefont {Bern}}, \bibinfo {author} {\bibfnamefont {L.~J.}\ \bibnamefont {Dixon}}, \bibinfo {author} {\bibfnamefont {D.~C.}\ \bibnamefont {Dunbar}}, \ and\ \bibinfo {author} {\bibfnamefont {D.~A.}\ \bibnamefont {Kosower}},\ }\href {\doibase 10.1016/0550-3213(94)90179-1} {\bibfield  {journal} {\bibinfo  {journal} {Nucl. Phys. B}\ }\textbf {\bibinfo {volume} {425}},\ \bibinfo {pages} {217} (\bibinfo {year} {1994})},\ \Eprint {http://arxiv.org/abs/hep-ph/9403226} {arXiv:hep-ph/9403226} \BibitemShut {NoStop}%
\bibitem [{\citenamefont {Bern}\ \emph {et~al.}(1995)\citenamefont {Bern}, \citenamefont {Dixon}, \citenamefont {Dunbar},\ and\ \citenamefont {Kosower}}]{Bern:1994cg}%
  \BibitemOpen
  \bibfield  {author} {\bibinfo {author} {\bibfnamefont {Z.}~\bibnamefont {Bern}}, \bibinfo {author} {\bibfnamefont {L.~J.}\ \bibnamefont {Dixon}}, \bibinfo {author} {\bibfnamefont {D.~C.}\ \bibnamefont {Dunbar}}, \ and\ \bibinfo {author} {\bibfnamefont {D.~A.}\ \bibnamefont {Kosower}},\ }\href {\doibase 10.1016/0550-3213(94)00488-Z} {\bibfield  {journal} {\bibinfo  {journal} {Nucl. Phys. B}\ }\textbf {\bibinfo {volume} {435}},\ \bibinfo {pages} {59} (\bibinfo {year} {1995})},\ \Eprint {http://arxiv.org/abs/hep-ph/9409265} {arXiv:hep-ph/9409265} \BibitemShut {NoStop}%
\bibitem [{\citenamefont {Britto}\ \emph {et~al.}(2005)\citenamefont {Britto}, \citenamefont {Cachazo},\ and\ \citenamefont {Feng}}]{Britto:2004nc}%
  \BibitemOpen
  \bibfield  {author} {\bibinfo {author} {\bibfnamefont {R.}~\bibnamefont {Britto}}, \bibinfo {author} {\bibfnamefont {F.}~\bibnamefont {Cachazo}}, \ and\ \bibinfo {author} {\bibfnamefont {B.}~\bibnamefont {Feng}},\ }\href {\doibase 10.1016/j.nuclphysb.2005.07.014} {\bibfield  {journal} {\bibinfo  {journal} {Nucl. Phys. B}\ }\textbf {\bibinfo {volume} {725}},\ \bibinfo {pages} {275} (\bibinfo {year} {2005})},\ \Eprint {http://arxiv.org/abs/hep-th/0412103} {arXiv:hep-th/0412103} \BibitemShut {NoStop}%
\bibitem [{\citenamefont {Britto}\ \emph {et~al.}(2006)\citenamefont {Britto}, \citenamefont {Feng},\ and\ \citenamefont {Mastrolia}}]{Britto:2006sj}%
  \BibitemOpen
  \bibfield  {author} {\bibinfo {author} {\bibfnamefont {R.}~\bibnamefont {Britto}}, \bibinfo {author} {\bibfnamefont {B.}~\bibnamefont {Feng}}, \ and\ \bibinfo {author} {\bibfnamefont {P.}~\bibnamefont {Mastrolia}},\ }\href {\doibase 10.1103/PhysRevD.73.105004} {\bibfield  {journal} {\bibinfo  {journal} {Phys. Rev. D}\ }\textbf {\bibinfo {volume} {73}},\ \bibinfo {pages} {105004} (\bibinfo {year} {2006})},\ \Eprint {http://arxiv.org/abs/hep-ph/0602178} {arXiv:hep-ph/0602178} \BibitemShut {NoStop}%
\bibitem [{\citenamefont {Ossola}\ \emph {et~al.}(2007)\citenamefont {Ossola}, \citenamefont {Papadopoulos},\ and\ \citenamefont {Pittau}}]{Ossola:2006us}%
  \BibitemOpen
  \bibfield  {author} {\bibinfo {author} {\bibfnamefont {G.}~\bibnamefont {Ossola}}, \bibinfo {author} {\bibfnamefont {C.~G.}\ \bibnamefont {Papadopoulos}}, \ and\ \bibinfo {author} {\bibfnamefont {R.}~\bibnamefont {Pittau}},\ }\href {\doibase 10.1016/j.nuclphysb.2006.11.012} {\bibfield  {journal} {\bibinfo  {journal} {Nucl. Phys. B}\ }\textbf {\bibinfo {volume} {763}},\ \bibinfo {pages} {147} (\bibinfo {year} {2007})},\ \Eprint {http://arxiv.org/abs/hep-ph/0609007} {arXiv:hep-ph/0609007} \BibitemShut {NoStop}%
\bibitem [{\citenamefont {Anastasiou}\ \emph {et~al.}(2007)\citenamefont {Anastasiou}, \citenamefont {Britto}, \citenamefont {Feng}, \citenamefont {Kunszt},\ and\ \citenamefont {Mastrolia}}]{Anastasiou:2006gt}%
  \BibitemOpen
  \bibfield  {author} {\bibinfo {author} {\bibfnamefont {C.}~\bibnamefont {Anastasiou}}, \bibinfo {author} {\bibfnamefont {R.}~\bibnamefont {Britto}}, \bibinfo {author} {\bibfnamefont {B.}~\bibnamefont {Feng}}, \bibinfo {author} {\bibfnamefont {Z.}~\bibnamefont {Kunszt}}, \ and\ \bibinfo {author} {\bibfnamefont {P.}~\bibnamefont {Mastrolia}},\ }\href {\doibase 10.1088/1126-6708/2007/03/111} {\bibfield  {journal} {\bibinfo  {journal} {JHEP}\ }\textbf {\bibinfo {volume} {03}},\ \bibinfo {pages} {111} (\bibinfo {year} {2007})},\ \Eprint {http://arxiv.org/abs/hep-ph/0612277} {arXiv:hep-ph/0612277} \BibitemShut {NoStop}%
\bibitem [{\citenamefont {Brandhuber}\ \emph {et~al.}(2021{\natexlab{b}})\citenamefont {Brandhuber}, \citenamefont {Chen}, \citenamefont {Travaglini},\ and\ \citenamefont {Wen}}]{Brandhuber:2021kpo}%
  \BibitemOpen
  \bibfield  {author} {\bibinfo {author} {\bibfnamefont {A.}~\bibnamefont {Brandhuber}}, \bibinfo {author} {\bibfnamefont {G.}~\bibnamefont {Chen}}, \bibinfo {author} {\bibfnamefont {G.}~\bibnamefont {Travaglini}}, \ and\ \bibinfo {author} {\bibfnamefont {C.}~\bibnamefont {Wen}},\ }\href {\doibase 10.1007/JHEP07(2021)047} {\bibfield  {journal} {\bibinfo  {journal} {JHEP}\ }\textbf {\bibinfo {volume} {07}},\ \bibinfo {pages} {047} (\bibinfo {year} {2021}{\natexlab{b}})},\ \Eprint {http://arxiv.org/abs/2104.11206} {arXiv:2104.11206 [hep-th]} \BibitemShut {NoStop}%
\bibitem [{\citenamefont {Tkachov}(1981)}]{Tkachov:1981wb}%
  \BibitemOpen
  \bibfield  {author} {\bibinfo {author} {\bibfnamefont {F.~V.}\ \bibnamefont {Tkachov}},\ }\href {\doibase 10.1016/0370-2693(81)90288-4} {\bibfield  {journal} {\bibinfo  {journal} {Phys. Lett. B}\ }\textbf {\bibinfo {volume} {100}},\ \bibinfo {pages} {65} (\bibinfo {year} {1981})}\BibitemShut {NoStop}%
\bibitem [{\citenamefont {Chetyrkin}\ and\ \citenamefont {Tkachov}(1981)}]{Chetyrkin:1981qh}%
  \BibitemOpen
  \bibfield  {author} {\bibinfo {author} {\bibfnamefont {K.~G.}\ \bibnamefont {Chetyrkin}}\ and\ \bibinfo {author} {\bibfnamefont {F.~V.}\ \bibnamefont {Tkachov}},\ }\href {\doibase 10.1016/0550-3213(81)90199-1} {\bibfield  {journal} {\bibinfo  {journal} {Nucl. Phys. B}\ }\textbf {\bibinfo {volume} {192}},\ \bibinfo {pages} {159} (\bibinfo {year} {1981})}\BibitemShut {NoStop}%
\bibitem [{\citenamefont {Laporta}(2000)}]{Laporta:2000dsw}%
  \BibitemOpen
  \bibfield  {author} {\bibinfo {author} {\bibfnamefont {S.}~\bibnamefont {Laporta}},\ }\href {\doibase 10.1142/S0217751X00002159} {\bibfield  {journal} {\bibinfo  {journal} {Int. J. Mod. Phys. A}\ }\textbf {\bibinfo {volume} {15}},\ \bibinfo {pages} {5087} (\bibinfo {year} {2000})},\ \Eprint {http://arxiv.org/abs/hep-ph/0102033} {arXiv:hep-ph/0102033} \BibitemShut {NoStop}%
\bibitem [{\citenamefont {Smirnov}(2008)}]{Smirnov:2008iw}%
  \BibitemOpen
  \bibfield  {author} {\bibinfo {author} {\bibfnamefont {A.~V.}\ \bibnamefont {Smirnov}},\ }\href {\doibase 10.1088/1126-6708/2008/10/107} {\bibfield  {journal} {\bibinfo  {journal} {JHEP}\ }\textbf {\bibinfo {volume} {10}},\ \bibinfo {pages} {107} (\bibinfo {year} {2008})},\ \Eprint {http://arxiv.org/abs/0807.3243} {arXiv:0807.3243 [hep-ph]} \BibitemShut {NoStop}%
\bibitem [{\citenamefont {Lee}(2012)}]{Lee:2012cn}%
  \BibitemOpen
  \bibfield  {author} {\bibinfo {author} {\bibfnamefont {R.~N.}\ \bibnamefont {Lee}},\ }\href@noop {} {\  (\bibinfo {year} {2012})},\ \Eprint {http://arxiv.org/abs/1212.2685} {arXiv:1212.2685 [hep-ph]} \BibitemShut {NoStop}%
\bibitem [{\citenamefont {Maierh\"ofer}\ \emph {et~al.}(2018)\citenamefont {Maierh\"ofer}, \citenamefont {Usovitsch},\ and\ \citenamefont {Uwer}}]{Maierhofer:2017gsa}%
  \BibitemOpen
  \bibfield  {author} {\bibinfo {author} {\bibfnamefont {P.}~\bibnamefont {Maierh\"ofer}}, \bibinfo {author} {\bibfnamefont {J.}~\bibnamefont {Usovitsch}}, \ and\ \bibinfo {author} {\bibfnamefont {P.}~\bibnamefont {Uwer}},\ }\href {\doibase 10.1016/j.cpc.2018.04.012} {\bibfield  {journal} {\bibinfo  {journal} {Comput. Phys. Commun.}\ }\textbf {\bibinfo {volume} {230}},\ \bibinfo {pages} {99} (\bibinfo {year} {2018})},\ \Eprint {http://arxiv.org/abs/1705.05610} {arXiv:1705.05610 [hep-ph]} \BibitemShut {NoStop}%
\bibitem [{\citenamefont {Peraro}(2019)}]{Peraro:2019svx}%
  \BibitemOpen
  \bibfield  {author} {\bibinfo {author} {\bibfnamefont {T.}~\bibnamefont {Peraro}},\ }\href {\doibase 10.1007/JHEP07(2019)031} {\bibfield  {journal} {\bibinfo  {journal} {JHEP}\ }\textbf {\bibinfo {volume} {07}},\ \bibinfo {pages} {031} (\bibinfo {year} {2019})},\ \Eprint {http://arxiv.org/abs/1905.08019} {arXiv:1905.08019 [hep-ph]} \BibitemShut {NoStop}%
\bibitem [{\citenamefont {Wu}\ \emph {et~al.}(2024)\citenamefont {Wu}, \citenamefont {Boehm}, \citenamefont {Ma}, \citenamefont {Xu},\ and\ \citenamefont {Zhang}}]{Wu:2023upw}%
  \BibitemOpen
  \bibfield  {author} {\bibinfo {author} {\bibfnamefont {Z.}~\bibnamefont {Wu}}, \bibinfo {author} {\bibfnamefont {J.}~\bibnamefont {Boehm}}, \bibinfo {author} {\bibfnamefont {R.}~\bibnamefont {Ma}}, \bibinfo {author} {\bibfnamefont {H.}~\bibnamefont {Xu}}, \ and\ \bibinfo {author} {\bibfnamefont {Y.}~\bibnamefont {Zhang}},\ }\href {\doibase 10.1016/j.cpc.2023.108999} {\bibfield  {journal} {\bibinfo  {journal} {Comput. Phys. Commun.}\ }\textbf {\bibinfo {volume} {295}},\ \bibinfo {pages} {108999} (\bibinfo {year} {2024})},\ \Eprint {http://arxiv.org/abs/2305.08783} {arXiv:2305.08783 [hep-ph]} \BibitemShut {NoStop}%
\bibitem [{\citenamefont {Matsumoto}(1998)}]{Matsumoto1998-2}%
  \BibitemOpen
  \bibfield  {author} {\bibinfo {author} {\bibfnamefont {K.}~\bibnamefont {Matsumoto}},\ }\href {http://www.math.kobe-u.ac.jp/~fe/xml/mr1662357.xml} {\bibfield  {journal} {\bibinfo  {journal} {Funkcial. Ekvac.}\ }\textbf {\bibinfo {volume} {41}},\ \bibinfo {pages} {291} (\bibinfo {year} {1998})}\BibitemShut {NoStop}%
\bibitem [{\citenamefont {Majima}\ \emph {et~al.}(2000)\citenamefont {Majima}, \citenamefont {Matsumoto},\ and\ \citenamefont {Takayama}}]{majima2000}%
  \BibitemOpen
  \bibfield  {author} {\bibinfo {author} {\bibfnamefont {H.}~\bibnamefont {Majima}}, \bibinfo {author} {\bibfnamefont {K.}~\bibnamefont {Matsumoto}}, \ and\ \bibinfo {author} {\bibfnamefont {N.}~\bibnamefont {Takayama}},\ }\href {\doibase 10.2748/tmj/1178207752} {\bibfield  {journal} {\bibinfo  {journal} {Tohoku Math. J. (2)}\ }\textbf {\bibinfo {volume} {52}},\ \bibinfo {pages} {489} (\bibinfo {year} {2000})}\BibitemShut {NoStop}%
\bibitem [{\citenamefont {Cacciatori}\ and\ \citenamefont {Mastrolia}(2022)}]{Cacciatori:2022mbi}%
  \BibitemOpen
  \bibfield  {author} {\bibinfo {author} {\bibfnamefont {S.~L.}\ \bibnamefont {Cacciatori}}\ and\ \bibinfo {author} {\bibfnamefont {P.}~\bibnamefont {Mastrolia}},\ }\href@noop {} {\  (\bibinfo {year} {2022})},\ \Eprint {http://arxiv.org/abs/2211.03729} {arXiv:2211.03729 [hep-th]} \BibitemShut {NoStop}%
\bibitem [{\citenamefont {Brunello}\ \emph {et~al.}(2024{\natexlab{a}})\citenamefont {Brunello}, \citenamefont {Crisanti}, \citenamefont {Giroux}, \citenamefont {Mastrolia},\ and\ \citenamefont {Smith}}]{Brunello:2023fef}%
  \BibitemOpen
  \bibfield  {author} {\bibinfo {author} {\bibfnamefont {G.}~\bibnamefont {Brunello}}, \bibinfo {author} {\bibfnamefont {G.}~\bibnamefont {Crisanti}}, \bibinfo {author} {\bibfnamefont {M.}~\bibnamefont {Giroux}}, \bibinfo {author} {\bibfnamefont {P.}~\bibnamefont {Mastrolia}}, \ and\ \bibinfo {author} {\bibfnamefont {S.}~\bibnamefont {Smith}},\ }\href {\doibase 10.1103/PhysRevD.109.094047} {\bibfield  {journal} {\bibinfo  {journal} {Phys. Rev. D}\ }\textbf {\bibinfo {volume} {109}},\ \bibinfo {pages} {094047} (\bibinfo {year} {2024}{\natexlab{a}})},\ \Eprint {http://arxiv.org/abs/2311.14432} {arXiv:2311.14432 [hep-th]} \BibitemShut {NoStop}%
\bibitem [{\citenamefont {Kosower}\ \emph {et~al.}(2022)\citenamefont {Kosower}, \citenamefont {Monteiro},\ and\ \citenamefont {O'Connell}}]{Kosower:2022yvp}%
  \BibitemOpen
  \bibfield  {author} {\bibinfo {author} {\bibfnamefont {D.~A.}\ \bibnamefont {Kosower}}, \bibinfo {author} {\bibfnamefont {R.}~\bibnamefont {Monteiro}}, \ and\ \bibinfo {author} {\bibfnamefont {D.}~\bibnamefont {O'Connell}},\ }\href {\doibase 10.1088/1751-8121/ac8846} {\bibfield  {journal} {\bibinfo  {journal} {J. Phys. A}\ }\textbf {\bibinfo {volume} {55}},\ \bibinfo {pages} {443015} (\bibinfo {year} {2022})},\ \Eprint {http://arxiv.org/abs/2203.13025} {arXiv:2203.13025 [hep-th]} \BibitemShut {NoStop}%
\bibitem [{\citenamefont {Schwinger}(1961)}]{Schwinger:1960qe}%
  \BibitemOpen
  \bibfield  {author} {\bibinfo {author} {\bibfnamefont {J.~S.}\ \bibnamefont {Schwinger}},\ }\href {\doibase 10.1063/1.1703727} {\bibfield  {journal} {\bibinfo  {journal} {J. Math. Phys.}\ }\textbf {\bibinfo {volume} {2}},\ \bibinfo {pages} {407} (\bibinfo {year} {1961})}\BibitemShut {NoStop}%
\bibitem [{\citenamefont {Keldysh}(1964)}]{Keldysh:1964ud}%
  \BibitemOpen
  \bibfield  {author} {\bibinfo {author} {\bibfnamefont {L.~V.}\ \bibnamefont {Keldysh}},\ }\href@noop {} {\bibfield  {journal} {\bibinfo  {journal} {Zh. Eksp. Teor. Fiz.}\ }\textbf {\bibinfo {volume} {47}},\ \bibinfo {pages} {1515} (\bibinfo {year} {1964})}\BibitemShut {NoStop}%
\bibitem [{\citenamefont {Jakobsen}\ \emph {et~al.}(2022{\natexlab{b}})\citenamefont {Jakobsen}, \citenamefont {Mogull}, \citenamefont {Plefka},\ and\ \citenamefont {Sauer}}]{Jakobsen:2022psy}%
  \BibitemOpen
  \bibfield  {author} {\bibinfo {author} {\bibfnamefont {G.~U.}\ \bibnamefont {Jakobsen}}, \bibinfo {author} {\bibfnamefont {G.}~\bibnamefont {Mogull}}, \bibinfo {author} {\bibfnamefont {J.}~\bibnamefont {Plefka}}, \ and\ \bibinfo {author} {\bibfnamefont {B.}~\bibnamefont {Sauer}},\ }\href {\doibase 10.1007/JHEP10(2022)128} {\bibfield  {journal} {\bibinfo  {journal} {JHEP}\ }\textbf {\bibinfo {volume} {10}},\ \bibinfo {pages} {128} (\bibinfo {year} {2022}{\natexlab{b}})},\ \Eprint {http://arxiv.org/abs/2207.00569} {arXiv:2207.00569 [hep-th]} \BibitemShut {NoStop}%
\bibitem [{\citenamefont {K\"alin}\ \emph {et~al.}(2023)\citenamefont {K\"alin}, \citenamefont {Neef},\ and\ \citenamefont {Porto}}]{Kalin:2022hph}%
  \BibitemOpen
  \bibfield  {author} {\bibinfo {author} {\bibfnamefont {G.}~\bibnamefont {K\"alin}}, \bibinfo {author} {\bibfnamefont {J.}~\bibnamefont {Neef}}, \ and\ \bibinfo {author} {\bibfnamefont {R.~A.}\ \bibnamefont {Porto}},\ }\href {\doibase 10.1007/JHEP01(2023)140} {\bibfield  {journal} {\bibinfo  {journal} {JHEP}\ }\textbf {\bibinfo {volume} {01}},\ \bibinfo {pages} {140} (\bibinfo {year} {2023})},\ \Eprint {http://arxiv.org/abs/2207.00580} {arXiv:2207.00580 [hep-th]} \BibitemShut {NoStop}%
\bibitem [{\citenamefont {Veneziano}\ and\ \citenamefont {Vilkovisky}(2022)}]{Veneziano:2022zwh}%
  \BibitemOpen
  \bibfield  {author} {\bibinfo {author} {\bibfnamefont {G.}~\bibnamefont {Veneziano}}\ and\ \bibinfo {author} {\bibfnamefont {G.~A.}\ \bibnamefont {Vilkovisky}},\ }\href {\doibase 10.1016/j.physletb.2022.137419} {\bibfield  {journal} {\bibinfo  {journal} {Phys. Lett. B}\ }\textbf {\bibinfo {volume} {834}},\ \bibinfo {pages} {137419} (\bibinfo {year} {2022})},\ \Eprint {http://arxiv.org/abs/2201.11607} {arXiv:2201.11607 [gr-qc]} \BibitemShut {NoStop}%
\bibitem [{\citenamefont {Georgoudis}\ \emph {et~al.}(2024)\citenamefont {Georgoudis}, \citenamefont {Heissenberg},\ and\ \citenamefont {Russo}}]{Georgoudis:2023eke}%
  \BibitemOpen
  \bibfield  {author} {\bibinfo {author} {\bibfnamefont {A.}~\bibnamefont {Georgoudis}}, \bibinfo {author} {\bibfnamefont {C.}~\bibnamefont {Heissenberg}}, \ and\ \bibinfo {author} {\bibfnamefont {R.}~\bibnamefont {Russo}},\ }\href {\doibase 10.1007/JHEP03(2024)089} {\bibfield  {journal} {\bibinfo  {journal} {JHEP}\ }\textbf {\bibinfo {volume} {03}},\ \bibinfo {pages} {089} (\bibinfo {year} {2024})},\ \Eprint {http://arxiv.org/abs/2312.07452} {arXiv:2312.07452 [hep-th]} \BibitemShut {NoStop}%
\bibitem [{\citenamefont {Bini}\ \emph {et~al.}(2024)\citenamefont {Bini}, \citenamefont {Damour}, \citenamefont {De~Angelis}, \citenamefont {Geralico}, \citenamefont {Herderschee}, \citenamefont {Roiban},\ and\ \citenamefont {Teng}}]{Bini:2024rsy}%
  \BibitemOpen
  \bibfield  {author} {\bibinfo {author} {\bibfnamefont {D.}~\bibnamefont {Bini}}, \bibinfo {author} {\bibfnamefont {T.}~\bibnamefont {Damour}}, \bibinfo {author} {\bibfnamefont {S.}~\bibnamefont {De~Angelis}}, \bibinfo {author} {\bibfnamefont {A.}~\bibnamefont {Geralico}}, \bibinfo {author} {\bibfnamefont {A.}~\bibnamefont {Herderschee}}, \bibinfo {author} {\bibfnamefont {R.}~\bibnamefont {Roiban}}, \ and\ \bibinfo {author} {\bibfnamefont {F.}~\bibnamefont {Teng}},\ }\href {\doibase 10.1103/PhysRevD.109.125008} {\bibfield  {journal} {\bibinfo  {journal} {Phys. Rev. D}\ }\textbf {\bibinfo {volume} {109}},\ \bibinfo {pages} {125008} (\bibinfo {year} {2024})},\ \Eprint {http://arxiv.org/abs/2402.06604} {arXiv:2402.06604 [hep-th]} \BibitemShut {NoStop}%
\bibitem [{\citenamefont {Damgaard}\ \emph {et~al.}(2019)\citenamefont {Damgaard}, \citenamefont {Haddad},\ and\ \citenamefont {Helset}}]{Damgaard:2019lfh}%
  \BibitemOpen
  \bibfield  {author} {\bibinfo {author} {\bibfnamefont {P.~H.}\ \bibnamefont {Damgaard}}, \bibinfo {author} {\bibfnamefont {K.}~\bibnamefont {Haddad}}, \ and\ \bibinfo {author} {\bibfnamefont {A.}~\bibnamefont {Helset}},\ }\href {\doibase 10.1007/JHEP11(2019)070} {\bibfield  {journal} {\bibinfo  {journal} {JHEP}\ }\textbf {\bibinfo {volume} {11}},\ \bibinfo {pages} {070} (\bibinfo {year} {2019})},\ \Eprint {http://arxiv.org/abs/1908.10308} {arXiv:1908.10308 [hep-ph]} \BibitemShut {NoStop}%
\bibitem [{\citenamefont {Bjerrum-Bohr}\ \emph {et~al.}(2022)\citenamefont {Bjerrum-Bohr}, \citenamefont {Plant\'e},\ and\ \citenamefont {Vanhove}}]{Bjerrum-Bohr:2021wwt}%
  \BibitemOpen
  \bibfield  {author} {\bibinfo {author} {\bibfnamefont {N.~E.~J.}\ \bibnamefont {Bjerrum-Bohr}}, \bibinfo {author} {\bibfnamefont {L.}~\bibnamefont {Plant\'e}}, \ and\ \bibinfo {author} {\bibfnamefont {P.}~\bibnamefont {Vanhove}},\ }\href {\doibase 10.1007/JHEP03(2022)071} {\bibfield  {journal} {\bibinfo  {journal} {JHEP}\ }\textbf {\bibinfo {volume} {03}},\ \bibinfo {pages} {071} (\bibinfo {year} {2022})},\ \Eprint {http://arxiv.org/abs/2111.02976} {arXiv:2111.02976 [hep-th]} \BibitemShut {NoStop}%
\bibitem [{\citenamefont {Bern}\ and\ \citenamefont {Kosower}(1992)}]{Bern:1991aq}%
  \BibitemOpen
  \bibfield  {author} {\bibinfo {author} {\bibfnamefont {Z.}~\bibnamefont {Bern}}\ and\ \bibinfo {author} {\bibfnamefont {D.~A.}\ \bibnamefont {Kosower}},\ }\href {\doibase 10.1016/0550-3213(92)90134-W} {\bibfield  {journal} {\bibinfo  {journal} {Nucl. Phys. B}\ }\textbf {\bibinfo {volume} {379}},\ \bibinfo {pages} {451} (\bibinfo {year} {1992})}\BibitemShut {NoStop}%
\bibitem [{\citenamefont {Siegel}(1979)}]{Siegel:1979wq}%
  \BibitemOpen
  \bibfield  {author} {\bibinfo {author} {\bibfnamefont {W.}~\bibnamefont {Siegel}},\ }\href {\doibase 10.1016/0370-2693(79)90282-X} {\bibfield  {journal} {\bibinfo  {journal} {Phys. Lett. B}\ }\textbf {\bibinfo {volume} {84}},\ \bibinfo {pages} {193} (\bibinfo {year} {1979})}\BibitemShut {NoStop}%
\bibitem [{\citenamefont {Anastasiou}\ \emph {et~al.}(2023)\citenamefont {Anastasiou}, \citenamefont {Karlen},\ and\ \citenamefont {Vicini}}]{Anastasiou:2023koq}%
  \BibitemOpen
  \bibfield  {author} {\bibinfo {author} {\bibfnamefont {C.}~\bibnamefont {Anastasiou}}, \bibinfo {author} {\bibfnamefont {J.}~\bibnamefont {Karlen}}, \ and\ \bibinfo {author} {\bibfnamefont {M.}~\bibnamefont {Vicini}},\ }\href {\doibase 10.1007/JHEP12(2023)169} {\bibfield  {journal} {\bibinfo  {journal} {JHEP}\ }\textbf {\bibinfo {volume} {12}},\ \bibinfo {pages} {169} (\bibinfo {year} {2023})},\ \Eprint {http://arxiv.org/abs/2308.14701} {arXiv:2308.14701 [hep-ph]} \BibitemShut {NoStop}%
\bibitem [{\citenamefont {Mastrolia}\ and\ \citenamefont {Mizera}(2019)}]{Mastrolia:2018uzb}%
  \BibitemOpen
  \bibfield  {author} {\bibinfo {author} {\bibfnamefont {P.}~\bibnamefont {Mastrolia}}\ and\ \bibinfo {author} {\bibfnamefont {S.}~\bibnamefont {Mizera}},\ }\href {\doibase 10.1007/JHEP02(2019)139} {\bibfield  {journal} {\bibinfo  {journal} {JHEP}\ }\textbf {\bibinfo {volume} {02}},\ \bibinfo {pages} {139} (\bibinfo {year} {2019})},\ \Eprint {http://arxiv.org/abs/1810.03818} {arXiv:1810.03818 [hep-th]} \BibitemShut {NoStop}%
\bibitem [{\citenamefont {Brunello}\ \emph {et~al.}(2024{\natexlab{b}})\citenamefont {Brunello}, \citenamefont {Chestnov}, \citenamefont {Crisanti}, \citenamefont {Frellesvig}, \citenamefont {Mandal},\ and\ \citenamefont {Mastrolia}}]{Brunello:2023rpq}%
  \BibitemOpen
  \bibfield  {author} {\bibinfo {author} {\bibfnamefont {G.}~\bibnamefont {Brunello}}, \bibinfo {author} {\bibfnamefont {V.}~\bibnamefont {Chestnov}}, \bibinfo {author} {\bibfnamefont {G.}~\bibnamefont {Crisanti}}, \bibinfo {author} {\bibfnamefont {H.}~\bibnamefont {Frellesvig}}, \bibinfo {author} {\bibfnamefont {M.~K.}\ \bibnamefont {Mandal}}, \ and\ \bibinfo {author} {\bibfnamefont {P.}~\bibnamefont {Mastrolia}},\ }\href {\doibase 10.1007/JHEP09(2024)015} {\bibfield  {journal} {\bibinfo  {journal} {JHEP}\ }\textbf {\bibinfo {volume} {09}},\ \bibinfo {pages} {015} (\bibinfo {year} {2024}{\natexlab{b}})},\ \Eprint {http://arxiv.org/abs/2401.01897} {arXiv:2401.01897 [hep-th]} \BibitemShut {NoStop}%
\bibitem [{\citenamefont {Anastasiou}\ and\ \citenamefont {Melnikov}(2002)}]{Anastasiou:2002yz}%
  \BibitemOpen
  \bibfield  {author} {\bibinfo {author} {\bibfnamefont {C.}~\bibnamefont {Anastasiou}}\ and\ \bibinfo {author} {\bibfnamefont {K.}~\bibnamefont {Melnikov}},\ }\href {\doibase 10.1016/S0550-3213(02)00837-4} {\bibfield  {journal} {\bibinfo  {journal} {Nucl. Phys. B}\ }\textbf {\bibinfo {volume} {646}},\ \bibinfo {pages} {220} (\bibinfo {year} {2002})},\ \Eprint {http://arxiv.org/abs/hep-ph/0207004} {arXiv:hep-ph/0207004} \BibitemShut {NoStop}%
\bibitem [{\citenamefont {Anastasiou}\ \emph {et~al.}(2004)\citenamefont {Anastasiou}, \citenamefont {Melnikov},\ and\ \citenamefont {Petriello}}]{Anastasiou:2003gr}%
  \BibitemOpen
  \bibfield  {author} {\bibinfo {author} {\bibfnamefont {C.}~\bibnamefont {Anastasiou}}, \bibinfo {author} {\bibfnamefont {K.}~\bibnamefont {Melnikov}}, \ and\ \bibinfo {author} {\bibfnamefont {F.}~\bibnamefont {Petriello}},\ }\href {\doibase 10.1103/PhysRevD.69.076010} {\bibfield  {journal} {\bibinfo  {journal} {Phys. Rev. D}\ }\textbf {\bibinfo {volume} {69}},\ \bibinfo {pages} {076010} (\bibinfo {year} {2004})},\ \Eprint {http://arxiv.org/abs/hep-ph/0311311} {arXiv:hep-ph/0311311} \BibitemShut {NoStop}%
\bibitem [{\citenamefont {Herrmann}\ \emph {et~al.}(2021)\citenamefont {Herrmann}, \citenamefont {Parra-Martinez}, \citenamefont {Ruf},\ and\ \citenamefont {Zeng}}]{Herrmann:2021lqe}%
  \BibitemOpen
  \bibfield  {author} {\bibinfo {author} {\bibfnamefont {E.}~\bibnamefont {Herrmann}}, \bibinfo {author} {\bibfnamefont {J.}~\bibnamefont {Parra-Martinez}}, \bibinfo {author} {\bibfnamefont {M.~S.}\ \bibnamefont {Ruf}}, \ and\ \bibinfo {author} {\bibfnamefont {M.}~\bibnamefont {Zeng}},\ }\href {\doibase 10.1103/PhysRevLett.126.201602} {\bibfield  {journal} {\bibinfo  {journal} {Phys. Rev. Lett.}\ }\textbf {\bibinfo {volume} {126}},\ \bibinfo {pages} {201602} (\bibinfo {year} {2021})},\ \Eprint {http://arxiv.org/abs/2101.07255} {arXiv:2101.07255 [hep-th]} \BibitemShut {NoStop}%
\bibitem [{\citenamefont {De~Angelis}\ \emph {et~al.}(2025)\citenamefont {De~Angelis}, \citenamefont {Kosower}, \citenamefont {Ma}, \citenamefont {Wu},\ and\ \citenamefont {Zhang}}]{DeAngelis:2025agn}%
  \BibitemOpen
  \bibfield  {author} {\bibinfo {author} {\bibfnamefont {S.}~\bibnamefont {De~Angelis}}, \bibinfo {author} {\bibfnamefont {D.~A.}\ \bibnamefont {Kosower}}, \bibinfo {author} {\bibfnamefont {R.}~\bibnamefont {Ma}}, \bibinfo {author} {\bibfnamefont {Z.}~\bibnamefont {Wu}}, \ and\ \bibinfo {author} {\bibfnamefont {Y.}~\bibnamefont {Zhang}},\ }\href@noop {} {\  (\bibinfo {year} {2025})},\ \Eprint {http://arxiv.org/abs/2508.04394} {arXiv:2508.04394 [hep-th]} \BibitemShut {NoStop}%
\bibitem [{\citenamefont {Blanchet}\ and\ \citenamefont {Damour}(1988)}]{Blanchet:1987wq}%
  \BibitemOpen
  \bibfield  {author} {\bibinfo {author} {\bibfnamefont {L.}~\bibnamefont {Blanchet}}\ and\ \bibinfo {author} {\bibfnamefont {T.}~\bibnamefont {Damour}},\ }\href {\doibase 10.1103/PhysRevD.37.1410} {\bibfield  {journal} {\bibinfo  {journal} {Phys. Rev. D}\ }\textbf {\bibinfo {volume} {37}},\ \bibinfo {pages} {1410} (\bibinfo {year} {1988})}\BibitemShut {NoStop}%
\bibitem [{\citenamefont {Weinberg}(1965)}]{Weinberg:1965nx}%
  \BibitemOpen
  \bibfield  {author} {\bibinfo {author} {\bibfnamefont {S.}~\bibnamefont {Weinberg}},\ }\href {\doibase 10.1103/PhysRev.140.B516} {\bibfield  {journal} {\bibinfo  {journal} {Phys. Rev.}\ }\textbf {\bibinfo {volume} {140}},\ \bibinfo {pages} {B516} (\bibinfo {year} {1965})}\BibitemShut {NoStop}%
\bibitem [{\citenamefont {Bini}\ \emph {et~al.}(2023)\citenamefont {Bini}, \citenamefont {Damour},\ and\ \citenamefont {Geralico}}]{Bini:2023fiz}%
  \BibitemOpen
  \bibfield  {author} {\bibinfo {author} {\bibfnamefont {D.}~\bibnamefont {Bini}}, \bibinfo {author} {\bibfnamefont {T.}~\bibnamefont {Damour}}, \ and\ \bibinfo {author} {\bibfnamefont {A.}~\bibnamefont {Geralico}},\ }\href {\doibase 10.1103/PhysRevD.108.124052} {\bibfield  {journal} {\bibinfo  {journal} {Phys. Rev. D}\ }\textbf {\bibinfo {volume} {108}},\ \bibinfo {pages} {124052} (\bibinfo {year} {2023})},\ \Eprint {http://arxiv.org/abs/2309.14925} {arXiv:2309.14925 [gr-qc]} \BibitemShut {NoStop}%
\bibitem [{\citenamefont {Owen}\ and\ \citenamefont {Sathyaprakash}(1999)}]{Owen:1998dk}%
  \BibitemOpen
  \bibfield  {author} {\bibinfo {author} {\bibfnamefont {B.~J.}\ \bibnamefont {Owen}}\ and\ \bibinfo {author} {\bibfnamefont {B.~S.}\ \bibnamefont {Sathyaprakash}},\ }\href {\doibase 10.1103/PhysRevD.60.022002} {\bibfield  {journal} {\bibinfo  {journal} {Phys. Rev. D}\ }\textbf {\bibinfo {volume} {60}},\ \bibinfo {pages} {022002} (\bibinfo {year} {1999})},\ \Eprint {http://arxiv.org/abs/gr-qc/9808076} {arXiv:gr-qc/9808076} \BibitemShut {NoStop}%
\bibitem [{\citenamefont {Blanchet}\ and\ \citenamefont {Schaefer}(1993)}]{Blanchet:1993ec}%
  \BibitemOpen
  \bibfield  {author} {\bibinfo {author} {\bibfnamefont {L.}~\bibnamefont {Blanchet}}\ and\ \bibinfo {author} {\bibfnamefont {G.}~\bibnamefont {Schaefer}},\ }\href {\doibase 10.1088/0264-9381/10/12/026} {\bibfield  {journal} {\bibinfo  {journal} {Class. Quant. Grav.}\ }\textbf {\bibinfo {volume} {10}},\ \bibinfo {pages} {2699} (\bibinfo {year} {1993})}\BibitemShut {NoStop}%
\bibitem [{\citenamefont {Bohnenblust}\ \emph {et~al.}(2025)\citenamefont {Bohnenblust}, \citenamefont {Ita}, \citenamefont {Kraus},\ and\ \citenamefont {Schlenk}}]{Bohnenblust:2025gir}%
  \BibitemOpen
  \bibfield  {author} {\bibinfo {author} {\bibfnamefont {L.}~\bibnamefont {Bohnenblust}}, \bibinfo {author} {\bibfnamefont {H.}~\bibnamefont {Ita}}, \bibinfo {author} {\bibfnamefont {M.}~\bibnamefont {Kraus}}, \ and\ \bibinfo {author} {\bibfnamefont {J.}~\bibnamefont {Schlenk}},\ }\href@noop {} {\  (\bibinfo {year} {2025})},\ \Eprint {http://arxiv.org/abs/2505.15724} {arXiv:2505.15724 [hep-th]} \BibitemShut {NoStop}%
\bibitem [{\citenamefont {Heissenberg}\ and\ \citenamefont {Russo}(2025)}]{Heissenberg:2025fcr}%
  \BibitemOpen
  \bibfield  {author} {\bibinfo {author} {\bibfnamefont {C.}~\bibnamefont {Heissenberg}}\ and\ \bibinfo {author} {\bibfnamefont {R.}~\bibnamefont {Russo}},\ }\href@noop {} {\  (\bibinfo {year} {2025})},\ \Eprint {http://arxiv.org/abs/2511.13835} {arXiv:2511.13835 [hep-th]} \BibitemShut {NoStop}%
\bibitem [{\citenamefont {Elkhidir}\ \emph {et~al.}(2025)\citenamefont {Elkhidir}, \citenamefont {O'Connell},\ and\ \citenamefont {Roiban}}]{Elkhidir:2024izo}%
  \BibitemOpen
  \bibfield  {author} {\bibinfo {author} {\bibfnamefont {A.}~\bibnamefont {Elkhidir}}, \bibinfo {author} {\bibfnamefont {D.}~\bibnamefont {O'Connell}}, \ and\ \bibinfo {author} {\bibfnamefont {R.}~\bibnamefont {Roiban}},\ }\href {\doibase 10.1103/fxdk-5qwc} {\bibfield  {journal} {\bibinfo  {journal} {Phys. Rev. Lett.}\ }\textbf {\bibinfo {volume} {135}},\ \bibinfo {pages} {151601} (\bibinfo {year} {2025})},\ \Eprint {http://arxiv.org/abs/2408.15961} {arXiv:2408.15961 [hep-th]} \BibitemShut {NoStop}%
\bibitem [{\citenamefont {Brunello}\ \emph {et~al.}(2025)\citenamefont {Brunello}, \citenamefont {Chestnov}, \citenamefont {Crisanti}, \citenamefont {Giroux},\ and\ \citenamefont {Smith}}]{Brunello:2025cot}%
  \BibitemOpen
  \bibfield  {author} {\bibinfo {author} {\bibfnamefont {G.}~\bibnamefont {Brunello}}, \bibinfo {author} {\bibfnamefont {V.}~\bibnamefont {Chestnov}}, \bibinfo {author} {\bibfnamefont {G.}~\bibnamefont {Crisanti}}, \bibinfo {author} {\bibfnamefont {M.}~\bibnamefont {Giroux}}, \ and\ \bibinfo {author} {\bibfnamefont {S.}~\bibnamefont {Smith}},\ }\href@noop {} {\  (\bibinfo {year} {2025})},\ \Eprint {http://arxiv.org/abs/2510.26874} {arXiv:2510.26874 [hep-th]} \BibitemShut {NoStop}%
\bibitem [{\citenamefont {Bini}\ and\ \citenamefont {Damour}(2024)}]{Bini:2024tft}%
  \BibitemOpen
  \bibfield  {author} {\bibinfo {author} {\bibfnamefont {D.}~\bibnamefont {Bini}}\ and\ \bibinfo {author} {\bibfnamefont {T.}~\bibnamefont {Damour}},\ }\href {\doibase 10.1103/PhysRevD.110.064005} {\bibfield  {journal} {\bibinfo  {journal} {Phys. Rev. D}\ }\textbf {\bibinfo {volume} {110}},\ \bibinfo {pages} {064005} (\bibinfo {year} {2024})},\ \Eprint {http://arxiv.org/abs/2406.04878} {arXiv:2406.04878 [gr-qc]} \BibitemShut {NoStop}%
\bibitem [{\citenamefont {Lebedev}(1976)}]{Lebedev:1976}%
  \BibitemOpen
  \bibfield  {author} {\bibinfo {author} {\bibfnamefont {V.}~\bibnamefont {Lebedev}},\ }\href {\doibase https://doi.org/10.1016/0041-5553(76)90100-2} {\bibfield  {journal} {\bibinfo  {journal} {USSR Computational Mathematics and Mathematical Physics}\ }\textbf {\bibinfo {volume} {16}},\ \bibinfo {pages} {10} (\bibinfo {year} {1976})}\BibitemShut {NoStop}%
\bibitem [{\citenamefont {Kovacs}\ and\ \citenamefont {Thorne}(1978)}]{Kovacs:1978eu}%
  \BibitemOpen
  \bibfield  {author} {\bibinfo {author} {\bibfnamefont {S.~J.}\ \bibnamefont {Kovacs}}\ and\ \bibinfo {author} {\bibfnamefont {K.~S.}\ \bibnamefont {Thorne}},\ }\href {\doibase 10.1086/156350} {\bibfield  {journal} {\bibinfo  {journal} {Astrophys. J.}\ }\textbf {\bibinfo {volume} {224}},\ \bibinfo {pages} {62} (\bibinfo {year} {1978})}\BibitemShut {NoStop}%
\bibitem [{\citenamefont {Besier}\ \emph {et~al.}(2020)\citenamefont {Besier}, \citenamefont {Wasser},\ and\ \citenamefont {Weinzierl}}]{Besier:2019kco}%
  \BibitemOpen
  \bibfield  {author} {\bibinfo {author} {\bibfnamefont {M.}~\bibnamefont {Besier}}, \bibinfo {author} {\bibfnamefont {P.}~\bibnamefont {Wasser}}, \ and\ \bibinfo {author} {\bibfnamefont {S.}~\bibnamefont {Weinzierl}},\ }\href {\doibase 10.1016/j.cpc.2020.107197} {\bibfield  {journal} {\bibinfo  {journal} {Comput. Phys. Commun.}\ }\textbf {\bibinfo {volume} {253}},\ \bibinfo {pages} {107197} (\bibinfo {year} {2020})},\ \Eprint {http://arxiv.org/abs/1910.13251} {arXiv:1910.13251 [cs.MS]} \BibitemShut {NoStop}%
\bibitem [{\citenamefont {Riva}\ and\ \citenamefont {Vernizzi}(2021)}]{Riva:2021vnj}%
  \BibitemOpen
  \bibfield  {author} {\bibinfo {author} {\bibfnamefont {M.~M.}\ \bibnamefont {Riva}}\ and\ \bibinfo {author} {\bibfnamefont {F.}~\bibnamefont {Vernizzi}},\ }\href {\doibase 10.1007/JHEP11(2021)228} {\bibfield  {journal} {\bibinfo  {journal} {JHEP}\ }\textbf {\bibinfo {volume} {11}},\ \bibinfo {pages} {228} (\bibinfo {year} {2021})},\ \Eprint {http://arxiv.org/abs/2110.10140} {arXiv:2110.10140 [hep-th]} \BibitemShut {NoStop}%
\end{thebibliography}%

\onecolumngrid
\newpage
\appendix
\section{The LO Waveform}
\label{app:treewaveform}
\begin{figure}[!t]
    \centering
\includegraphicsbox[scale=1.5]{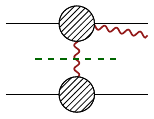} \>
\includegraphicsbox[scale=1.5]{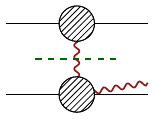} \>
    \caption{The singularities in the $q_i^2$ channels are 
    probed by generalized unitarity cuts in complex 
    kinematics. At tree-level we have contributions only
    from the one-graviton exchange.
    \label{fig:TL_cuts}}
\end{figure}

\begin{figure}[!t]
    \centering
    \includegraphicsbox[width=0.4\textwidth]{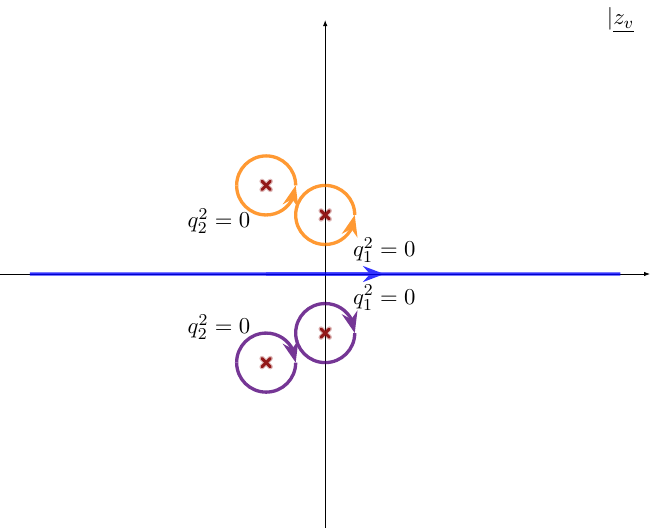}\>
    \caption{The analytic structure of the $z_v$ integrals 
    appearing in the LO waveform. In general, each integral 
    has four poles corresponding to the configurations in 
    which $q_1^2=0$ and $q_2^2=0$. We can apply Cauchy's 
    residue theorem to deform the integration contour in the
    upper- or lower-half plane in order to have it wrap around
    the poles.  This simplifies the computation to 
    a sum of residues.}
    \label{fig:complex_zv_tree}
\end{figure}

Here we present an analytic derivation of the LO
gravitational waveform in the frequency domain. 
The derivation illustrates the main features of the framework 
that underpins our one-loop computation. 
At this order, the waveform receives contributions only
from tree-level amplitudes.  The result was originally 
obtained in the time domain by Kovacs and 
Thorne~\cite{Kovacs:1978eu},
and more recently using 
WQFT~\cite{Jakobsen:2021smu,Mougiakakos:2021ckm,%
Jakobsen:2021lvp} and QFT 
techniques~\cite{DeAngelis:2023lvf,Brandhuber:2023hhl,%
Aoude:2023dui}.
\paragraph{Integrand and Generalized Unitarity.}
We can construct the integrand for the tree-level amplitude 
using generalized unitarity cuts in the $q_i^2$ channels.  In
this case each of these cuts correspond to a single on-shell graviton 
exchange, as shown in Fig.~\ref{fig:TL_cuts},
\begin{equation}
    \Integrand^{(0)} =  
    \operatorname{Cut}_{[q_2]}\bigl[
        \ampl_4^{(0)}( \bp_1;-q_2, k )  \otimes_1\ampl_3^{(0)}( \bp_2;q_2)
        \bigr]\, \cup \,
    \operatorname{Cut}_{[q_1]}\bigl[
        \ampl_3^{(0)}( \bp_1;q_1) \otimes_1 \ampl_4^{(0)}( \bp_2;-q_1,k) 
        \bigr]\,.
    \label{TreeLevelIntegrand}
\end{equation}
The symbol $\otimes_i$ denotes the sewing of $i$ internal graviton lines, 
and $\operatorname{Cut}_{[j]}$ indicates the generalized unitarity cut in the $[j]$ legs.
The symbol $\cup$ stands for the merging the contributions of the two factorization channels; this is necessary to 
avoid double counting contributions which have non-vanishing residues at both $q_1^2=0$ and $q_2^2=0$.
The three-point amplitudes we need to build the integrand are,
\begin{equation}
    \ampl_3^{(0)}( p_i; k ) = 
    \kappa (\varepsilon_k \cdot p_i)^2\,,\qquad 
    \ampl_3^{(0)}( k_1, k_2, k_3) = 
    \kappa (\varepsilon_{1}\cdot \varepsilon_{2}\,
    \varepsilon_{3}\cdot k_1 
    + \varepsilon_{2}\cdot \varepsilon_{3}\,
    \varepsilon_{1}\cdot k_2
    + \varepsilon_{3}\cdot \varepsilon_{1}\, 
    \varepsilon_{2}\cdot k_3)^2\,,
\end{equation}
written in a double-copy form.
From these amplitudes we can build the 
quantum Compton amplitude, 
\begin{equation}
    \label{eq:tree_quantum_compton}
    \ampl_4^{(0)}( p_i \to p^\prime_i, k_1, k_2 ) = 
    \kappa^2 
    \frac{( p_i \cdot F_{1} \cdot F_{2} \cdot p_i 
    + p_i^\prime \cdot F_{1} \cdot F_{2} \cdot p_i^\prime )^2}
    {(2 k_1 \cdot k_2) (2 p_i \cdot k_1) 
    (2 p_i \cdot k_2)}\,,
\end{equation}
where the vector-field strength tensor is
$F_{i}^{\mu\nu} = k_i^\mu \varepsilon_{i}^\nu 
- k_i^\nu \varepsilon_{i}^\mu$ and the Riemann tensor
can be built out of it using a double copy:
$R_{i}^{\mu \nu \rho \sigma} = F_{i}^{\mu\nu} F_{i}^{\rho \sigma}$. We expand the Compton amplitude
in the heavy-mass limit,
\begin{equation}
    \label{eq:tree_heavy_compton}
    \ampl_4^{(0)}( \bp_i ; k_1 k_2 ) = 
    \kappa^2 \left[ i \pi \bM_i^3 \delta(\bu_i \cdot k_1) (\varepsilon_{k_1} \cdot \bu_i)^2 (\varepsilon_{k_2} \cdot \bu_i)^2 - \bM_i^2 \frac{(\bu_i \cdot F_{1} \cdot F_{2} \cdot \bu_i )^2}{(k_1 \cdot \bu_i)^2 q^2} \right] \ ,
\end{equation}
where as in Eq.~\eqref{BarredMomentum} we have expanded around 
$\bp_i^\mu = \bM_i \bu_i^\mu = p_i^\mu + \frac{q^\mu}{2}$, 
with $q^\mu = k_1^\mu + k_2^\mu$, $\bu_i^2 = 1$ and 
$\bu_i \cdot q = 0$, as prescribed by the LIPS 
measure~\eqref{eq:on_shell_measure}.
After summing over the internal graviton helicities using
Eq.~\eqref{eq:graviton_sum} and combining the various cuts,
the integrand is,  
\begin{equation}
\begin{aligned}
    \Integrand^{(0)} =  \frac{m_1^2 m_2^2 \kappa^3}
    { q_1^2 q_2^2 w_1^2}\biggl[& (q_1\cdot F_k \cdot u_1)^2 
    \bigl(\gamma^2-\frac{1}{D_s-2}\bigr)
    +\frac{q_1\cdot F_k \cdot u_1 u_1\cdot F_k \cdot u_2}{w_2} 
    \left(-\gamma (q_2^2 \gamma +2 w_1 w_2)+\frac{1}{D_s-2} q_2^2\right) \\
   & +\frac{(u_1\cdot F_k \cdot u_2)^2}{4 w_2^2} 
   \left((q_2^2 \gamma+2 w_1 w_2)^2
   -\frac{1}{D_s-2} q_2^4\right)\biggr]\,, 
\end{aligned}
\end{equation}
where $D_s-2$ is the number of graviton spin degrees of
freedom in the chosen variant of
dimensional regularization.
\paragraph{Integrand reduction}
The frequency space LO waveform~\eqref{eq:frequencydomain}
is given by the Fourier transform of the tree-level
integrand~\eqref{TreeLevelIntegrand},
\begin{equation}
     \cW_h^{(0)}(\omega,\unitn) = 
    \kappa \frac{\, e^{i \omega n\cdot b_2}}
     {4\pi  r }
     \int_{\hat{q}}\hdelta(2m_1 u_1\cdot q)
     \hdelta(2m_2 u_2\cdot (k-q)) \,e^{i b\cdot q}\,\Integrand^{(0)} \,. 
    \label{eq:integrand_tree}
\end{equation}
Here, $n^\mu=(1,\unitn)$.
We want to treat this object as a generalized one-loop 
integral. We can decompose the `tensor' numerators of the 
form $q_1\cdot F_k \cdot u_1$ in terms of form factors. 
The required form factor can be obtained from the generic 
rank-$N$ relation presented in Appendix~\ref{app:tensor}. 
The scalar integrals present in the form factors then belong 
to a single generalized integral family, of the form,
\def\tinvariant#1{D_{t:#1}}
\begin{equation}
    I_{a_1 1 1 a_4 a_{5}} = 
    \int_{\hat{q}}
    \ \frac{e^{\tinvariant1}\hdelta(\tinvariant2)
    \hdelta(\tinvariant3) \tinvariant1^{-a_1}}
           {\tinvariant4^{a_4}\tinvariant5^{a_5}}\,.
\end{equation}
The invariants $\tinvariant{i}$ are defined as follows,
\begin{equation}
     \tinvariant1  = i\,q\cdot b\,,\quad 
     \tinvariant2 = \bu_1 \cdot q\,,\quad
     \tinvariant3 =  \bu_2 \cdot (k-q)\,,\quad 
     \tinvariant4 =   q^2\,,\quad
     \tinvariant5  =   (q-k)^2\,.
\end{equation}
Just as for the integrals discussed in main text,
relations between these integrals arise from
generalized IBP relations, here effectively
in $q$-space,
 \begin{equation}
 \int_{\hat{q}} \frac{\partial}{\partial q^\mu }
 \biggl(\frac{v^\mu\,e^{\tinvariant1}}{\prod_{i=1}^{5} \tinvariant{i}^{a_i}}\biggr)= 0\,. 
\end{equation} 
Here the relations allow us to rewrite the LO waveform
in terms of just six Fourier MIs, of which three are
base integrals and three are `derivative' integrals.
(As in the text, the latter may be obtained from the former
by taking a derivative with respect to the impact parameter $b$.)  The basis integrals
are,
\begin{equation}
    J_1= I_{0 1 1 1 0}\,,\quad
    J_2=\derivlabel{I_{-1 1 1 1 0}}\,,\quad
    J_3=I_{0 1 1 0 1}\,,\quad
    J_4=\derivlabel{I_{-1 1 1 0 1}}\,,\quad
    J_5=I_{0 1 1 1 1}\,,\quad
    J_6=\derivlabel{I_{-1 1 1 1 1}}\,. 
\end{equation}
The waveform can be written as a linear combination of 
these integrals,
\begin{align}
    \cW_h^{(0)}(\omega,\unitn)\;  = \; 
    \sum_{i=1}^6    \; c_i\;  J_i \ , 
    \label{eq:waveform_tree}
\end{align}
where the coefficients $c_i$ are rational functions of the 
kinematic invariants; we give their expressions
in the ancillary file \texttt{coefficients\_tree.m}.
\paragraph{Integration.} 
Analytic expressions for the MIs can be obtained from the 
general results in Appendix~\ref{app:Fouriertransform},
obtained via Schwinger parametrization,
Eqs.~\eqref{eq:FTqq}, 
by setting $\alpha_1 = \alpha_2 = 1\,$. 
For completeness, we also present expressions for the MIs
in $D=4$ obtained using the contour deformation method
explained in Appendix~\ref{app:Fouriertransform}. 
In this case, the complex structure in the $z_v$ plane is 
simpler. The integrands contain only poles corresponding to
$q_1^2=0$ and $q_2^2=0$~\cite{DeAngelis:2023lvf}, 
as shown in Fig.~\ref{fig:complex_zv_tree}.
Performing the $z_v$ integration by contour deformation 
we obtain, 
\begin{equation}
\begin{aligned}
    J_1& = -\frac{1}{4 \pi\  \bM_1 \bM_2 \sqrt{\gamma^2-1}} 
    \int_{-\infty}^{+\infty} \! \dd z_b \; 
    e^{-i \normb\,z_b}\,\frac{1}{\sqrt{z_b^2+\hat{w}_2^2}}\,,
    \\
    J_3& = -\frac{1}{4 \pi\  \bM_1 \bM_2 \sqrt{\gamma^2-1}} 
    \int_{-\infty}^{+\infty} \! \dd z_b \; 
    e^{-i \normb\,z_b}\, \frac{1}{\sqrt{\left({b\cdot k}/
    {\normb} +z_b\right)^2+\hat{w}_1^2}}\,, \\
    J_5& = \frac{1}{4 \pi\  \bM_1 \bM_2 \sqrt{\gamma^2-1}} 
    \int_{-\infty}^{+\infty} \! \dd z_b \; 
    e^{-i \normb\,z_b}\,
    \frac{{\bigl[\left({b \cdot k}/{\normb}
                   +z_b\right)^2+\hat{w}_1^2\bigr]^{-1/2}}
    +{\bigl[z_b^2+\hat{w}_2^2\bigr]^{-1/2}}}
    {\left({b \cdot k}/{\normb}\right)^2
    +\left[\sqrt{z_b^2+\hat{w}_2^2}
    +\sqrt{\left({b \cdot k}/{\normb} 
    +z_b\right)^2+\hat{w}_1^2}\right]^2} \,.
    \label{eq:integrals_tree_num}
\end{aligned}
\end{equation}
The $z_b$ integrals can then be evaluated numerically after 
contour deformation.

\paragraph{Time domain.}
Starting from the frequency-space waveform,
we can obtain the time-domain one as in Eq.~\eqref{eq:timedomain}. 
It suffices to rescale $z_b$ and re-express $k^\mu$,
\begin{equation}
     z_b \to \omega\,\bar{z}_b\,,\qquad 
     k^\mu \to \omega \, n^\mu \,,
\end{equation}
to see that the overall dependence of the amplitude is
$\cW_h^{(0)}(\omega,\unitn)\sim \omega^0$. As a result,
we can combine the two exponentials to form a delta function, 
rendering the transformation to the time domain trivial,
\begin{align}
    J_i(t) & = \int \hd\omega\, e^{-i \omega u} J_i(w) = 
    f_i(\bar{z}_b)\big\vert_{\bar{z}_b\to -{u}/{\sqrt{-b^2}}}
    \,, 
\end{align}
with $u=t-r$ the retarded time, just as in the main text.

\section{The One-Loop Gravitational Compton Amplitude}
\label{app:compton}
\begin{figure}[!ht]
    \centering
    \includegraphicsbox[scale=1.5]{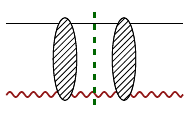} \>
    \includegraphicsbox[scale=1.5]{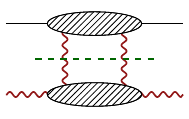} \>
     \caption{
        Generalized unitarity cuts in the $(\bp,k_1)$ 
     and $q$ channels used to reconstruct 
     the one-loop gravitational Compton integrand; 
     the $(\bp,k_2)$ cut can be obtained by exchanging 
     $k_1\leftrightarrow k_2$.}
    \label{fig:1loop_compton}
\end{figure}

The one-loop gravitational Compton integrand can be reconstructed
from its generalized unitarity cuts (see Fig.~\ref{fig:1loop_compton}) 
in the $(\bp_i, k_1)$ channel 
and in the $q$ channel (the expression in the
$(\bp_i, k_2)$ channel can be obtained by exchanging
$k_1\leftrightarrow k_2$), as:
\begin{equation}
    \ampl_4^{(1)} ( \bp_i ; k_1, k_2) = 
    \cC_{\bp_i,k_1}\cup \cC_{\bp_i,k_2}\cup\cC_{q} \, ,
\end{equation}
where $\cC_i$ denotes the contribution from the cut in the $i$ channel,
and the symbol $\cup$ stands for the merging of the contributions.
The ingredients are the tree-level Compton amplitude, 
in Eq.~\eqref{eq:tree_quantum_compton},
and the four-graviton amplitude:
\begin{equation}
    \ampl_4^{(0)}( k_1, k_2, k_3, k_4 ) = 
    \kappa^2 
    \frac{\left( F_{1} \cdot F_{2} 
    \cdot F_{3} \cdot F_{4} 
    - {F_{1} \cdot F_{2} \, F_{3} \cdot F_{4}}/{4} 
    + (3 \leftrightarrow 4) 
    + (2 \leftrightarrow 3)\right)^2}
    {(2 k_1 \cdot k_2) (2 k_2 \cdot k_3) 
    (2 k_1 \cdot k_3)}\,.
\end{equation}
The $(\bp_i, k_1)$ cut is obtained by gluing two tree-level 
Compton amplitudes at order $\bM_i^2$:
\def\loint#1{I^{0:#1}}
\begin{equation}
\begin{split}
    \ampl_4^{(1)} ( \bp_i ; k_1, k_2) \Big|_{(\bp_i, k_1)} = 
    \kappa^4 \bM_i^3 & 
    \biggl[ - \left( 1 - \frac{1}{D_s-2} \right) 
    \frac{\bu_i \cdot F_{1, \mu_1}\, \bu_i 
    \cdot F_{1, \mu_2}\, \bu_i \cdot F_{2, \mu_3}\, 
    \bu_i \cdot F_{2, \mu_4}}
    {(\bu_i \cdot k_1)^4}
    \loint{\mu_1\mu_2\mu_3\mu_4}_{1,1,1,1}\Big|_{(\bp_i, k_1)} 
    \\
    &\hphantom{\biggl[}+ 2 \bu_i \cdot F_{1} \cdot F_{2} 
    \cdot \bu_i \, \frac{\bu_i \cdot F_{1, \mu_1}\,
    \bu_i \cdot F_{1, \mu_2}}
    {(\bu_i \cdot k_1)^2} 
    \loint{\mu_1 \mu_2}_{1,1,1,1}\Big|_{(\bp_i, k_1)} \\
    &\hphantom{\biggl[}
    - \left(\bu_i \cdot F_{1} \cdot F_{2} \cdot \bu_i\right)^2 
    \loint{}_{1,1,1,1}\Big|_{(\bp_i, k_1)} \biggr]\,,
\end{split}
\end{equation}
where `$\big|_{(\bp_i, k_1)}$' denotes the cut in the
${(\bp_i, k_1)}$ channel, and,
\begin{equation}
    \loint{\mu_1 \dots \mu_n}_{1,1,1,1} = 
    \int_{\hat{\ell}} \ell^{\mu_1} \cdots \ell^{\mu_n} 
    \frac{i \pi \delta((\ell+k_1) \cdot \bu_i)}
    {(k_1+\ell)^2 \ell^2 (\ell-k_2)^2}\,.
\end{equation}
\def\lointq#1{I^{1:#1}}
\def\lointqq#1{I^{2:#1}}
The $q$-channel cut is obtained instead by sewing the 
Compton amplitude at order $\bM_i^3$ with the four-graviton 
amplitude, and is given by:
\begin{equation}
    \begin{split}
       \ampl_4^{(1)} ( \bp_i ; k_1, k_2) \Big|_{q} = 
       \kappa^4 \bM_i^3 
       & 
    \biggl\{ 
       -\left(1-\frac{1}{D_s-2}\right)\frac{4F_{1\, \mu_1}\cdot F_{2\, \mu_2}
       F_{1\, \mu_3}\cdot F_{2\, \mu_4}}{q^4}\lointq{\mu_1\mu_2\mu_3\mu_4}_{1,1,1,1}\Big|_{(q)} 
        \\ 
        &
        \frac{-2(\bu_i \cdot F_{2, \mu_1})\left(
            k_2 \cdot F_{1}\cdot \bu_i F_{1\, \mu_2}\cdot F_{2\, \mu_3} 
            -k_2 \cdot F_{1\, \mu_2} F_{1\, \mu_3}\cdot F_{2}\cdot \bu_i
            \right)}{(D_s-2)q^4}\lointq{\mu_1\mu_2\mu_3}_{1,1,1,1}\Big|_{(q)}
        \\
        &
        \frac{1}{4(D_2-2)q^4}\biggl[
        8 F_{1\, \mu_1}\cdot F_{2\, \mu_2} (2 (D_s-2)-5) q^2 \bu_i\cdot F_{1}\cdot F_{2}\cdot \bu_i
         -4 k_1\cdot F_{2}\cdot \bu_i k_2\cdot F_{1}\cdot \bu_i
        \\
        &
        +2 k_1\cdot \bu_i  \left(5 k_2\cdot F_{1\, \mu_1}  F_{2\, \mu_2}\cdot \bu_i-6 k_1\cdot F_{2\, \mu_1} F_{1\, \mu_2}\cdot \bu_i\right)
        \\ 
        &
        +4 k_2\cdot F_{1\, \mu_1} k_1\cdot F_{2}\cdot \bu_i \left(4  F_{1\, \mu_2}\cdot F_{2}\cdot \bu_i-3 F_{2\, \mu_2}\cdot F_{1}\cdot \bu_i\right)
        \\ 
        &
        8 q^2 \left(F_{1\, \mu_1}\cdot F_{2}\cdot \bu_i F_{1\, \mu_2}\cdot F_{2}\cdot \bu_i
        +3  F_{2\, \mu_1}\cdot F_{1}\cdot \bu_i F_{1\, \mu_2}\cdot F_{2}\cdot \bu_i
        +F_{2\, \mu_1}\cdot F_{1}\cdot \bu_iF_{2\, \mu_2}\cdot F_{1}\cdot \bu_i\right)
        \\ 
        &
        -8 k_1\cdot F_{2,\mu_1}\left(k_2\cdot F_{1}\cdot \bu_i 
        \left(F_{1\, \mu_2}\cdot F_{2}\cdot \bu_i-2 \cdot F_{2\, \mu_2}\cdot F_{1}\cdot \bu_i\right)
        +k_2\cdot F_{1,\mu_2} \bu_i\cdot F_{1}\cdot F_{2}\cdot \bu_i\right)
        \\ 
        &
        -5 q^2 F_{1\, \mu_1}\cdot \bu_i F_{2\, \mu_2}\cdot \bu_i 
        \biggr]
        \lointq{\mu_1\mu_2}_{1,1,1,1}\Big|_{(q)}
        \\
        &
        \frac{\text{Tr}[F_{1}\cdot F_{2}]}{4 (D_s-2)q^2}
        \biggl[
           2 k_1\cdot \bu_i \left(
            F_{1\, \mu_1}\cdot F_{2}\cdot \bu_i
           - F_{2\, \mu_1}\cdot F_{1}\cdot \bu_i
           \right) 
        \\ 
        & 
           +F_{1\, \mu_1}\cdot \bu_i k_1\cdot F_{2}\cdot \bu_i 
           -6F_{2 \, \mu_1}\cdot \bu_i k_2\cdot F_{1}\cdot \bu_i
        \biggr]
        \lointq{\mu_1}_{1,1,1,1}\Big|_{(q)}
        \\
        &
        \left(
            \frac{3 \text{Tr}[F_{1}\cdot F_{2}] \bu_i\cdot F_1 \cdot F_2 \cdot \bu_i}{4 (D_s-2)} 
            - (\bu_i\cdot F_1\cdot F_2\cdot\bu_i)^2 
            \right)\lointq{}_{1,1,1,1}\Big|_{(q)}
        \\ 
        & 
        + (1 \leftrightarrow 2)
        \biggr\} \, ,
    \end{split}
\end{equation}
where the integral families $\lointq{\mu_1 \dots \mu_n}_{1,1,1,1}$, $\lointqq{\mu_1 \dots \mu_n}_{1,1,1,1}$ are defined as,
\begin{equation}
    \lointq{\mu_1 \dots \mu_n}_{1,1,1,1} = 
    \int_{\hat{\ell}} \ell^{\mu_1} \cdots \ell^{\mu_n} 
    \frac{i \pi \delta(\ell \cdot \bu_i)}
    {(\ell)^2 (\ell+k_1)^2 (\ell+q)^2}\,, \qquad 
     \lointqq{\mu_1 \dots \mu_n}_{1,1,1,1} = 
    \int_{\hat{\ell}} \ell^{\mu_1} \cdots \ell^{\mu_n} 
    \frac{i \pi \delta(\ell \cdot \bu_i)}
    {(\ell)^2 (\ell+k_2)^2 (\ell+q)^2}\,.
\end{equation}
Here, $(1 \leftrightarrow 2)$ is the crossed term obtained by exchanging 
$k_1$ and $k_2$ (and their polarizations), 
wherein the loop integrals are those of the $\lointqq{\mu_1 \dots \mu_n}_{1,1,1,1}$ family.

To avoid double counting of contributions with non‑vanishing residues in more than one channel,
the three generalized cuts must be merged at the integrand level. We proceed by building
a permutation–symmetric, gauge–invariant ansatz composed of all independent tensor structures 
up to rank four, containing exactly four linearized field–strength tensors (two $F_1$ and two $F_2$),
contracted with $\{k_1,k_2,\ell,\bu_i\}$. 
The scalar denominator factors are organized into three structures, 
corresponding to different trivalent graphs,
given by:
\begin{equation}
    \cD_1 = \ell^2 (\ell+k_1)^2 (\ell+q)^2\,,\quad
    \cD_2 = \ell^2 (\ell+k_2)^2 (\ell+q)^2 \, , \quad
    \cD_3 = \ell^2  (\ell+q)^2 \, .
\end{equation}
Redundancies are eliminated using Bianchi identities, 
momentum conservation, and imposing symmetry under $k_1\leftrightarrow k_2$ exchange.
The remaining undetermined coefficients—rational functions of the external
kinematic invariants—are then fixed by imposing the simultaneous matching of the
$(\bar{p},k_1)$, $(\bar{p},k_2)$ and $q$-channel cuts, yielding a unique non‑redundant integrand.

The final compact expression for the one-loop Compton amplitude 
in the classical limit is provided in the ancillary file \texttt{compton.m}.
\section{Merging Generalized Cuts for the One-Loop Gravitational Waveform}
\label{app:integrand}
\begin{figure}[!t]
    \centering
    \includegraphicsbox[width=0.2\textwidth]{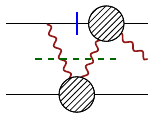}\>
    \includegraphicsbox[width=0.2\textwidth]{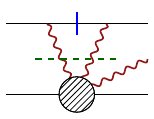}\>
    \includegraphicsbox[width=0.2\textwidth]{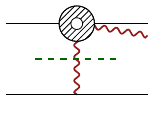}\>
    \caption{
        The three generalized unitarity cuts yielding the $\bM_1^3 \bM_2^2$ 
        contribution to the one-loop waveform integrand. Blobs denote tree-level amplitudes 
        in the heavy-mass expansion; the one-loop classical Compton subdiagram 
        contains a single localized (delta-function) matter propagator.  }
    \label{fig:cut_classical}
\end{figure}

The one-loop integrand for the gravitational waveform
is obtained by consistently merging the generalized unitarity
cut contributions shown in Fig.~\ref{fig:1loop_cuts}, together
with the analogous terms generated by exchanging the two
massive lines.
The integrand is given by,
\begin{equation}
    \Integrand^{(1)} =  \operatorname{Cut}_{[\ell,\ell-q_2]}\bigl[\ampl_5^{(0)}( \bp_1;-\ell,\ell-q_2, k ) \otimes_2 
    \ampl_4^{(0)}( \bp_2;\ell,-\ell+q_2)\bigr]\,\cup \,
    \operatorname{Cut}_{[q_2]}\bigl[\ampl_4^{(1)}( \bp_1;-q_2,k)  \otimes_1 \ampl_3^{(0)}( \bp_2;q_2)\bigr] + (1 \leftrightarrow 2)\,,
\end{equation}
where $(1 \leftrightarrow 2)$ denotes the exchange of the two massive lines. 
As in Appendix~\ref{app:treewaveform},
the operator $\otimes_i$ denotes sewing across $i$ on-shell internal graviton lines, 
and $\operatorname{Cut}_{[j]}$ denotes the generalized unitarity cut in the denomnators $[j]$. 
The required building blocks are: the one-loop Compton amplitude (Appendix~\ref{app:compton}); 
the tree-level five-point amplitude~\cite{Brandhuber:2021kpo}; 
and the tree-level four-point amplitude in Eq.~\eqref{eq:tree_heavy_compton}. 
In the heavy-mass expansion, classical one-loop contributions scale as $\bM_1^3 \bM_2^2$ and $\bM_1^2 \bM_2^3$. 
The $\bM_1^3 \bM_2^2$ term (Fig.~\ref{fig:cut_classical}) expressed in terms of classical amplitudes reads:
\begin{equation}
\begin{split}
    \Integrand^{(1)}\big\vert_{\bM_1^3 \bM_2^2} &
    = \cC_1 \cup \cC_2 \cup \cC_3  \, , 
\end{split}
\end{equation}
where:
\begin{equation}
\begin{split}
     \cC_1&  
     = \operatorname{Cut}_{[\ell,\ell-q_2]}\left[\ampl_{5,\bM_1^3}^{(0)}\big( \bp_1;-\ell,\ell-q_2, k )\otimes_{2}  
    \ampl_{4,\bM_2^2}^{(0)}\big( \bp_2;\ell,-\ell+q_2)\right]\, , \\ 
   \cC_2  & =  
    \operatorname{Cut}_{[\ell,\ell+q_1]}\left[\ampl_{4,\bM_1^3}^{(0)} \big( \bp_1;-\ell,\ell+q_1)\otimes_2
    \ampl_{5,\bM_2^2}^{(0)}\big( \bp_2;\ell,-\ell-q_1,k) \right]\, , \\
   \cC_3 & = \operatorname{Cut}_{[q_2]}\left[ 
    \ampl_{4,\bM_1^3}^{(1)}\big( \bp_1;-q_2,k)\otimes_{1}\ampl_{3,\bM_2^2}^{(0)}\big( \bp_2;q_2)
    \right]\,.
\end{split}
\end{equation}
Tree-level contributions proportional to $\bM_1^3$ contain a localized matter 
propagator on line 1 and thus carry the overall factor 
$\hat{\delta}(\ell \cdot \bu_1)$.
Cut merging is required to avoid double counting terms with 
non-vanishing residues in more than one channel. 
In this case the merge is simple: using Bianchi identities 
(and trivial momentum relations) the tensor structures are rewritten into 
a basis where the overlap can be removed automatically. A convenient choice is the 
following set of six independent structures:
\begin{equation}
    (\bu_1 \cdot F_k \cdot \bu_2)^2\,,\quad
   (q_1 \cdot F_k \cdot \bu_1)^2\,,\quad
     (\ell \cdot F_k \cdot \bu_1)^2\,,\quad
     (\bu_1 \cdot F_k \cdot \bu_2)(q_1 \cdot F_k \cdot \bu_1)\,,\quad
    (\bu_1 \cdot F_k \cdot \bu_2)(\ell \cdot F_k \cdot \bu_1)\,,\quad
    (q_1 \cdot F_k \cdot \bu_1) (\ell \cdot F_k \cdot \bu_1)\,.
\end{equation}
A compact gauge-invariant representation 
of the complete one-loop waveform integrand 
is supplied in the ancillary file \texttt{integrand.m}.

We refer to Refs.~\cite{Brandhuber:2023hhy,Herderschee:2023fxh} 
for a detailed discussion of the cut-merging procedure. 
After performing a loop-momentum IBP reduction we verified that 
our merged integrand is compatible with that of
Refs.~\cite{Brandhuber:2023hhy,Herderschee:2023fxh} up to terms 
polynomial in $q_i^2$, which give only contact (impact-parameter local) 
contributions and therefore drop out after the Fourier transform.
\section{Brown--Feynman--Passarino--Veltman reduction}
\label{app:tensor}
We perform tensor reduction following the algorithm presented 
in Ref.~\cite{Anastasiou:2023koq}. The space of external 
momenta is spanned by the vectors 
$v_i^\mu \in \{ u_1^\mu, u_2^\mu, b^\mu, k^\mu \}$. 
We define the projector,
\def\proj{P}
\begin{equation}
\proj^{\mu \nu} = \sum_{i=1}^4 \hat{v}_{i}^{\mu} v_{i}^{\nu}\,,
\end{equation}
where $\hat{v}_{i}^{\mu}$ is the vector dual to $v_i^\mu$, 
\textit{i.e.\/} $v_i^\mu \hat{v}_{j \mu} = \delta_{i j}$. A 
closed form of these vectors is given in terms of the Gram 
matrix $G_{i j} = v_i \cdot v_j$ and its inverse:
\begin{equation}
    \hat{v}_i^\mu = G^{-1}_{i j} v_j^\mu\, .
\end{equation}
As we work with four-dimensional external momenta,
we have,
\begin{equation}
    \eta^{\mu \nu} = \proj^{\mu \nu} \,,
\end{equation}
when contracted with an external momentum.
For a generic tensor of rank $N$, we write,
\begin{equation}
    T^{\mu_1 \dots \mu_N} = T_{\nu_1 \dots \nu_N} \prod_{i=1}^{N} \proj^{\mu_i \nu_i} \, .
\end{equation}
In particular, we have:
\begin{equation}
    \label{eq:duals_vectors}
    \begin{split}
        \Delta \,\hat{k}^\mu & = 
        (w_2 \gamma-w_1) b^2 \bu_1^{\mu }
        +(w_1 \gamma-w_2) b^2 \bu_2^{\mu } 
        + (1-\gamma^2) b^2 k^{\mu }
        +(\gamma^2-1) b\cdot k\, b^{\mu } \,,
        \\
        \Delta \, \hat{b}^\mu & = 
        (w_1-w_2 \gamma) b\cdot k\, \bu_1^{\mu } 
        +(w_2-w_1 \gamma) b\cdot k\, \bu_2^{\mu }
        +(\gamma^2-1) b\cdot k\, k^{\mu } 
        -(w_1^2-2 w_1 w_2\gamma+w_2^2) b^{\mu }\,,
        \\
        \Delta \, \hat{\bu}_1^\mu & = 
        (w_2^2 \left(-b^2\right)-b\cdot k^2) \bu_1^{\mu } 
        + (\gamma b\cdot k^2+w_1 w_2 b^2) \bu_2^{\mu } 
        + (w_2 \gamma-w_1) b^2 k^{\mu}
        +(w_1-w_2 \gamma) b\cdot k\, b^{\mu }\, ,
        \\
        \Delta \,\hat{\bu}_2^\mu & = 
        (\gamma b\cdot k^2+w_1 w_2 b^2) \bu_1^{\mu } 
        + (w_1^2 \left(-b^2\right)-b\cdot k^2)\bu_2^{\mu } 
        +(w_1 \gamma-w_2) b^2 k^{\mu}
        +(w_2-w_1 \gamma) b\cdot k\, b^{\mu }\,,
    \end{split}
\end{equation}
where
\begin{equation}
    \Delta = \det G_{i j} = (\gamma^2-1) b\cdot k^2
    -(w_1^2-2 w_1 w_2 \gamma+w_2^2) b^2\,.
    \label{DeltaDefinition}
\end{equation}

\section{Loop Integrals}
\label{app:loopintegrals}
All one-loop integrals we need belong to the same integral 
family, characterized by a single cut on a matter line imposed 
by the classical limit. 
When the cut is imposed on $\bu_1$, the integral family is 
defined by,
\def\oinvariant#1{D_{o:#1}}
\def\loopint#1{I^{[#1]}}
\begin{equation}
    \loopint1_{1,a_2,a_3,a_4,a_5} := 
    - \int_{\hat{\ell}} 
    \frac{i \pi \delta(\oinvariant1)}
    {(\oinvariant2+\alpha \, \varepsilon)^{a_2}
    \oinvariant3^{a_3}\oinvariant4^{a_4}
    \oinvariant5^{a_5}}\,, 
\end{equation}
where the invariants are,
\begin{equation}
    \oinvariant1 =  \bu_1\cdot \ell\,,\qquad 
    \oinvariant2  = \bu_2 \cdot \ell\,,\qquad
    \oinvariant3  = \ell^2\,,\qquad
    \oinvariant4 = (\ell+q_1)^2 \,,\qquad
    \oinvariant5= (\ell-q_2)^2  \,. 
\end{equation}
The expression $1/(\oinvariant2+i\alpha \,\varepsilon)^{a_2}$
should be understood as a distribution, defined as the 
$(a_2-1)$-th derivative of the simple pole with 
an $i \varepsilon$ prescription (when $\alpha = 1$) 
or via a principal value prescription ($\alpha = 0$):
\begin{align}
    \frac{1}{(\oinvariant2+i\alpha \,\varepsilon)^{a_2} }& =
    \frac{(-1)^{n-1}}{(n-1)!} 
    \frac{\partial^{n-1}}{\partial \oinvariant2^{n-1}} 
    \biggl(\frac{1}{\oinvariant2 + i\alpha\,\varepsilon}
    \biggr)\,,
\end{align}
\def\principal{\mathcal{P}}
where for $a_2=1$, we define,
\begin{equation}
    \frac{1}{D_2 + i\alpha\,\varepsilon} = 
    \principal\biggl( \frac{1}{\oinvariant2} \biggr) 
    - i\alpha \,\pi \delta(\oinvariant2)\,.
\end{equation}
We denote the principal value 
by $\principal(1/\oinvariant2)$.
 We thus have,
\begin{equation}
    \frac{1}{(\oinvariant2 + i\alpha \,\varepsilon)^n}
    = \principal \biggl( \frac{1}{\oinvariant2^n} \biggr) 
    - i\alpha\,\pi \frac{(-1)^{n-1}}{(n-1)!} 
    \delta^{(n-1)}(\oinvariant2)\,,
\end{equation}
where \( \delta^{(r)} \) denotes the \(r^{\textrm{th}}\)
derivative of the Dirac delta distribution.

Using IBPs and the method of canonical DEs as in 
Refs.~\cite{Caron-Huot:2023vxl,Bohnenblust:2023qmy,
Brunello:2024ibk}, we find 10 MIs, whose analytic expressions 
up to the relevant order in $\eps$ is,
\begin{equation}
\begin{aligned}
    \loopint1_{1,0,0,0,1} & = 
    -\frac{w_1}{8 \pi }\left[1+(i \pi + \log (\pi ) + 2 
    - 2 \log(w_1)) \eps\right]+\Ord(\eps^2)\,,
    \\
    \loopint1_{1,0,1,1,0} & = 
    -\frac{i}{16 \sqrt{-q_1^2}}+\mathcal{O}(\epsilon)\,,
    \\
    \loopint1_{1,0,1,0,1} & = 
    \frac{-i \pi+2 \log \bigl({w_1+\sqrt{w_1^2-q_2^2}}/
                              {\sqrt{-q_2^2}}\bigr)}
    {16 \pi\sqrt{w_1^2-q_2^2}}+\mathcal{O}(\eps)\,,
    \\
    \loopint1_{1,0,1,1,1} & = 
    \frac{1}{32 \pi  (-q_1^2) w_1 \,\eps}+\frac{i \pi 
    + 2 \log ({q_2^2}/{q_1^2})-2 \log w_1+\log \pi }
    {32 \pi(-q_1^2) w_1}+\mathcal{O}(\eps)\,,
    \\
    \loopint1_{1,1,0,1,0} & = 
    \frac{\alpha}{16 \pi  \sqrt{\gamma^2-1} \,\eps}
    +\frac{-i \pi + \alpha \bigl[\log (\gamma^2-1)-2\log w_2 
    +\log (4\pi )\bigr]}
    {16 \pi  \sqrt{\gamma^2-1}}+\Ord(\eps)\,,
    \\
    \loopint1_{1,1,0,0,1} & = 
    \frac{\alpha}{16 \pi \sqrt{\gamma^2-1} \,\eps}
    +\frac{-i \pi +2 \log \bigl(\gamma+\sqrt{\gamma^2-1}\bigr) 
    + \alpha \bigl[\log(\gamma^2-1)-2\log w_1 
    +\log (4\pi )\bigr]}{16 \pi\sqrt{\gamma^2-1}}
    +\Ord(\eps)\,,
    \\
    \loopint1_{1,1,0,1,1} & = 
    \frac{1}{32 \pi  w_1 w_2 \,\eps}
    +\frac{i \pi -2\log w_2 +\log \pi
    -2 \alpha \log \bigl(\gamma+\sqrt{\gamma^2-1}\bigr)}
    {32 \pi w_1 w_2}+\Ord(\eps)\,,
    \\
    \loopint1_{1,1,1,1,0} & = 
    \frac{\alpha}{16 \pi  (-q_1^2) 
         \sqrt{\gamma^2-1}\,\eps}
    +\frac{i \pi+ \alpha \bigl[2 \log (w_2)-2 \log (-q_1^2) 
    -\log (\gamma^2-1)+\log (4\pi )\bigr]}
    {16 \pi(-q_1^2) \sqrt{\gamma^2-1}}+\Ord(\eps)\,,
    \\
    \loopint1_{1,1,1,0,1} & = 
    \frac{\alpha}{16 \pi (-q_2^2) 
      \sqrt{\gamma^2-1}\,\eps}
    \\ &\hphantom{=}\,
    +\frac{i \pi -2 \log \bigl(\gamma+\sqrt{\gamma^2-1}\bigr)
    + \alpha \bigl[2 \log (w_1)-2 \log (-q_2^2)
    -\log (\gamma^2-1) +\log (4\pi )\bigr]}{16 \pi (-q_2^2) \sqrt{\gamma^2-1}}+\Ord(\eps)\,. 
\end{aligned}
\end{equation}
The parameter $\alpha$ again specifies the $i\varepsilon$
prescription for the propagator: here, $\alpha = 0$
corresponds to the principal-value prescription for the
massive propagator, while $\alpha = - 1$ corresponds to the 
retarded prescription.

The pentagon integral through $\Ord(\eps^0)$ can be written 
as a linear combination of four-point functions given
above,
\begin{equation}
\begin{aligned}
    \loopint1_{1,1,1,1,1}  &= 
    \frac{1}{2 \bigl[(q_1^2)^2 w_1^2
    -2 q_1^2 q_2^2 w_1 w_2 \gamma
    +(q_2^2)^2 w_2^2\bigr]}
    \\ &\hspace*{10mm}
    \times\Bigl[ 2\, \bigl(
    -q_1^2 w_1 \gamma (q_1^2-q_2^2)
    +q_2^2 w_2 (q_1^2-q_2^2)-2q_1^2 w_1^2 w_2\bigr) 
    \,\loopint1_{1,0,1,1,1} \\
    &\hphantom{\times\Bigl[}\,\hspace*{10mm}
    -2\, \bigl(w_1^2 (-q_1^2+2 
       w_2^2)+w_1 w_2 \gamma (q_1^2+q_2^2)
       -q_2^2 w_2^2\bigr) \,\loopint1_{1,1,0,1,1} \\
    &\hphantom{\times\Bigl[}\,\hspace*{10mm}
    +\left((q_1^2)^2-q_1^2 \gamma^2 
                (q_1^2-q_2^2)-q_1^2 q_2^2+2 q_2^2 
                 w_2^2-2 q_1^2 w_1 w_2 \gamma\right)
                 \,\loopint1_{1,1,1,0,1} \\
    &\hphantom{\times\Bigl[}\,\hspace*{10mm}
        + \left((q_2^2)^2+q_2^2 \gamma^2 
            (q_1^2-q_2^2)-q_1^2 q_2^2+2 q_1^2 w_1^2-2 
               q_2^2 w_1 w_2 \gamma\right) 
            \,\loopint1_{1,1,1,1,0} 
          \Bigr] + \Ord(\eps) \, . 
\end{aligned}
\end{equation}

Similarly $\loopint2_{a_1,1,a_3,a_4,a_5}$ is defined 
with a matter cut on the leg carrying $\bu_2$.

\section{Fourier Transform Details}
\label{app:Fouriertransform}

In this Appendix, we provide details of our method for 
computing the Fourier transform of Feynman integrals. 
We will discuss three different methods: using contour 
deformation; using Schwinger parametrization;
and resumming the post-Newtonian expansion. 
The first method has the widest range of applicability to 
the integrals of interest. We have found the other two 
useful for some of the integrals and as a cross-check.

\begin{figure}[!t]
    \centering
    \includegraphicsbox[width=0.4\textwidth]{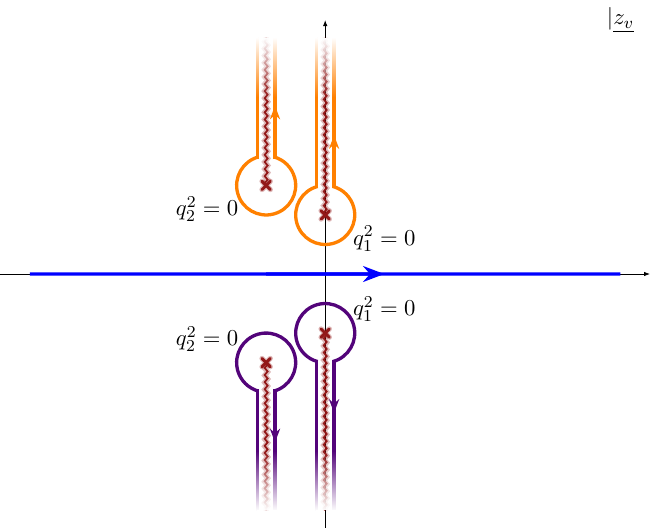}\>
    \caption{The analytic structure of the $z_v$ integrals. In 
    general each integral has four singularities corresponding 
    to the configurations in which $q_1^2=0$ and $q_2^2=0$. 
    We may choose a representation of the function so that 
    all the branch cuts are parallel to the imaginary axis. 
    We can apply Cauchy's residue theorem to deform the
    integration contour in the upper- or lower-half plane
    to encircle the singularities.}
    \label{fig:complex_zb}
\end{figure}

\paragraph{Contour deformation} We illustrate
the general approach by carrying out the Fourier transform 
in two examples:
\begin{equation}
\begin{aligned}
    f_1 (w_1,w_2,\gamma,b\cdot k, b^2)& = 
    \!-\int\! \hd^D q \, 
    \hdelta(2\bar{p}_1\cdot q)
    \hdelta(2\bar{p}_2\cdot (k-q))\,e^{i b\cdot q} 
    \frac{\log ({-q^2}/{\mu^2})}{(k-q)^2}\,,  \\
    f_2 (w_1,w_2,\gamma,b\cdot k, b^2)&= 
    \!-\int\! \hd^D q \, \hdelta(2\bar{p}_1\cdot q)
    \hdelta(2\bar{p}_2\cdot (k-q))\,e^{i b\cdot q} 
    \frac{\log ({-q^2}/{\mu^2})}{q^2}\,.
\end{aligned}
\end{equation}
All the essential features can be understood in these
examples. We begin by choosing a convenient 
parametrization for $q$ in terms of the 
external momenta~\cite{Cristofoli:2021vyo,Brunello:2024ibk},
\def\vs{{\vphantom{\mu}}}
\begin{equation}
    \label{eq:qparam}
	q^{\mu}= z_1^\vs \bu_1^{\mu} + z_2^\vs \bu_2^{\mu} 
     + z_b^\vs \frac{b^{\mu}}{\normb} 
     + z_v^\vs \tilde{v}^{\mu}_\perp\,,
\end{equation}
where $v_\perp^\mu$ is a unit vector orthogonal to $u_1^\mu$, 
$u_2^\mu$ and $b^\mu$:
\begin{align}
 v_\perp^2 = -1\,,\quad 
 v_\perp \cdot b = v_\perp \cdot u_1 = v_\perp \cdot u_2 = 0\,.
\end{align}
As we are restricting to four-dimensional kinematics, the contraction of this vector with $k^\mu$ is also simple,
\begin{equation}
    \label{eq:identity_vperp}
    (k\cdot v_\perp)^2 = - (\hat{w}_1^2 - 
    2 \hat{w}_1 \hat{w}_2 \gamma + \hat{w}_2^2) 
    + \frac{(b\cdot k)^2}{b^2} = 
    \frac{\Delta}{b^2 (\gamma^2-1)}\ ,
\end{equation}
where $\Delta$ is given in Eq.~\eqref{DeltaDefinition}, and
we define,
\begin{equation}
    \hat{w}_i = \frac{w_i}{\sqrt{\gamma^2-1}}\,.
\end{equation}
For generic $D$, the measure reads
\begin{equation}
    \label{eq:Ddim_measure}
    \hat{\dd}^D q \, \hdelta(2  \bp_1\cdot q_1) 
    \hdelta(2  \bp_2\cdot ( k - q) ) = 
    \dd^4 z\, \dd^{D-4} v_\perp 
    \frac{\sqrt{\gamma^2-1}\, z_v^{D-4}}{4(2\pi)^{D-2}
    \bM_1 \bM_2} \, \delta\left(z_1 + \gamma z_2\right)
    \delta\left(z_2 (\gamma^2-1) + w_2\right)\,,
\end{equation}
where the integration over $z_v$ is along the positive real 
axis and $\dd^{D-4} v_\perp$ is a $(D-4)$-dimensional 
spherical integral. In $D=4$, we recover the parametrization 
presented in Ref.~\cite{Cristofoli:2021vyo},
\begin{equation}
    \label{eq:vperp}
    v_\perp^\mu = \frac{\epsilon^{\mu \nu \rho \sigma} u_{1 \nu}  u_{2 \rho}  b_{\sigma}}
    {\sqrt{\gamma^2-1}\, \normb}\,,
\end{equation}
and the $z_v$ integration runs along the whole real axis. 
For the finite part of the amplitude, the integration 
in $D=4$ suffices. Here, we find,
\begin{align}
    f_{i} (w_1,w_2,\gamma,b\cdot k, b^2) &= 
    \frac{1}{16 \pi^2 \bM_1 \bM_2 \sqrt{\gamma^2-1}} 
    \!\int_{-\infty}^{+\infty} \! \dd z_b 
    \!\int_{-\infty}^{\infty}
    \!\dd z_v\;  F_{i}(z_b,z_v)\,,
\end{align}
where,
\begin{equation}
\begin{aligned}
    F_1(z_b,z_v) &= 
    \frac{e^{-i \normb z_b} \log \left(z_v^2 + z_b^2  + \hat{w}_2^2\right)}
    {(z_v + k\cdot v_{\perp})^2 
    + (z_b + {b\cdot k}/{\normb})^2  + \hat{w}_1^2}\,, 
    \\ \qquad F_2(z_b,z_v) &= \frac{e^{-i \normb z_b} 
    \log (z_v^2 + z_b^2  + \hat{w}_2^2)}
    {z_v^2 + z_b^2  + \hat{w}_2^2} \,. 
\end{aligned}
\end{equation}
In general, we expect to find four poles and four branch 
points of the integrand in the complex $z_v$ plane (see 
Fig.~\ref{fig:complex_zb}), corresponding to singularities at
$q_{1,2}^2=0$. In the $z_v$ plane, these singularities 
will be found at,
\begin{align}
    \hat{z}_1^{\pm} = 
    \pm i \sqrt{z_b^2  + \hat{w}_2^2}\,,\qquad 
    \hat{z}_2^{\pm} = -  k\cdot v_{\perp}  \pm i 
    \biggl[\biggl(z_b + \frac{b\cdot k}{\normb} \biggr)^2
    + \hat{w}_1^2 \biggr]^{1/2}\,.
\end{align}
 In the specific cases under consideration, the first 
 integrand $F_1$ has a pole at $q_2^2=0$ and a branch cut 
 starting at $q_1^2=0$, whereas the second integrand $F_2$
 has a pole overlapping with a branch cut at $q_1^2=0$. 

We apply Cauchy's theorem to deform the integration contour 
and encircle the singularities either in the upper- or 
lower-half planes (UHP and LHP, respectively). 
The contour must be considered carefully, as the integration 
along the discontinuity is divergent whenever the branch
points merge with poles. Accordingly, we need to split the 
contour into two pieces, as shown in Fig.~\ref{fig:complex_zb}:
\begin{itemize}
    \item A (small) circular contour with radius $\rho$ nearly 
    encircling the pole (and possibly branch point) and 
    ending on the two sides of the branch cut.
    \item The integration over the discontinuity from a 
    distance $\rho$ from the singularity to infinity along the
    pure imaginary direction (corresponding to our choice of
    branch cuts). 
    For this contribution we change integration variable
    from $z_v$ to $x$ via $z_v = \hat{z}_{1,2}^+ +i x+i \rho$.
\end{itemize}
If we have a pole and branch point which merge, the integrals
along the two contours are separately divergent as 
$\rho \to 0$, but the sum of the two integrals is finite. 
The discontinuity integral of interest can be written as the 
integration of a rational function on a line.
We can rewrite the integral of a rational function $R(x)$ 
along a path between two complex points $a$ and $b$ as a
contour integral of the function $R(x) \log(\frac{x-a}{x-b})$, 
with a branch cut along a path from $a$ to $b$.
We can then evaluate the integral using Cauchy's residue 
theorem,
\def\Res{\mathop{\rm Res}}
\def\DiscA{\mathop{\rm Disc}}
\begin{equation}
    \begin{aligned}
        \!\int_a^b \! \dd x\ R(x) &= 
        -\frac{1}{2\pi i}\!\int_\gamma\!\dd x \ R(x)\ \log\biggl(\frac{x-a}{x-b}\biggr) \\
        &= -\mathlarger{\sum_{i}}\, \Res_{x=z_i} R(x)\, \log\biggl(\frac{x-a}{x-b}\biggr)\,,
    \end{aligned} 
\end{equation} 
where \( \gamma \) is a positively oriented contour encircling
the branch cut from $a$ to $b$ and \( \{z_i\} \) are the poles
of $R(x)$ in the complex plane.
If $R(x)$ also contains square roots, we must first
rationalize the integrand (for this purporse, we used the 
\textsl{Mathematica\/} package \textsf{RationalizeRoots} 
\cite{Besier:2019kco}). It may happen that the discontinuity
contribution has integrable singularities along the
integration path (see for example, 
Eq.~\eqref{eq:disc_hard_triangle}). Should this happen,
we must split the integration contour, and apply the rewriting
technique piecewise.

For the first integral, the $z_v$ integration reads,
\begin{equation}
    \begin{split}
        \!\int_{-\infty}^{\infty}\! \dd z_v\; F_1(z_b,z_v) &=  
        \int_{\rotatebox{0}{$\circlearrowright$}_\rho[ 
        \hat{z}_2^+]} \dd  z_v \ F_1(z_b,z_v)
        +i  \int_0^\infty \dd x\,
        \DiscA_{\hat{z}_1^+>0}\,F_1(z_b,z_v) \\
        & = \frac{\pi  \log \biggl[\biggl(\sqrt{\left(z_b + 
        {b\cdot k}/\normb\right)^2+\hat{w}_1^2}
        +\sqrt{\vphantom{A^2}z_b^2+\hat{w}_2^2}\biggr)^2
        -{(b\cdot k)^2}/{b^2}\biggr]}{\sqrt{\left(z_b 
        + {b\cdot k}/{\normb}\right)^2+\hat{w}_1^2}}\ ,
    \end{split}
\end{equation}
while for the second integral we find,
\begin{equation}
    \begin{aligned}
        \int_{-\infty}^{\infty} \dd z_v\; F_2(z_b,z_v) & = 
        \int_{\rotatebox{0}{$\circlearrowright$}_\rho[
        \hat{z}_1^+] } \dd  z_v \, F_2(z_b,z_v)
        +i  \int_\rho^\infty \dd x\, 
        \DiscA_{\hat{z}_1>0} F_2(z_b,z_v)  \\
        & = \frac{2 \pi  \log \bigl(2 
        \sqrt{\vphantom{A^2}z_b^2+\hat{w}_2^2}\bigr)}
        {\vphantom{A^2}\sqrt{z_b^2+\hat{w}_2^2}}\,.
    \end{aligned}
\end{equation}
In the second integration we needed to keep the cut-off in 
$\rho$ explicit as the two contributions diverge separately as 
$\log\rho$.  The divergence cancels in the sum.

At this point, we are left with a one-dimensional Fourier
transform. For waveform contributions up to one-loop order,
the integrand will involve square roots and logarithms with
nested square roots. In the complex plane, the singularities
in $z_b$ are located at,
\begin{align}
    {z}_{b} = \pm i \hat{w}_2\ , \qquad z_{b} = -\frac{b \cdot k}{\normb} \pm i \hat{w}_1 \ .
\end{align}
We choose a representation of the function such that all
branch cuts are parallel to the imaginary axis. The 
integration over $z_b$ can then be performed numerically 
efficiently by contour deformation.  We deform the contour 
in the lower-half plane to encircle the singularities so that 
$e^{-i \normb z_b}$ gives an exponential suppression at large 
$|z_b|$. This allows for a numerically stable and efficient
integration strategy.

We can also aim at a fully analytic evaluation of the waveform
integrals in the time domain. Jakobsen \textit{et al.\/}
already showed at tree-level~\cite{Jakobsen:2021lvp} that the
one-dimensional Fourier transform in $z_b$ can be bypassed by
combining it with the Fourier transform from frequency 
to the time domain (for a detailed discussion, see also 
Refs.~\cite{Herderschee:2023fxh,DeAngelis:2023lvf}). 
At one-loop, such a simplification persists, though not to
the same extent. Together with the trivial delta function and
derivatives thereof, we would find an extra integration with
an extra principal-value extraction or logarithmic
`propagator', instead of the exponential factor. Then, even 
though the space of functions reduces to a familiar class 
(\textit{e.g.\/} polylogarithms and square roots), the result
becomes harder to handle analytically and numerically. Thus, we did not 
pursue this direction.

The contour-integration method is tailored for the one-loop
waveform computation as the loop integrals involve at most 
logarithms and square roots. However, it is unlikely to be an 
optimal approach for higher-order computations. 
We expect that a more unified treatment of Fourier and 
loop integrals will be needed to compute the required
integrals efficiently.

\paragraph{Schwinger parametrization\/} An alternative integration method for integrals of the form,
\begin{equation}
\label{eq:SchwingerIntegral}
    S_{\alpha_1, \alpha_2} = \!\int\! \hd^D q\; 
    \hdelta(2\bar{p}_1\cdot q)\hdelta(2\bar{p}_2\cdot (k-q)) 
    \frac{e^{i b\cdot q}}
    {(-q^2)^{\alpha_1} \left[-(k-q)^2\right]^{\alpha_2}}\,,
\end{equation}
is a generalization of the Schwinger-parametrization
approach introduced by Vernizzi~and~Riva~\cite{Riva:2021vnj}. 
We can rewrite,
\begin{equation}
    \frac{e^{i b\cdot q}}
    {(-q^2)^{\alpha_1} \left[-(k-q)^2\right]^{\alpha_2}} = 
    \frac{1}{\Gamma(\alpha_1)\Gamma(\alpha_2)}
    \int_0^{\infty} \dd x_1\; x_1^{\alpha_1-1}
    \int_0^{\infty} \dd x_2\; x_2^{\alpha_2-1}\, 
    e^{i b\cdot q + x_1 q^2 + x_2 (k-q)^2}\,.
\end{equation}
Parametrize the $q$ integration as described in the previous
subsection and notice that the $z_b$ integral is now
a simple Gaussian integration,
\begin{equation}
    \begin{split}
        S_{\alpha_1, \alpha_2} = & \frac{(2\pi)^{2-d} 
        \sqrt{\pi}}{4 \bM_1 \bM_2 \sqrt{\gamma^2-1} 
        \Gamma(\alpha_1) \Gamma(\alpha_2)} \int\! \dd^2 x \;
        \frac{x_1^{\alpha_1-1} x_2^{\alpha_2-1}}
        {\sqrt{\vphantom{A}x_1+x_2}} 
        \exp \biggl[-\hat{w}_2^2 (x_1 - x_2) 
        - 2 \hat{w}_1 \hat{w}_2 \gamma x_2 + 
        \frac{i \normb + 2 x_2 {b\cdot k}/{\normb}}
        {4(x_1 + x_2)}\biggr]\\
        &\hspace*{43mm}\times \int\! \dd z_v\, z_v^{D-4} e^{- (x_1+x_2) z_v^2} \!\int\! \dd^{D-4}v_\perp e^{-2 x_2 z_v k \cdot v_\perp}\ .
    \end{split}
\end{equation}
First, perform the angular integration
\begin{equation}
    \begin{split}
        \int\! \dd^{D-4}v_\perp\;
        e^{-2 x_2 z_v k \cdot v_\perp} &= 
        \frac{2 \pi^{({D-4})/{2}}}
        {\Gamma\bigl(\frac{D-4}{2}\bigr)} 
        \int_{-1}^{+1} \dd\cos\theta \, 
        \sin^{({D-6})/{2}}\theta\;
        \exp\biggl[2 x_2 z_v \cos\theta
        \sqrt{\frac{\Delta}{b^2 (\gamma^2-1)}}\biggr]\\
        & = 2 \pi^{({D-3})/{2}} 
        \biggl[2 x_2 z_v 
        \sqrt{\frac{\Delta}
        {2 b^2 (\gamma^2-1)}}\biggr]^{({5-D})/{2}} 
        I_{({D-5})/{2}}\biggl( 2 x_2 z_v
        \sqrt{\frac{\Delta}
        {b^2 (\gamma^2-1)}}\biggr)\\
        & \overset{D=4}{=} 2 \cosh\biggl(2 x_2 z_v \sqrt{\frac{\Delta}{b^2 (\gamma^2-1)}}\biggr)\,,
    \end{split}
\end{equation}
where $I_\alpha (x)$ is a modified Bessel function of the 
first kind. In the first line, we split the angular 
integration into two: a direction which is longitudinal to 
$k^\mu$, whose projection is fixed in
Eq.~\eqref{eq:identity_vperp}, and an additional angular 
integration which factorizes. Note that the sign in front of 
the square root in the exponent is irrelevant. As we take 
the limit $D\rightarrow4$, we recover the sum of two 
exponentials, with exponents which differ by a sign, 
\textit{i.e.\/} we recover the $z_v$ integration along the 
full real axis. The $z_v$ integration is a known integral:
\begin{equation}
    \int_0^\infty \dd z_v \, e^{-\alpha z_v^2}\, 
    z_v^{({D-3})/{2}}\, I_{({D-5})/{2}}(z_v) = 
    e^{{1}/{\alpha}} (2 \alpha)^{({3-D})/{2}}\,.
\end{equation}
The full result then takes the form,
\begin{equation}
    S_{\alpha_1,\alpha_2} = 
    \frac{(4\pi)^{1-{D}/{2}}}{4 \bM_1 \bM_2 \sqrt{\gamma^2-1} 
    \Gamma(\alpha_1) \Gamma(\alpha_2)} 
    \int\! \dd^2 x \frac{x_1^{\alpha_1-1} x_2^{\alpha_2-1}}
    {(x_1+x_2)^{{D}/{2}-1}} \,
    \exp\biggl[-\frac{x_1^2 \hat{w}_2^2 + x_2^2 \hat{w}_1^2 
    + 2 x_1 x_2 \hat{w}_1 \hat{w}_2 \gamma 
    - {b^2}/{4} - i x_2 b \cdot k}{(x_1+x_2)}\biggr]\,.
\end{equation}
Finally, we perform the change of variables suggested in 
Ref.~\cite{Riva:2021vnj}:
\begin{equation}
    x_1 = -\frac{b^2 (1-x)}{4 \lambda}\,, \qquad x_2 = -\frac{b^2 x}{4 \lambda}\,,
\end{equation}
and recognize the following integral representation of the 
modified Bessel function of the second kind:
\begin{equation}
    \int_0^\infty\!\dd\lambda\, e^{-{a}/{\lambda} - \lambda} 
    \lambda^b = 2\, a^{({b+1})/{2}} K_{-1-b}(2\sqrt{a})\,.
\end{equation}
The final result for $S$ is,
\begin{equation}
\label{eq:alpha12_final}
    S_{\alpha_1,\alpha_2} = 
    \frac{(4\pi)^{1-{D}/{2}}
    ({\normb}/{2})^{\alpha_1+\alpha_2+1-{D}/{2}}}
    {2 \bM_1 \bM_2 \sqrt{\gamma^2-1} 
    \,\Gamma(\alpha_1) \Gamma(\alpha_2)} 
    \int_0^1\! \dd x \, e^{i x b \cdot k} 
    (1-x)^{\alpha_1-1} x^{\alpha_2-1} 
    \Omega(x)^{{D}/{2}-\alpha_1 -\alpha_2-1} 
    K_{\alpha_1 + \alpha_2 + 1 -{D}/{2}} (\normb\Omega(x))\,,
\end{equation}
in which,
\begin{equation}
    \Omega(x)= \sqrt{ x^2 \hat{w}_1^2 + (1-x)^2 \hat{w}_2^2 
    + 2 (1-x) x \hat{w}_1 \hat{w}_2 \gamma}\,.
\end{equation}
We observe that most of the Fourier-loop CMIs can be written
in this form.  Indeed, the Fourier transforms of all the
loop MIs except $j^{L,(1)}_{1,0,1,0,1}$ and 
$j^{L,(1)}_{1,1,1,1,1}$ can be written in this form;
see Ref.~\cite{Brunello:2024ibk}. 
The cases $\alpha_1 = 0$ or $\alpha_2=0$ cannot be obtained
trivially from the representation~\eqref{eq:alpha12_final}, 
but a parallel yet simpler derivations gives,
\begin{equation}
\begin{aligned}
    S_{\alpha,0} &= \frac{(4\pi)^{1-{D}/{2}}
    ({2 \hat{w}_2}/{\normb})^{{D}/{2}-(\alpha+1)}}{
    2 \bM_1 \bM_2 \sqrt{\gamma^2-1}\, \Gamma(\alpha)} 
    K_{\alpha + 1 -{D}/{2}} \bigl(\normb\hat{w}_2\bigr)\,,\\
    S_{0,\alpha} &= \frac{(4\pi)^{1-{D}/{2}} 
    ({2 \hat{w}_1}/{\normb})^{{D}/{2}-(\alpha+1)}}{
    2 \bM_1 \bM_2 \sqrt{\gamma^2-1}\, 
    \Gamma(\alpha)} 
    \,e^{i b \cdot k} K_{\alpha + 1 -{D}/{2}} 
    \bigl(\normb\hat{w}_1\bigr)\,.
\end{aligned}
    \label{eq:FTqq}
\end{equation}

\paragraph{Tree-level integrals to $\cO(\epsilon)$} In this 
paragraph, we continue our consideration of the integrals in the 
family~\eqref{eq:SchwingerIntegral}. In principle, we can 
expand the results~(\ref{eq:alpha12_final},\ref{eq:FTqq}) 
around $D=4$ to obtain the $\frac{\epsilon}{\epsilon}$ 
contributions to the Fourier integrals, as the divergent 
contributions are always proportional to the tree-level 
waveform.  We can write the latter in terms of the integrals
$S_{1,0}$, $S_{0,1}$ and $S_{1,1}$, along with their 
derivatives with respect to $\normb$. On the other hand, 
it is useful to avoid the appearance of Bessel functions,
and write these terms as a one-dimensional Fourier transform
just of logarithms. We start from the $D$-dimensional
parametrization of the integration~\eqref{eq:Ddim_measure},
without assuming any particular form of the orthogonal vector
$v_\perp^\mu$ spanning the $(D-3)$-dimensional space 
orthogonal to the scattering plane. We can further split the 
angular integration,
\begin{equation}
    v_\perp^\mu = z \, \sqrt{\frac{\Delta}{-b^2 (\gamma^2-1)}}
    \,\hat{k}^\mu + \sqrt{1-z^2} \, \hat{v}_\perp^\mu\,,
\end{equation}
where $\hat{k}^\mu$ is the dual vector associated to $k^\mu$,
defined in Eq.~\eqref{eq:duals_vectors}, and 
$\hat{v}_\perp^\mu$ is a unit vector orthogonal to the 
four-dimensional physical space. The integration over the 
$(D-4)$-dimensional angle factorizes to yield,
\begin{equation}
    \int \dd^{D-5} \hat{v}_\perp = 
    \frac{2 \pi^{{D}/{2}-2}}
    {\Gamma\bigl(\frac{D}{2}-2\bigr)}\,.
\end{equation}
The full integrals are then,
\begin{equation}
    S_{\alpha_1, \alpha_2} = 
    \frac{2}{(4\pi)^{{D}/{2}} \Gamma\bigl(\frac{D}{2}-2\bigr)
    \sqrt{\gamma^2-1}\, \bM_1 \bM_2} 
    \int_{-\infty}^{+\infty}\! \dd z_b\; 
    e^{- i \normb z_b}\int_0^\infty\! \dd z_v\, z_v^{D-4}
    \int_{-1}^{+1}\! \dd z\, (1-z^2)^{{D}/2-3} 
    F_{\alpha_1, \alpha_2}(z_b,z_v,z)\,,
\end{equation}
in which,
\begin{equation}
    F_{\alpha_1, \alpha_2}(z_b,z_v,z) = 
    \frac{1}{\left(z_b^2+z_v^2+\hat{w}_2^2\right)^{\alpha_1} 
    \Bigl[(z_b + {b\cdot k}/{\normb})^2
    +z_v^2 +2 z z_v \sqrt{\frac{\Delta}{(-b^2) (\gamma^2-1)}} 
    +\frac{\Delta}{(-b^2) (\gamma^2-1)} 
    +\hat{w}_1^2\Bigr]^{\alpha_2}}\,.
\end{equation}
The $z$ integration can be computed in terms of an 
${}_2 F_1$ hypergeometric function, which we can then
expand for small $\eps$. We then need to perform 
the $z_v$ integral analytically. As we already know
the analytic results for the cases in which only one of the
two denominators appear, we can focus on 
the remaining case, $(\alpha_1,\alpha_2)=(1,1)$. 
The $z_v$ integration can be computed in terms of dilogarithms
and logarithms (possibly squared), 
using a strategy similar to the one presented above.

\paragraph{Resummed small-velocity expansion} 
We present last an additional strategy that has proven 
particularly useful in finding a compact result for the 
Fourier transform of $j^{L,\bu_1}_{1,0,1,0,1}$ in terms of 
a Struve-H function:
\begin{equation}
\begin{split}
    \tilde{S} & = e^{i b\cdot k} \int\! \hd^4 q \;
    \hdelta(2\bar{p}_1\cdot (k-q))\,
    \hdelta(2\bar{p}_2\cdot q)\, e^{- i b\cdot q} 
    \frac{2 \log \bigl(({w_1+\sqrt{\vphantom{A_1^2}w_1^2-q^2}})/{\sqrt{-q^2}}\bigr)-i \pi}
    {16 \pi\sqrt{\vphantom{A_1^2}w_1^2-q^2}} \\
    &= e^{i b\cdot k} \int\! \hd^4 q \; 
    \hdelta(2\bar{p}_1\cdot (k-q))\,
    \hdelta(2\bar{p}_2\cdot q)\, e^{- i b\cdot q} 
    \biggl[-i\mathlarger{\sum}_{n=0}^\infty\, 
    \frac{\left(\frac{1}{2}\right)_n}{16 n!\, \sqrt{-q^2}} 
    \biggl(\frac{w_1^2}{q^2}\biggr)^n 
    - \mathlarger{\sum}_{n=0}^\infty\, \frac{n!\, w_1}
    {8 \pi \left(\frac{3}{2}\right)_n q^2} 
    \biggl(\frac{w_1^2}{q^2}\biggr)^n \biggr]\,.
\end{split}
\end{equation}
In these expressions, $(a)_n$ denotes the Pochhammer
symbol, $(a)_n=\Gamma(n+a)/\Gamma(a)$.
Each term in the sums is within the integral families of 
Eq.~\eqref{eq:FTqq}:
\begin{equation}
        \tilde{S}  = 
        \frac{\sqrt{\pi} e^{i b\cdot k}}
        {4 \bM_1 \bM_2 (4\pi)^2 \normb} 
        \sum_{n=0}^\infty (1-\gamma^2)^{n} 
    \biggl(\frac{\hat{w}_1 \normb}{2}\biggr)^{n+1} 
        \biggl[ - \frac{\sqrt2 i \,K_{n-\frac{1}{2}} 
        (\hat{w}_1 \normb)}
        {n!\, \sqrt{(\gamma^2-1) \hat{w}_1 \normb}} 
        + \frac{K_{n} (\hat{w}_1 \normb)}
        {\Gamma(n+\frac{3}{2})} \biggr]\,.
\end{equation}
The first series of terms can be resummed into a simple 
exponential. In the second term in the sum, we substitute 
the following integral representation of the Bessel function,
\begin{equation}
    K_{n} (z) = \frac{\sqrt{\pi } z^n}
    {2^n \Gamma (n+\frac{1}{2})} 
    \int_1^\infty\! \dd t\, e^{-t z} 
    (t^2-1)^{n-{1}/{2}}\ .
\end{equation}
Upon exchanging the order of integration and summation,
we can recognize the expansion of a Struve function,
\begin{equation}
    \mathbf{H}_{-1} (z) = \sum_{n=0}^\infty 
    \frac{\left(\frac{z}{2}\right)^{2n}}
    {\Gamma(n+\frac{3}{2})\, \Gamma(n+\frac{1}{2})}\,.
\end{equation}
Finally, using the change of variables $t=\cosh x$, we obtain
the result,
\begin{equation}
\label{eq:hard_triangle_result_1}
    \tilde{S} = \frac{e^{i b\cdot k}}{128 \pi \bM_1 \bM_2}
    \biggl[\int_0^\infty\! \dd x \, 
    e^{-\hat{w}_1 \normb \cosh{x} } \hat{w}_1 
    \mathbf{H}_{-1} (w_1 \normb \sinh{x}) 
    - i \frac{e^{-\hat{w}_1 \normb \gamma }}
    {\normb \sqrt{\gamma ^2-1}}\biggr] \,,
\end{equation}
where the representation of the second term is particularly 
well-suited to numerical evaluation thanks to the 
double-exponential suppression for $x\gtrsim 0$.
We can compare this result with the contour deformation 
strategy. We notice that along $q^2 > 0$, the discontinuity 
of the triangle integral takes a discontinuous form,
\begin{equation}
\label{eq:disc_hard_triangle}
    \DiscA_{q^2>0} 
    \frac{2 \log\bigl(({w_1+\sqrt{\vphantom{A_1^2}w_1^2-q^2}})/
    {\sqrt{-q^2}}\bigr)-i \pi}
    {16 \pi\sqrt{w_1^2-q^2}} = \begin{cases}
        - \frac{i}{8 \sqrt{\vphantom{A_1^2}w_1^2 - q^2}}\,,
        &\quad 0 < q^2 < w_1^2 \\
        - \frac{1}{8 \sqrt{\vphantom{A_1^2}q^2 - w_1^2}}\,,
        &\quad q^2 > w_1^2
    \end{cases}\,.
\end{equation}
The integration in $z_v$ along the discontinuity of 
these expressions yields the same splitting presented in
Eq.~\eqref{eq:hard_triangle_result_1}\footnote{The $z_v$ 
integration of the second expression does not converge. 
We may, however, observe, that with the use
of any regulator (for example, $z_v^{-\rho}$ or, 
equivalently performing the Fourier integral in 
dimensional regularization), the divergent part as
we remove the regulator yields a result polynomial in $z_b$
--- ultimately integrating to a localized contribution 
in impact parameter space which we drop.}.
In particular, one obtains,
\begin{equation}
    \begin{split}
        \tilde{S} &= \frac{e^{i b \cdot k}}
        {16 \pi^2 \bM_1 \bM_2}\int_{-\infty}^{+\infty}\!\dd z_b\;
        e^{i z_b \normb} \,\biggl[\frac{1}{8} 
        \arcsin\frac{w_1}{\sqrt{\vphantom{A^2}z_b^2 + \hat{w}_1^2 \gamma^2}} 
        + \frac{i}{16}\log ( z_b^2 + \hat{w}_1^2 \gamma^2) 
        \biggr]\\
        & = \frac{e^{i b \cdot k}}
        {128 \pi \bM_1 \bM_2 \sqrt{\gamma ^2-1}} 
        \biggl[\frac{1}{\pi} \int_{-\infty}^{+\infty}\! 
        \dd z_b\; e^{i z_b \normb} 
        \arcsin\frac{w_1}{\sqrt{\vphantom{A^2}z_b^2 + \hat{w}_1^2 \gamma^2}} 
        - i \frac{e^{-\hat{w}_1 \normb \gamma }}
        {\normb}\biggr] \,,
    \end{split}
\end{equation}
where the two terms are ordered accordingly to the splitting 
in Eq.~\eqref{eq:disc_hard_triangle}.  The result matches 
the small-velocity-resummed 
result~\eqref{eq:hard_triangle_result_1} numerically.

\end{document}